\documentclass[prd,aps,
tightenlines,floats,nofootinbib,preprintnumbers,onecolumn]{revtex4}

\usepackage{graphicx}
\usepackage{amsmath,amssymb,color}
\usepackage{grffile}
\usepackage{ulem}

\newcommand{\xx}{\boldsymbol x}

\newcommand{\kk}{\boldsymbol{k}}

\newcommand{\beq}{\begin{equation}}
\newcommand{\eeq}{\end{equation}}

\newcommand{\HH}{\mathcal{H}}

\newcommand{\Mpl}{M_{\rm pl}}

\begin{document}

\title{
Gravitational wave spectra from oscillon formation after inflation }

\author{
Takashi Hiramatsu$^1$,
Evangelos I. Sfakianakis$^{2,3,4}$,
Masahide Yamaguchi$^5$
}

\affiliation{
$^1$ Department of Physics, Rikkyo University, Toshima, Tokyo 171-8501, Japan\\
$^2$ Nikhef, Science Park 105, 1098 XG Amsterdam, The Netherlands\\
$^3$ Lorentz Institute for theoretical physics, University of Leiden, 2333CA Leiden, The Netherlands\\
$^4$ Institut de F\'isica d'Altes Energies (IFAE), The Barcelona Institute of
Science and Technology (BIST), Campus UAB, 08193 Bellaterra, Barcelona
\\
$^5$ Department of Physics, Tokyo Institute of Technology, 2-12-1 Ookayama, Meguro-ku, Tokyo 152-8551, Japan
}

\preprint{Nikhef 2020-028, RUP-20-33}

\begin{abstract}
We systematically investigate the preheating behavior of single field inflation with an oscillon-supporting potential. We compute both the properties of the emitted gravitational waves as well as the number density and characteristics of the produced oscillons. 
By performing numerical
simulations for a variety of potential types, we divide the analyzed potentials in two families, each of them containing potentials with varying large- or small-field dependence. 
We find that the shape of the spectrum and the amplitude of emitted gravitational waves have a universal feature with the peak around the physical wavenumber $k/a \sim m$ at the inflaton oscillation starting period,
irrespective of the exact potential shape. 
This can be used as a smoking-gun for deducing the existence of a violent preheating phase and possible oscillon formation after inflation. Despite this apparent universality, we also find  differences in the shape of the spectrum of emitted gravitational waves between the two families of potentials, leading to discriminating features between them.
In particular, all potentials show the emergence of a two-peak structure in the gravitational wave spectrum, arising at the time of oscillon formation.
However, potentials that exhibit efficient parametric resonance tend to smear out this structure and by the end of the simulation the two-peak structure is replaced by one broad peak in the GW spectrum.
We further compute the number density and properties of the produced oscillons for each potential choice, finding  differences in the number density and size distribution of stable oscillons and transient overdensities.
We also perform a linear fluctuation analysis and use the corresponding Floquet charts to relate the results of our simulations to the structure of parametric resonance for the various potential types. We find that the growth rate of the scalar perturbations and the associated oscillon formation time are sensitive to the small-field shape of a potential while the macroscopic physical properties of oscillons such as the total number depend on the large-field shape of a potential.

\end{abstract}

\maketitle

\makeatletter
\let\toc@pre\relax
\let\toc@post\relax
\makeatother

\section{Introduction}
\label{sec:intro}

Inflation \cite{Starobinsky:1980te,Sato:1980yn,Guth:1980zm}, a period of
accelerated expansion in the very early Universe, has been receiving increasingly strong
support by several observations. Inflation makes  the Universe (almost) spatially flat, and provides a mechanism for generating not only
primordial curvature perturbations but also primordial gravitational
waves. The amplitude of gravitational waves is model-dependent and in the simplest models it reveals the energy scale of inflation.  In fact, the
observations of the cosmic microwave background (CMB) anisotropies
\cite{Bennett:1996ce,Bennett:2012zja,Akrami:2018odb} detected 
primordial curvature perturbations, being almost scale invariant and
Gaussian, and confirmed the spatial flatness of the Universe, as
predicted by inflation. Thus, even though primordial tensor perturbations and
small scale curvature perturbations generated during inflation have unfortunately not
yet been detected, the slow-roll dynamics during inflation, responsible for the CMB-relevant fluctuations, is well understood and tested.

On the other hand, the transition from the inflationary epoch to the hot
big-bang, a radiation dominated epoch, is much less known.
Reheating is needed to bring the Universe into a state filled with a thermal plasma, as required by Big Bang Nucleosynthesis.
Originally,  the reheating process was assumed to be solely controlled by a perturbative decay of the inflaton to radiative degrees of freedom.
The importance of non-linear dynamics was later recognized 
\cite{Traschen:1990sw,Kofman:1994rk,Shtanov:1994ce,Kofman:1997yn} and has  since received significant attention, both analytically and numerically (see e.g. Ref.~\cite{Amin:2014eta} for a review of preheating).

Among such non-linear dynamics, oscillons
\cite{Bogolyubsky:1976nx,Bogolyubsky:1976sc,Gleiser:1993pt,Copeland:1995fq,Kasuya:2002zs,Amin:2010jq, vanDissel:2020zje},
localized long-lived objects, are now attracting increasing attention,
partly because, some inflation models
\cite{Kallosh:2013hoa,Kallosh:2013yoa,Galante:2014ifa,Broy:2015qna}
preferred by observations and the axion potentials suggested by the
string axiverse \cite{Arvanitaki:2009fg} easily lead to oscillon
formation \cite{Amin:2011hj}\footnote{The existence conditions and the
lifetimes of oscillons are discussed e.g. in
Refs.~\cite{Kawasaki:2015vga,Amin:2013ika,Ibe:2019vyo,
Sfakianakis:2012bq}. A simple intuitive criterion is a potential that is
locally quadratic near its minimum and becomes ``flatter'' at large
field values.}.  Furthermore, the formation of such oscillons can be a
powerful source of gravitational waves. In fact, gravitational waves
might be the only tool to directly probe the dynamics and the non-linear
nature of the reheating epoch.

A lot of studies on gravitational waves emitted during the formation of
oscillons can be found in the recent literature
\cite{Zhou:2013tsa,Antusch:2016con,Liu:2017hua,Lozanov:2017hjm,Amin:2018xfe,Kitajima:2018zco,Liu:2018rrt,Lozanov:2019ylm}.
In this paper, we try to address the following question: How much do the
properties of gravitational waves emitted from the formation of
oscillons, such as the shape and the amplitude of the power spectrum,
depend on the potential of a source scalar field. If the resulting gravitational wave power spectrum
has an almost universal shape, irrespective of the details of the scalar potential, it can be
a smoking-gun for  gravitational waves associated with efficient preheating and oscillon
formation. If, on the other hand, the gravitational waves retain a memory of the inflaton potential, this spectral information will be
useful for discriminating between  different potential shapes and thus probing the inflaton potential at small field values. A similar idea was recently proposed
in Ref.~\cite{Lozanov:2017hjm}, though we systematically examine a larger variety
 of potential types and classify them based on their small-field and large-field shape.

The organization of this paper is as follows. In the next section, basic
equations to describe oscillon formation and to estimate gravitational
waves are given. In Section~\ref{sec:models}, the potentials we consider in the paper
are listed and classified. In Section IV, our numerical setup is
given. In Section~\ref{sec:results}, the results of numerical simulations on oscillon
formation and gravitational waves emitted from such processes are given
and discussed. In Section~\ref{sec:linear}, we perform a linear analysis of the system, in order
to understand the numerical results qualitatively and build physical intuition about the various contributing factors to oscillon formation. We offer our conclusions and prospects for future work in Section~\ref{sec:conc}. 

\section{Basic equations}
\label{sec:basic}

We consider a canonical scalar field coupled minimally to gravity. The relevant action is given as
%
\begin{equation}
 S_\phi = \int\!d^4x\,\sqrt{-g}\left(
  {1\over 2} M_{\rm pl}^2 R- \frac{1}{2}g^{\mu\nu}\partial_\mu\phi\partial_\nu\phi
   - V(\phi)\right),
 \label{eq:action}
\end{equation}
%
with $ds^2=a^2(-d\eta^2+dx^2)$, where $\eta$ is the conformal time, which is related to cosmic time $t$ as $dt=a\, d\eta$. The scalar field satisfies the Klein-Gordon
equation,
%
\begin{align}
 \phi'' + 2\HH\phi'
 -\triangle \phi
  = -a^2\frac{dV}{d\phi},
\end{align}
%
where $\HH=aH$ is the reduced Hubble parameter.  
Throughout this work primes represent derivatives with respect to conformal
time $\eta$, and $\triangle=\delta^{ij}\partial_i\partial_j$ is the spatial Laplacian.
To remove the first time-derivative, we redefine the field as 
$\psi = a\phi$, leading to 
\begin{align}
 \psi''
 -  \frac{a''}{a}\psi
 -\triangle\psi
 = -a^3\frac{dV}{d\phi}\, .
 \label{eq:eq_psi}
\end{align}
%
The perturbed gravitational field {$h_{ij}$ satisfying
$h_{ii}=0=\partial_i h_{ij}$} obeys the equation,
%
\begin{align}
 h''_{ij} + 2\HH h'_{ij} - \triangle h_{ij} = {\frac{2}{\Mpl^2}\Pi_{ij}^{\rm TT}},
\end{align}
%
where $\Mpl^{-2}=8\pi G$ is the reduced Planck mass, { $\Pi_{ij} \equiv T_{ij} - g_{ij}
\langle p \rangle$ is the anisotropic stress and $\langle p
\rangle$ denotes  the background homogeneous pressure. The superscript TT
represents the transverse-traceless part of the anisotropic stress tensor}. Defining $\chi_{ij}=a\,
h_{ij}$, we have
%
\begin{align}
 \chi''_{ij}  -\frac{a''}{a}\chi_{ij}
  - \triangle \chi_{ij} = {\frac{2a}{\Mpl^2}\Pi_{ij}^{\rm TT}.}
 \label{eq:eq_h}
\end{align}
%
The energy-momentum tensor of the scalar field is given as
%
\begin{align}
 T_{\mu\nu} = \partial_\mu\phi\partial_\nu\phi
-g_{\mu\nu}\left(\frac{1}{2}\partial^\lambda\phi\partial_\lambda\phi+V\right).
\label{eq:def_T}
\end{align}
%
The possible components sourcing the gravitational waves are
%
\begin{align}
 {\Pi_{ij}}
 &= \frac{1}{a^2}\partial_i\psi\partial_j\psi \, ,
\end{align}
%
whereas the other terms are dropped when we perform the transverse-traceless
projection. The details for the evaluation of the gravitational wave 
spectrum are given for completeness in Appendix~\ref{subsec:GW}.

\section{models}
\label{sec:models}

\subsection{Models of systematic survey for small-field shape dependence}
\label{subsec:mimic}

In Ref.~\cite{Amin:2011hj}, the authors studied  oscillon formation during preheating in a one-parameter family of models, in which the inflaton potential is
\begin{align}
V_A(\phi) &= 
 \frac{m^2M^2}{2\alpha_A}\left[\left(1+\frac{\phi^2}{M^2}\right)^{\alpha_A}-1\right].
 \label{eq:VA}
\end{align}
%
This potential class has the necessary feature to allow for the existence of oscillons: it is locally quadratic around the minimum and shallower than quadratic at larger field values. Furthermore, for $\alpha_A=1/2$, one recovers the well-known axion monodromy potential, which is linear at large field values, $V_A\sim m^2 M |\phi|$.

In Ref.~\cite{Zhou:2013tsa} this potential was revisited and the authors obtained a gravitational wave spectrum possessing multiple peaks,
which 
are related to the higher harmonics present in the oscillon itself.
The higher harmonics of an oscillon are related to the Taylor expansion of the scalar potential (see Ref.~\cite{Zhou:2013tsa} and references therein) and thus 
they are determined from the features of the potential
at  small-field values. 
In order for oscillons to form after inflation, parametric resonance must be efficient enough to allow for certain wavenumbers to grow enough to probe the non-linear structure of the potential.
After the oscillons form, 
the oscillon itself must be supported
by  non-linear effects. 
In particular, the effects of dispersion, that would make the oscillon dissipate into radiative modes, are counter-acted by the non-linearity of the potential. In another --albeit equivalent-- description, the frequency of the oscillon is smaller than the mass of the free particles in the theory, due to the fact that the potential is flatter than quadratic at larger field values. Hence, the oscillon does not decay into free particles, because they are held together by an interaction energy, which makes the decay kinematically suppressed (see e.g. Ref.~\cite{Hertzberg:2010yz} for quantum effects on oscillon decay).
The size of the resulting oscillons and the 
peaks of gravitational wave spectrum can be related to both the small-field
and the large-field features of the potential.

\begin{figure}[h]
 \centering
 \includegraphics[width=6cm]{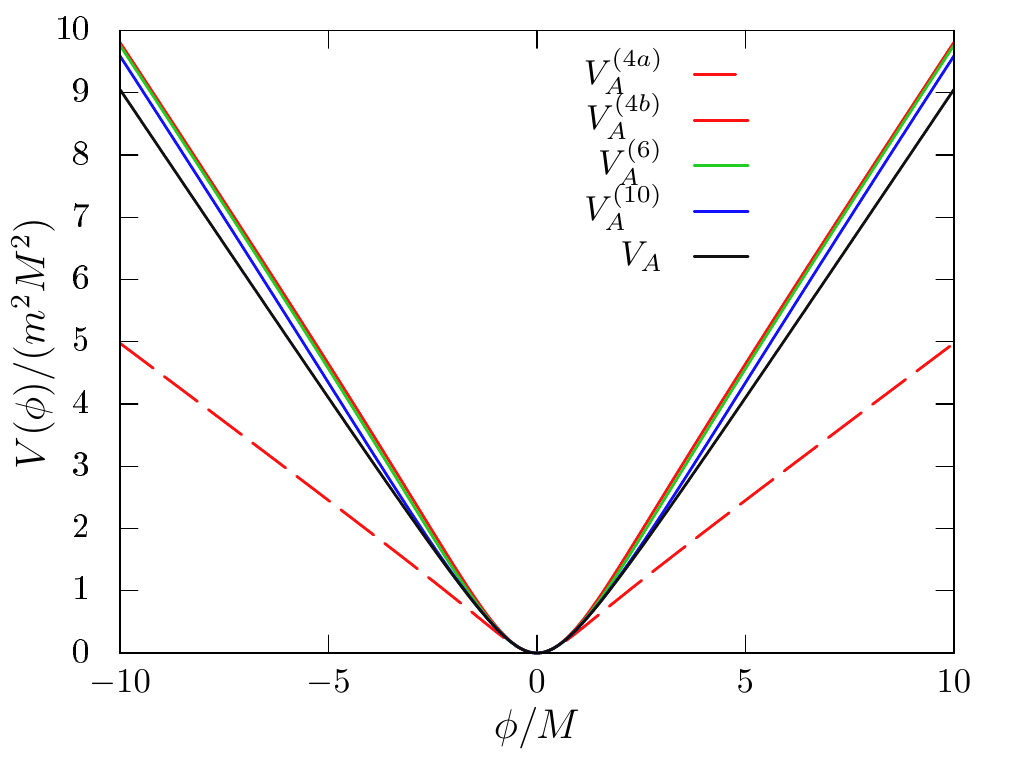}
 \includegraphics[width=6cm]{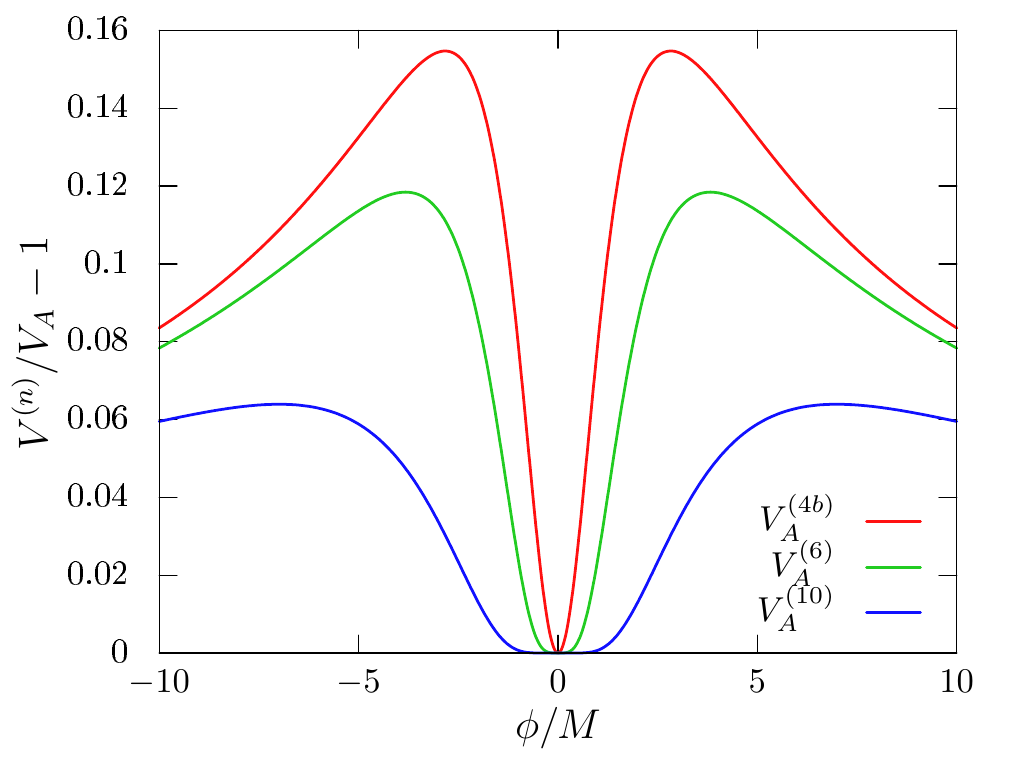}
 \caption{The left panel shows the deformed potentials defined in Eqs.~(\ref{eq:VAapp4a})-(\ref{eq:VAapp10}).
 The right panel shows the relative difference to the potential $V_A(\phi)$ of Eq.~\eqref{eq:VA}.
We do not show the relative difference of $V_A^{(4a)}$ to $V_A$, since  they have different slopes for large field values, as explained in the main text.
}
 \label{fig:VAapp}
\end{figure}

In our present study, we first investigate how the small-field shape of the scalar potential
affects  oscillon formation and GW emission. To see this, we introduce a series of model
potentials which asymptotically behave similarly to the potential $V_A(\phi)$
given in Eq.~\eqref{eq:VA}, with $\alpha_A=1/2$, namely,
$V\propto \phi$ for $\phi \gg M$ and whose structure around the origin
is deformed by approximating $V_A$ with the Pad\'e approximation
starting with the functional form, $x^2/\sqrt{1+x^2}$,
%
\begin{align}
 V_{A}^{(4a)}(\phi) &= 
 m^2M^2\frac{x^2}{2\sqrt{1+x^2}} = V_A(\phi)+\mathcal{O}(x^4)\label{eq:VAapp4a},\\
 V_{A}^{(4b)}(\phi) &= 
 m^2M^2\frac{x^2}{\sqrt{4+x^2}} = V_A(\phi)+\mathcal{O}(x^4)\label{eq:VAapp4b},\\
 V_{A}^{(6)}(\phi) &= 
 m^2M^2\frac{x^2}{\sqrt{1+x^2}}\frac{2+x^2}{4+x^2}=V_A(\phi)+\mathcal{O}(x^6)\label{eq:VAapp6},\\
 V_{A}^{(10)}(\phi) &= 
 m^2M^2\frac{x^2}{\sqrt{1+x^2}}\frac{8+8x^2+x^4}{16+12x^2+x^4}=V_A(\phi)+\mathcal{O}(x^{10})\label{eq:VAapp10},
\end{align}
%
where $x:=\phi/M$ and the number in parentheses indicates the order of the approximant.
These shapes are shown in Fig.~\ref{fig:VAapp}.
The lowest-order approximation is $V_A^{(4a)}$.
However, as shown in the left panel of Fig.~\ref{fig:VAapp}, the asymptotic shape of $V_A^{(4a)}$ 
is far from that of $V_A(\phi)$. So we 
use $V_A^{(4b)}$ instead of $V_A^{(4a)}$ {and set $V_A^{(4)}= V_A^{(4b)}$}.

\subsection{Models of systematic survey for large field shape dependence}
\label{subsec:survey}

Many oscillon-supporting potentials exhibit a shallow growth or a flat ``plateau'' at large field values and a quadratic minimum, joined together through a transitional regime. These potentials are also observationally favored for inflation, since they lead to small values of the tensor-to-scalar ratio $r$, as required by the latest CMB measurements.
We define three potential types, in order to 
examine the relation of  the exact potential shape to the emitted 
gravitational wave spectra and the corresponding  formation efficiency of large overdensities. We first introduce
a generic  four-parameter model,
%
\begin{align}
 V_n(\phi) = 
  \frac{1}{2}m^2M^2\frac{|\phi/M|^\alpha}{\left[1+\beta\left|\phi/M\right|^\gamma\right]^\delta} \, .
\end{align}
%
The potential $V_X(\phi)$ behaves as
$\propto |\phi|^\alpha$ at $\phi\to 0$, and as $\propto
|\phi|^{\alpha-\gamma\delta}$ at $\phi\to\infty$.
We fix the behavior around the minimum to that of a massive scalar field and
 focus our attention on three kinds of one-parameter families in which we restrict
the variation of the parameters as 
$(\alpha,\beta,\gamma,\delta)=(2,1,\alpha_1,1),(2,\alpha_2,2,1),(2,1,\alpha_3,2/\alpha_3)$, namely,
%
\begin{equation}
\begin{aligned}
 V_1(\phi) &= 
  \frac{1}{2}m^2M^2\frac{(\phi/M)^2}{1+\left|\phi/M\right|^{\alpha_1}},
  &\quad
 V_2(\phi) &= 
   \frac{1}{2}m^2M^2\frac{(\phi/M)^2}{1+\alpha_2\left(\phi/M\right)^2},\\
 V_3(\phi) &= 
  \frac{1}{2}m^2M^2\frac{(\phi/M)^2}{\left[1+\left|\phi/M\right|^{\alpha_3}\right]^{2/\alpha_3}}.
\end{aligned}
\label{eq:V1-3}
\end{equation}

%

\begin{figure}[h]
 \centering
 \includegraphics[width=5.5cm]{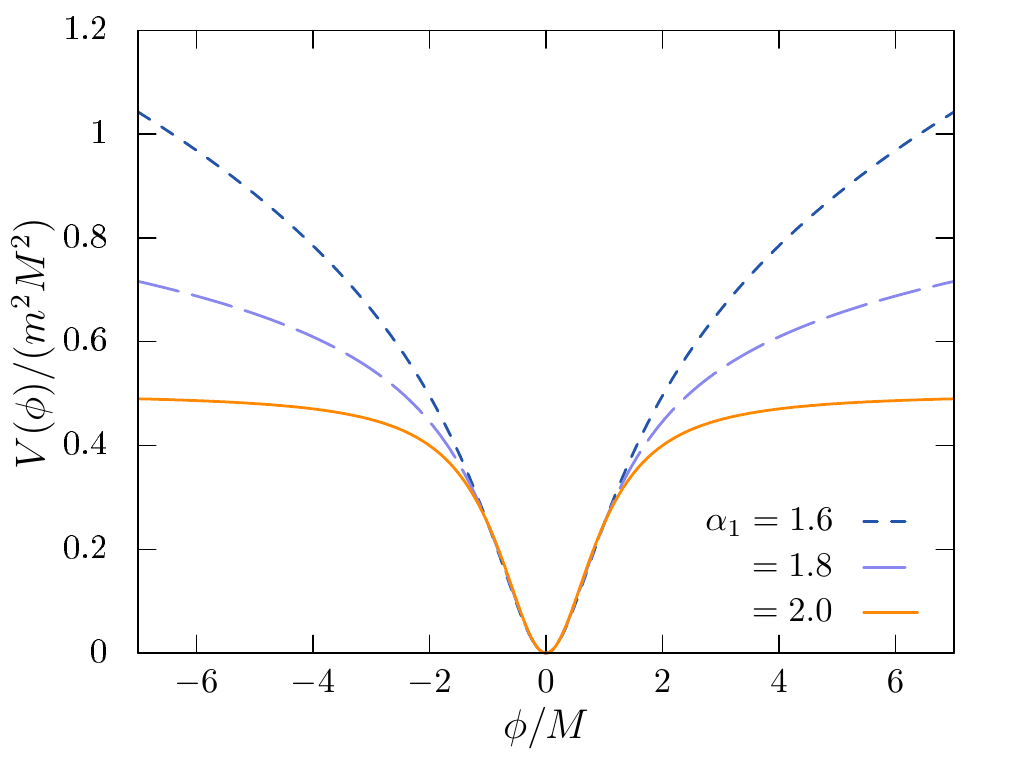}
 \includegraphics[width=5.5cm]{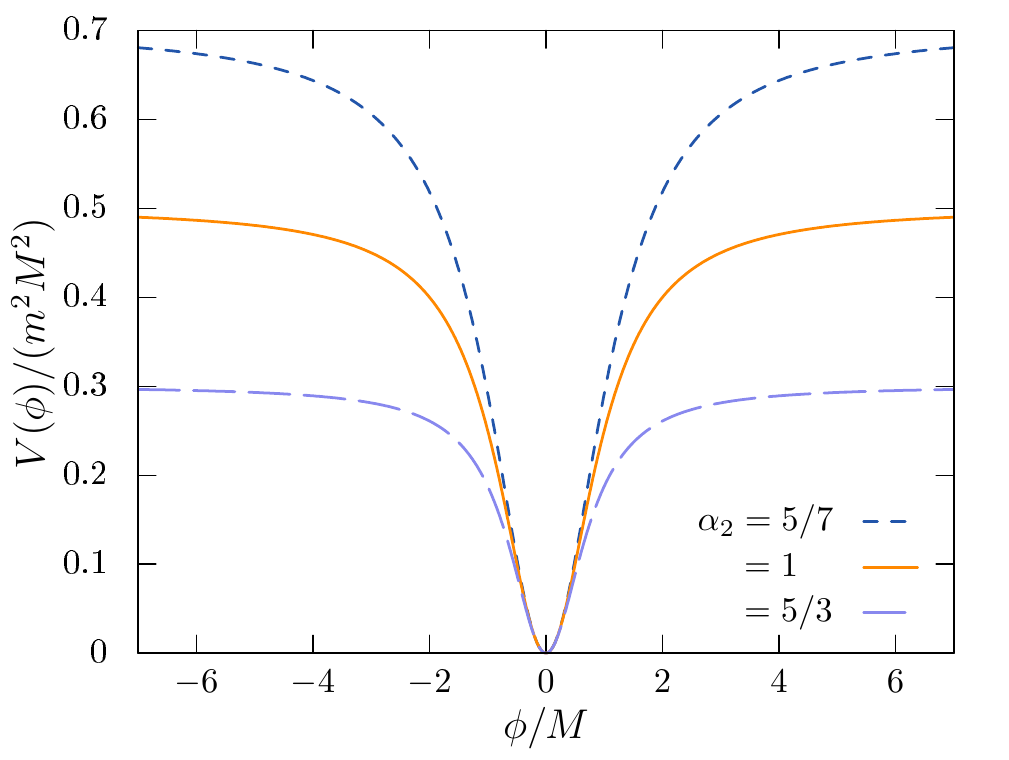}
 \includegraphics[width=5.5cm]{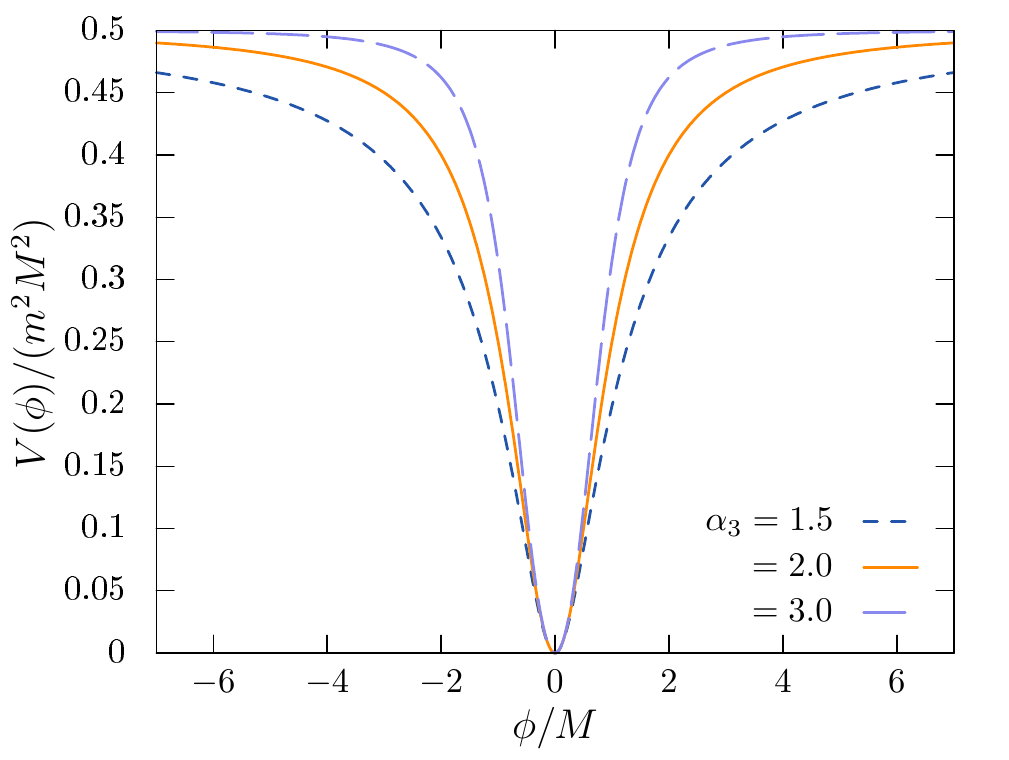}
 \caption{Potential shape of $V_1(\phi)$ with
 ${\alpha_1}=1.6, 1.8$ and $2.0$ (left), and $V_2(\phi)$ with
 ${\alpha_2}=5/7, 1$ and $5/3$ (middle), and $V_3(\phi)$ with
 ${\alpha_3}=1.5, 2$ and $3$ (right). The orange solid lines {represent}
 the same shape. The vertical axis is normalised as $V(\phi)/(m^2M^2)$ ,
 and the horizontal axis is $\phi/M$.}
 \label{fig:V1-3}
\end{figure}

The shapes of these potentials are shown in Fig.~\ref{fig:V1-3} and are chosen to model basic features of shallow inflationary potentials 
\begin{itemize}
\item $V_1$ describes potentials with a variety of asymptotic power-law growth at large field values.
\item $V_2$ describes plateau potentials with varying asymptotic amplitude.
\item  $V_3$ describes plateau potentials with varying effective width around the minimum at $\phi=0$, defined as the field amplitude at which the potential approaches the plateau value. In other words, the potentials of the type $V_3$ correspond to the potentials with the same large- and small-field behavior and differ by the size of the transitional region between the two.
\end{itemize}
It is important that all potentials are locally quadratic near the origin, hence describe free massive particles at small field values, while they become ``flatter'' for larger field values, thus in principle supporting the formation of oscillons.

Notice that they are related to each other; in fact they coincide for a particular parameter choice
$V_1(\phi;\alpha_1=2)=V_2(\phi;\alpha_2=1)=V_3(\phi;\alpha_3=2)
= 
{1\over 2} m^2 {M^2} \phi^2 / ( M^2  + \phi^2) 
$.

\section{Numerical setup}
\label{sec:numerical}

We impose periodic boundary conditions on the boundaries of the computational domain, and we choose the initial conditions,\footnote{Since we are interested in sourced gravitational waves, we do not initialize $h_{ij}$ in its quantum vacuum.}
%
\begin{align}
 \phi(\xx,\eta_0) &= \phi_0 + \delta\phi(\xx), \\
 h_{ij}(\xx,\eta_0) &= 0,
\end{align}
%
where $\eta_0$ is the initial conformal time and $\delta\phi(\xx)$ is a Gaussian random field so that the power spectrum is
equivalent to that given in Minkowski spacetime,
%
\begin{align}
 P_{\delta\phi}(k) = \frac{1}{2k} \, .
\end{align}
%
This is consistent with the Universe at the end of inflation, given that the size of the computational domain is sufficiently less than the horizon scale, so that the relevant quantum fluctuations do not ``feel'' the space-time curvature, at least initially.

The initial scale factor is set to be
$a(\eta_0)=1$,
and its time-evolution is governed by the Friedmann equation,
%
\begin{align}
 \HH^2 = \frac{a^2}{3\Mpl^2}\langle\rho_\phi\rangle
\end{align}
 where $\langle\rho_\phi\rangle$ is the averaged energy density of the scalar field $\phi$
\begin{align}
 \rho_\phi = \frac{1}{2a^2}\phi'{}^2 + \frac{1}{2a^2}(\partial\phi)^2+V.
\end{align}
%
We  impose an initial condition for the amplitude of the scalar field 
$\phi_0$ such that $V_i(\phi_0)=m^2M^2/4$.
For the time-derivative, we impose 
$\phi'(\xx,\eta_0)=0$ and $h'_{ij}(\xx,\eta_0) = 0$.
Therefore the Hubble parameter at the beginning of simulations is 
$H_{\rm in}=\sqrt{V_{\rm in}/3\Mpl^2}=mM/\sqrt{12}\Mpl$.

We redefine the scalar field $\phi$ as $\widetilde{\phi}=\phi/M$
to make it dimensionless, and then the gravitational coupling constant
in the right-hand side of Eq.~(\ref{eq:eq_h}) is normalised such that
%
\begin{align}
 \epsilon_G \equiv \frac{M}{\Mpl} \, .
\end{align}
%

We assume $m=10^{-2}M$ 
and the coupling parameter $\epsilon_G$ is set to be $\epsilon_G=10^{-2}$.
The field equations are solved with the Leap-frog method 
in a three-dimensional box with $256^3$ grid points whose (comoving) size is chosen as $L=60\,m^{-1}$.
Note that the box size is less than the initial horizon scale, $L/a \ll H_{\rm in}^{-1}$.
The time-interval is $d\eta=0.05\,m^{-1}$ 
and we perform simulations until $\eta=600\,m^{-1}$.
The spatial derivatives are approximated as the second-order finite differences.

For all cases, we evaluate the gravitational wave spectrum using Eq.~(\ref{eq:Omega_GW_final}) and the power spectrum of the scalar
fluctuations using Eq.~(\ref{eq:def_P}). Furthermore we compute the time-evolution of
number density of oscillons, and their size distribution. 
To reduce the 
variance
from the initial random field, possibly leading to unphysical artifacts, we perform $10$  simulations for each model parameter and average them.

\section{Results}
\label{sec:results}

\subsection{Axion monodromy $V_A$ and small-field dependence}
\label{subsec:VAapp}

\begin{figure}[!ht]
 \centering
 \includegraphics[width=.3\textwidth]{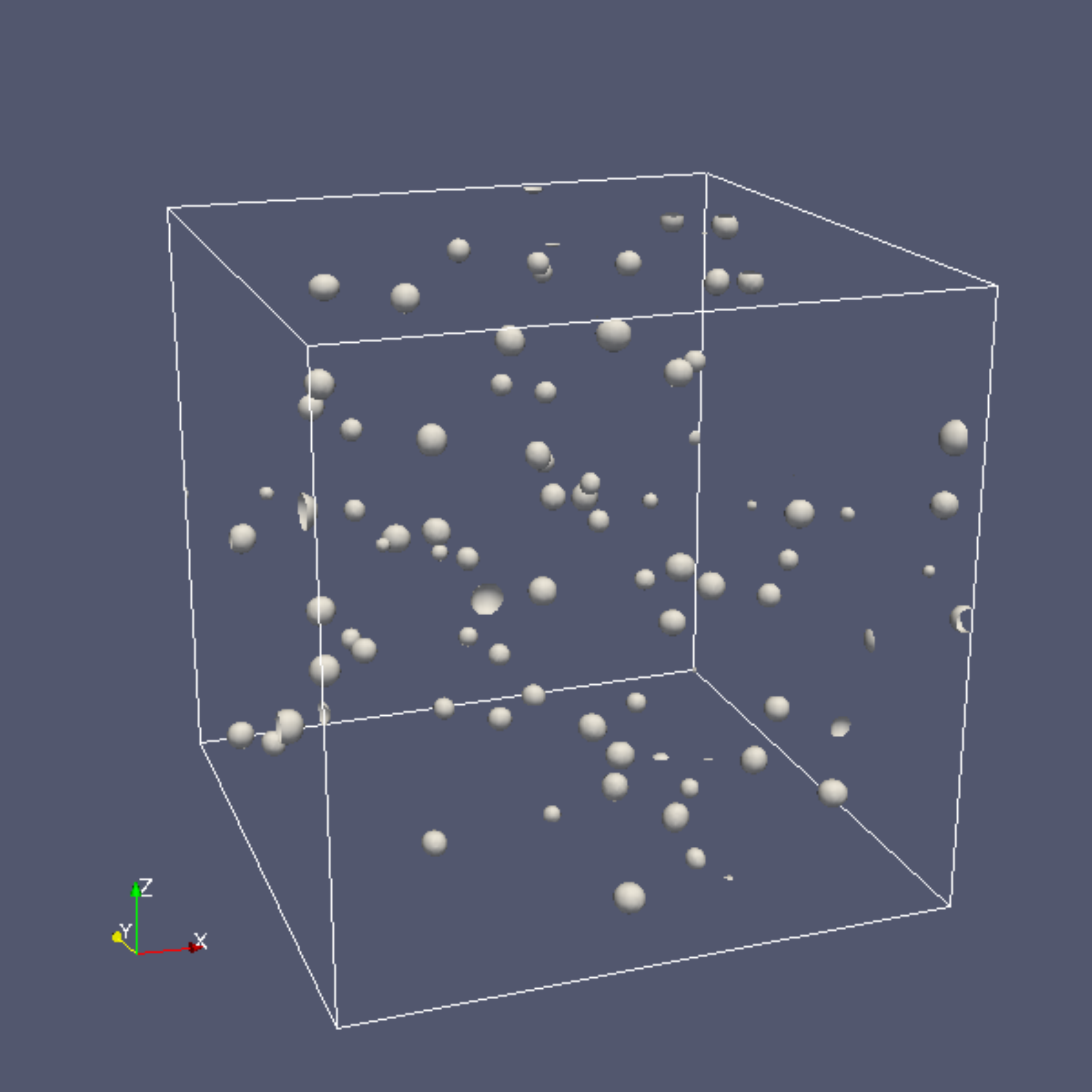}
 \caption{Oscillons in the simulations for $V_A(\phi;\alpha_1=1/2)$
 with $\epsilon_G=10^{-2}$. The
 isosurface with $\rho=1m^2M^2$ at $\eta=400m^{-1}$ is shown.}
 \label{fig:VA}
\end{figure}

We start by performing the simulation for the axion monodromy potential, $V_A(\phi)$ with $\alpha_A=1/2$, in order to make contact with the results of Ref.~\cite{Zhou:2013tsa}.
In Fig.~\ref{fig:VA}, we show the isosurface of the energy density
at $\eta=400\,m^{-1}$ with $\epsilon_G=10^{-2}$, which is the
fiducial parameter in Ref.~\cite{Zhou:2013tsa}.
We find the formation of multiple oscillons in our  box, which  are stable in the time-scale of the simulation.

\begin{figure}[!ht]
 \centering
 \includegraphics[width=.45\textwidth]{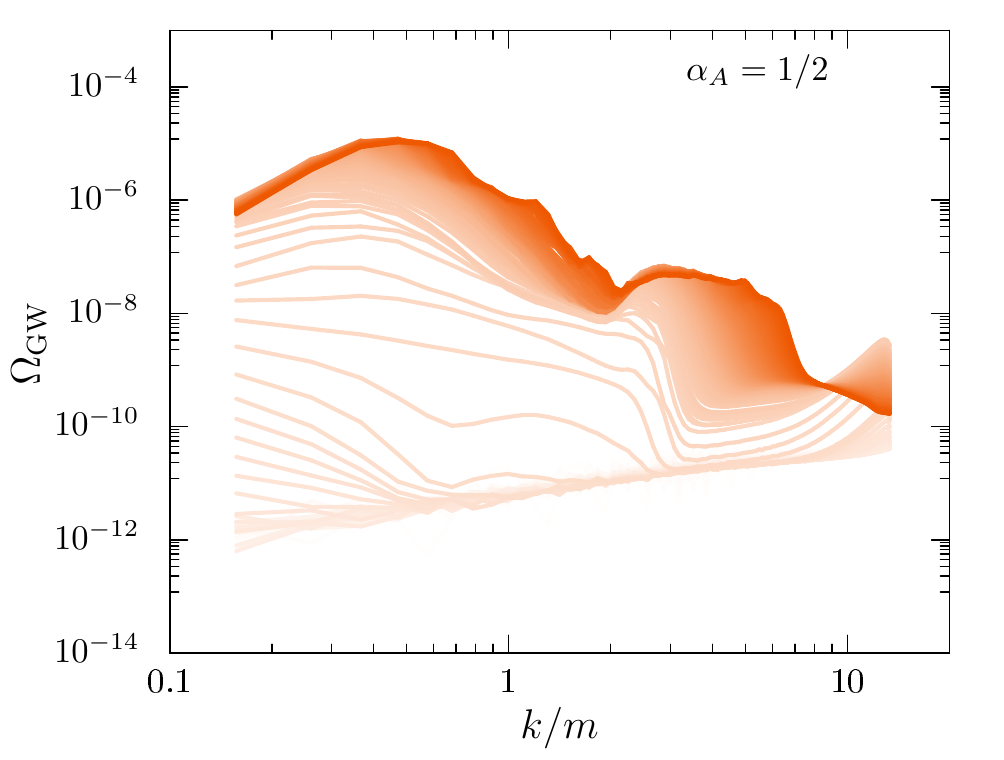}
  \includegraphics[width=.45\textwidth]{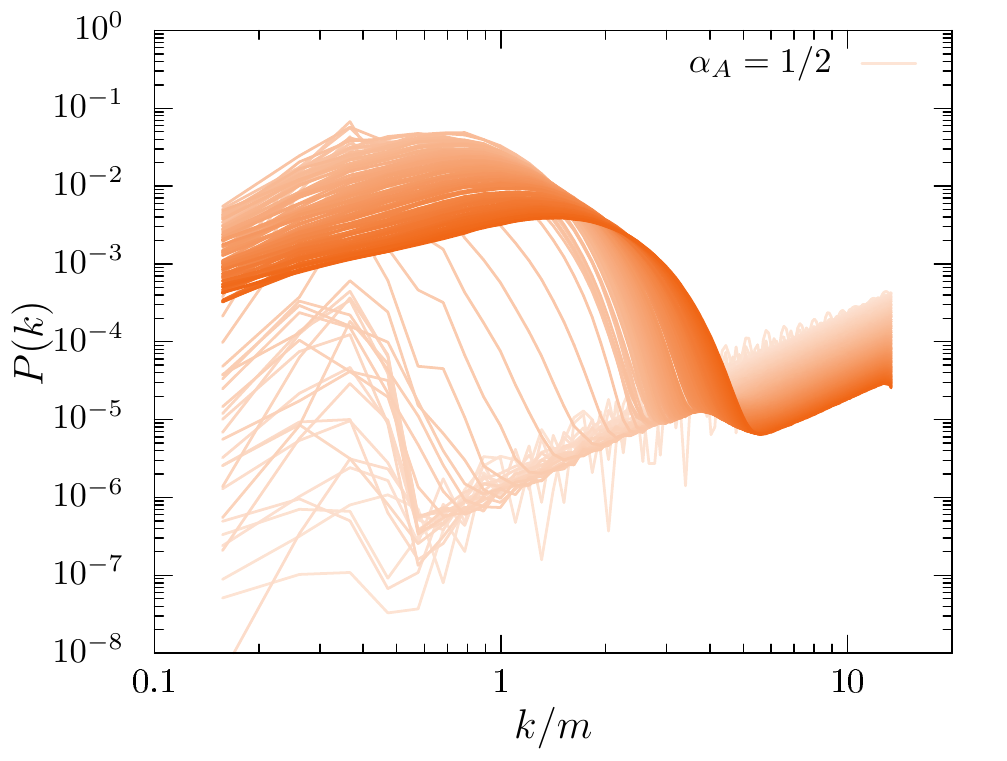}
 \caption{Time evolution of GW power spectrum 
 (left) and the power spectrum of the inflaton fluctuations $\delta\phi$ (right) evaluated during simulation time. The colour gradation represents the time evolution of spectrum, with darker colours representing later times. The time spacing between two consecutive curves corresponds to $\Delta\eta =5m^{-1}$. Notice that the present value of $\Omega_{\rm GW}$ is given by multiplying by $\Omega_{r,0}=0.916\times
 10^{-4}$ and $(g_{*,f}/g_{*,0})^{1/3}$ (see Eq.~\eqref{eq:GWfactor}). $k$ is the comoving wavenumber with $a(\eta_0)=1$. 
 }
 \label{fig:VA_gw}
\end{figure}

The time-evolution of the corresponding GW spectrum is shown in Fig.~\ref{fig:VA_gw} in
which the line color becomes thicker as time goes forward. 
We confirmed  the existence of distinct peaks at the final time
as reported in Ref.~\cite{Zhou:2013tsa}.
The amplitude of the GW spectrum at the end of simulation is given as 
$\Omega_{\rm GW,f}\approx 10^{-6}$ at most. If the early matter-dominated
phase is terminated at a time corresponding to the end of the simulation, after which the universe quickly transitions to the radiation-dominated epoch, the present value of $\Omega_{\rm GW}$ is given by multiplying
by $\Omega_{r,0}=0.916\times 10^{-4}$ and $(g_{*,f}/g_{*,0})^{1/3}$.
Taking these factors into account, we compute the current GW amplitude
$\Omega_{\rm GW,0}\approx (10^{-10}\,-\,10^{-11})$. This
amplitude is slightly larger than the value shown in Ref.~\cite{Zhou:2013tsa} because $m=10^{-2}M$ is slightly larger than the mass used in Ref.~\cite{Zhou:2013tsa} and the initial condition is also slightly different. Given these well understood differences, our results are consistent with those of Ref.~\cite{Zhou:2013tsa}.


After {mostly} recovering the results of the axion monodromy
potential $V_A$, we focus on models $V^{(n)}_{A}$ deformed from $V_{A}$.
As we mentioned, $V^{(n)}_{A}$ asymptotically behaves as $V_{A}$, while
the small-field shape is deformed from $V_{A}$ by using the Pad\'e
approximants. In this section, we investigate the impact of the
{(small-field)} deformation on the oscillon formation and the
resulting gravitational wave spectra.

\begin{figure}[h]
 \centering{
 \includegraphics[width=.45\textwidth]{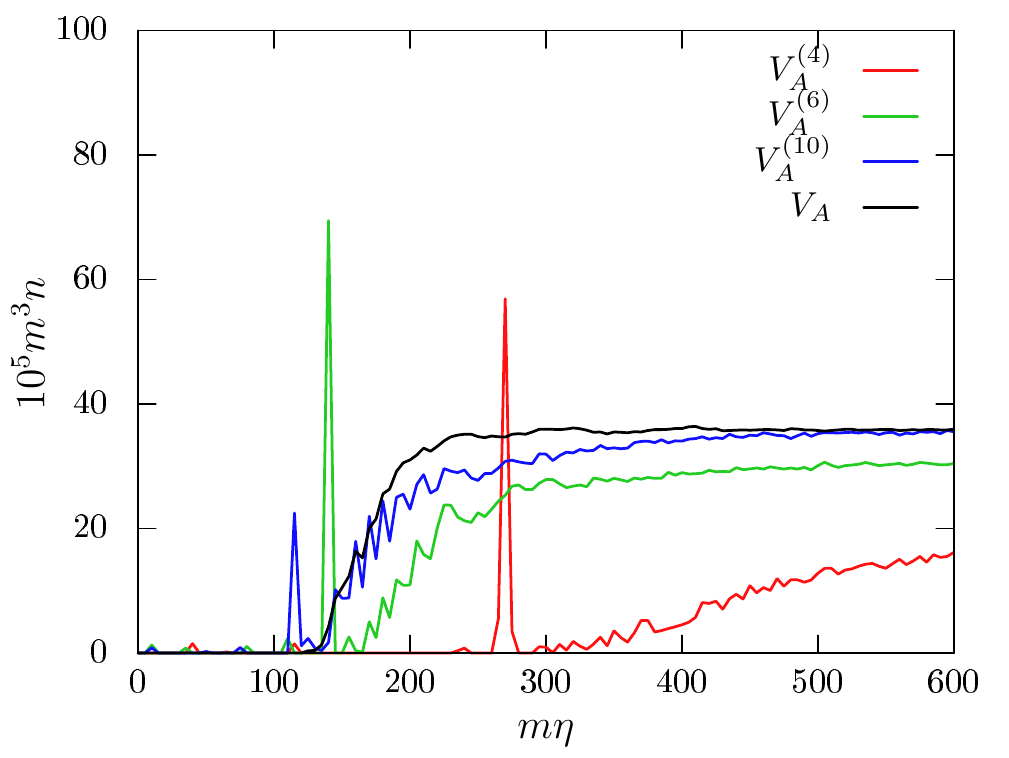}
 \includegraphics[width=.45\textwidth]{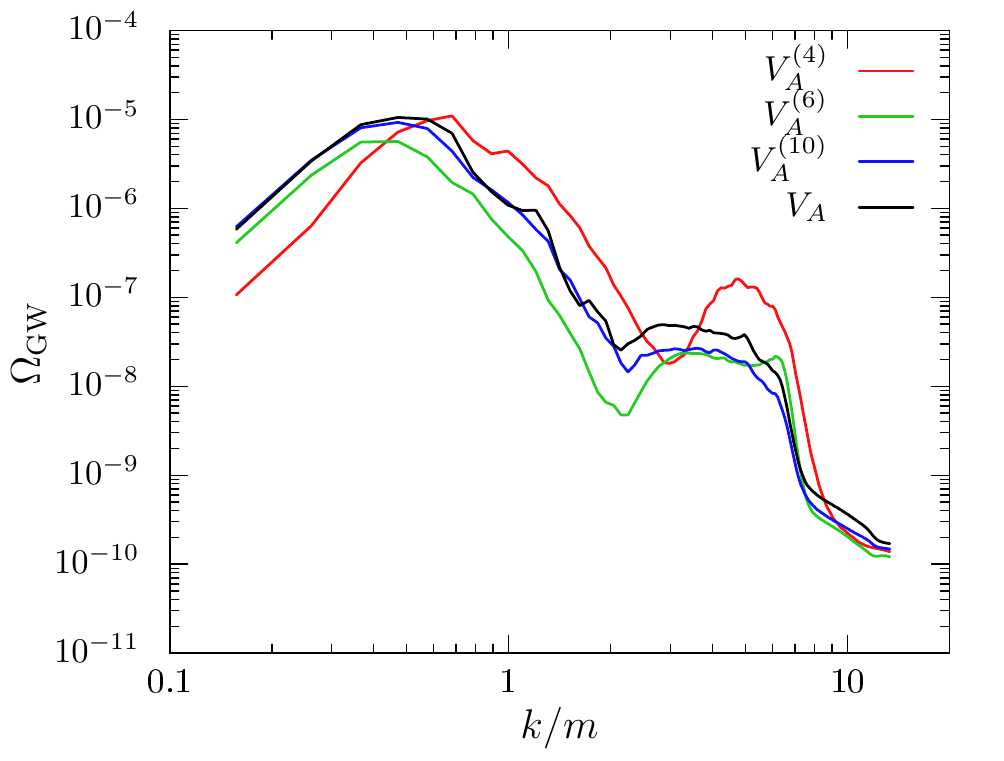}
 }
 \caption{Time evolution of the number density of oscillons (left) and
gravitational wave spectrum at the final time of simulations (right).
}
 \label{fig:res_VAapp}
\end{figure}

In the left panel of Fig.~\ref{fig:res_VAapp}, we show the time-evolution
of the number of oscillons. To display them clearly, we omit the
error bars 
resulting from averaging over 10 realisations.
The simulation results contain transient objects, local over-densities that do not possess the longevity of  oscillons. To remove them from our counting, we used a simple criterion of only considering over-densities whose width $w$ exceeds a cutoff value $w>w_c$.
 In the left panel of Fig.~\ref{fig:res_VAapp}, we set $w_c=3.5\,m^{-1}$. We discuss  the selection criterion as well as the oscillon identification algorithm that we used in Appendix~\ref{sec:ident}. We see that for the case $V_A$ the number density of oscillons starts growing at $\eta \sim 110\,m^{-1}$ monotonically and reach an asymptotic value of $n\simeq 35\times 10^{-5}m^{-3}$. On the other hand, for the approximated cases, the number densities of oscillons grow later as the approximations get farther away from the original monodromy potential. In fact, for the case $V_A^{(10)}$, oscillons starts appearing at $\eta \sim 110\,m^{-1}$, which is  almost the same time as for the case $V_A$, and for the case $V_A^{(6)}$ oscillons starts appearing slightly later at $\eta \sim 125\,m^{-1}$. In the least approximated case $V_A^{(4)}(=V_A^{(4b)})$, early oscillon production is severely suppressed until $\eta \lesssim 250\,m^{-1}$, which indicates that the inflaton field does not undergo efficient parametric resonance. The instability bands of this system depend on the details of the functional shape of the given potential, as explained in detail in Section~\ref{subsec:VAappFloq}. Hence, in the less approximated cases, fluctuations need more time to grow, enter the non-linear regime and ultimately form oscillons. Then, even in the approximated cases, though the number densities of oscillons slightly oscillate and have spikes, they eventually grow and reach  almost the same asymptotic values as that of  case $V_A$.
It should be noted that the spikes correspond to the amplification of fluctuations, leading to transient inhomogeneities, which are picked up by our oscillon detection algorithm.\footnote{The spikes show very brief increases in the number density of oscillons. In this sense, even though we introduced a selection rule for ``filtering out" transient overdensities, we see that our counting is still somewhat susceptible to them.} Thus, even though the formation times were delayed in the approximated cases, the total number of oscillons at late times is largely insensitive to the exact form of the potential that we use. Apparently, at the final time of the simulation, the number of the oscillon in the least approximated case $V_A^{(4)}$ is still growing, but we expect that it will saturate around the same number of oscillons with the other cases. This fact implies that the number is insensitive to the small-field shape of potential. In fact, after an oscillon forms, the field value in an oscillon becomes larger and the simple picture of parametric resonance breaks down. The physical properties of an oscillon are controlled by non-linearities, probing the potential beyond its local shape near the minimum.

In the right panel of Fig.~\ref{fig:res_VAapp}, we show the
gravitational wave spectra for each case evaluated at the end of
the simulation. The amplitude and the detailed structure of the
gravitational wave spectrum are found to be (slightly) sensitive to
the small-field shape of potential {though its shape is almost
universal}. The gravitational waves are produced most efficiently when
the oscillons are formed and non-spherical structures appear. After
that, the oscillons become spherical, which stops the production of
gravitational waves. For example, the slight shift of the peaks and troughs of 
$\Omega_{\rm GW}$ for $V_A^{(4)}$ can be attributed to the later fragmentation of the inflaton field,
occurring closer to $\eta \simeq 250 \,m^{-1}$, rather than $\eta\simeq 100 \,m^{-1}$, which is the case for the other three potential shapes. Therefore the final amplitude of gravitational
waves is sensitive to the time when oscillons form,
which is, in turn, sensitive to the small-field shape of the potential.

{In summary, in all cases but one oscillons are efficiently formed and the final number density is almost universal though the formation time is sensitive to the shape around the origin and oscillon formation occurs earlier for potentials that contain more terms and more closely approximate the  monodromy potential of Eq.~\eqref{eq:VA}. This fact indicates that the initial growth of oscillons strongly depends on the instability bands of the given potential which are determined by the details of the functional form around the origin. These issues will be  discussed further in detail in Section~\ref{subsec:VAappFloq}.}
Furthermore, the final GW spectra are similar in all cases {(except the slight shift of the peaks and troughs which depends on the time of inflaton fragmentation)} and --as expected-- approach the form of $V_A$ for potentials $V_A^{(n)}$ with larger values of $n$, hence potentials that approximate $V_A$ more closely.
We leave a more thorough analytical and numerical investigation of the type and longevity of created oscillons in each case for future work.

\subsection{large-field dependence: systematic studies on $V_n(\phi)$}
\label{subsec:Vn}

\begin{figure}[t]
 \centering
 \includegraphics[width=.3\textwidth]{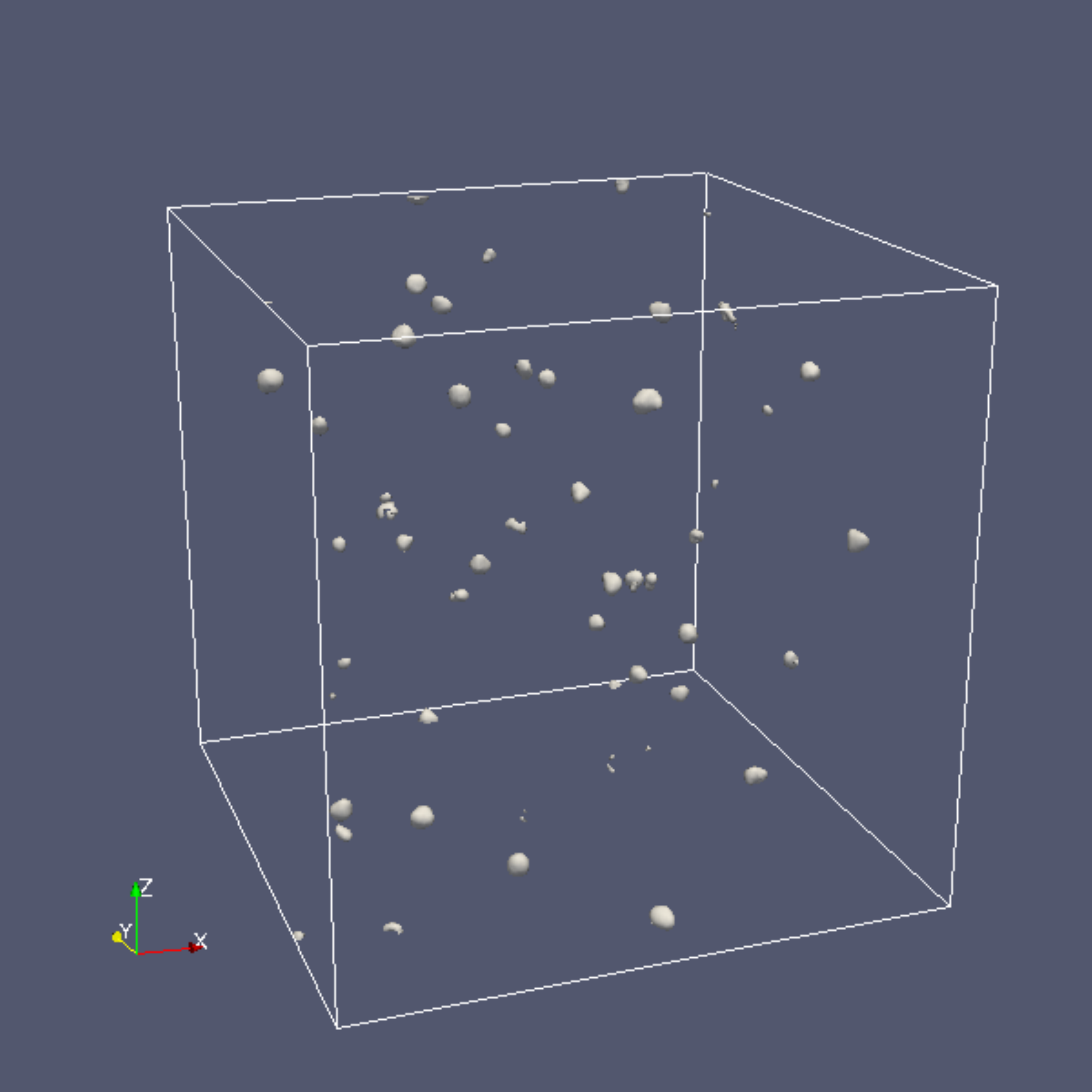}
 \caption{Oscillons in the simulations for $V_1$ with $\alpha_1=2$. The
 isosurface with $\rho=1m^2M^2$ at $\eta=400\,m^{-1}$ is shown.}
 \label{fig:res_V1_3d}
 \centering{
 \includegraphics[width=.45\textwidth]{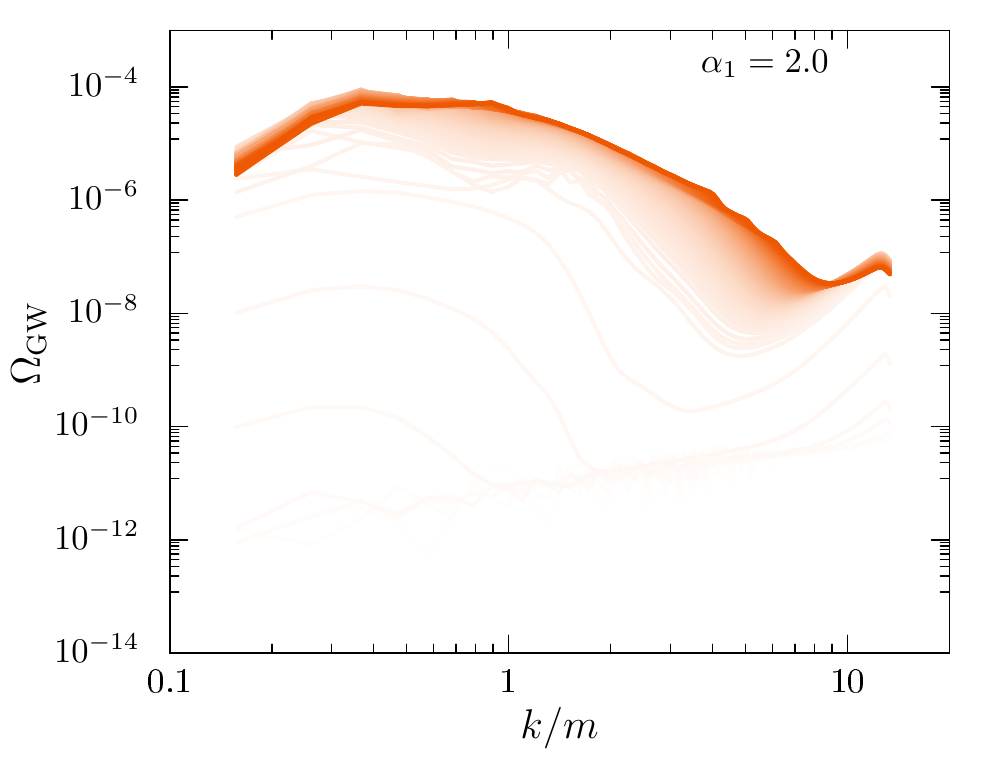}
 \includegraphics[width=.45\textwidth]{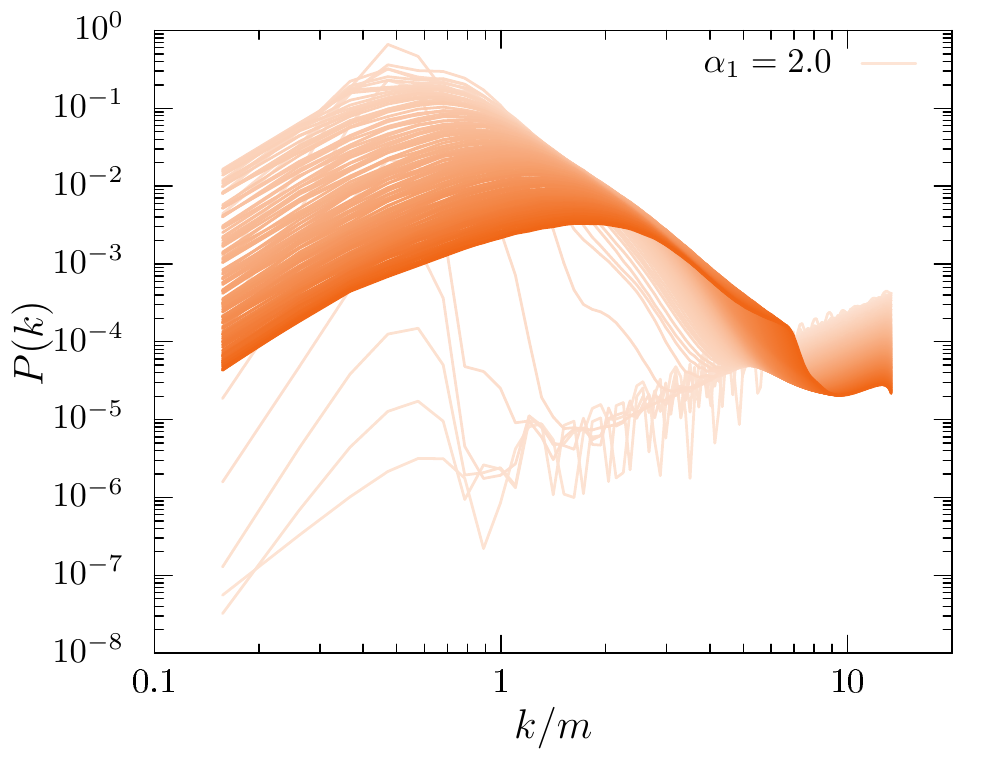}
 }
 \caption{Gravitational wave spectrum (left) and the power spectrum of
 inflaton fluctuations $\delta\phi$ (right) for $V_1(\phi)$ with $\alpha_1=2$. The colour gradation 
 represents the time evolution of spectrum, with darker colours representing later times.
}
 \label{fig:res_V1_GW}
\end{figure}

We move on to study the impact of the large-field shape of potential
on oscillon formation and the resulting gravitational wave spectrum.
In order to disentangle the various potential features, we start with the fiducial potential $V_1(\phi;\alpha_1=2) ={1\over 2} m^2  \phi^2 / (1+\phi^2/M^2)$
given in Eq.~(\ref{eq:V1-3}) and consider deformed potentials  given in Eqs.~(\ref{eq:V1-3}), having for example different asymptotic behavior, while sharing a similar small-field shape.

The isosurface at $\rho=1m^2M^2$ for $V_1(\alpha_1=2;\phi)$ is shown in 
Fig.~\ref{fig:res_V1_3d}.  This corresponds to the fiducial model
in the present study, as shown by the orange solid line in Fig.~\ref{fig:V1-3}. The oscillons do not exhibit a perfectly spherical shape, but we see the appearance of a number of  ``spikes'' on each individual oscillon caused by unstable modes on small
scales. The time-evolution of the power spectrum of $\delta\phi$ is
shown in the right panel in Fig.~\ref{fig:res_V1_GW}.
There, we initially see a well-defined range of wavenumbers that become unstable, given by $k\lesssim m$. However, at late times
 we see a broad range of wavenumbers growing and the resulting spectrum is  featureless. 
 This is reminiscent of preheating in other models, where lattice simulations showed significant  re-scattering between the modes, leading to a UV cascade of power (see e.g. Ref.~\cite{Nguyen:2019kbm} for a recent study, albeit in a different model).
 As a
result, the gravitational wave spectrum shown in the left panel in
Fig.~\ref{fig:res_V1_GW} is almost flat over an order of
magnitude in $k$ space and the features reported in
Ref.~\cite{Zhou:2013tsa} do not emerge.

\begin{figure}[h]
 \centering{
 \includegraphics[width=.24\textwidth]{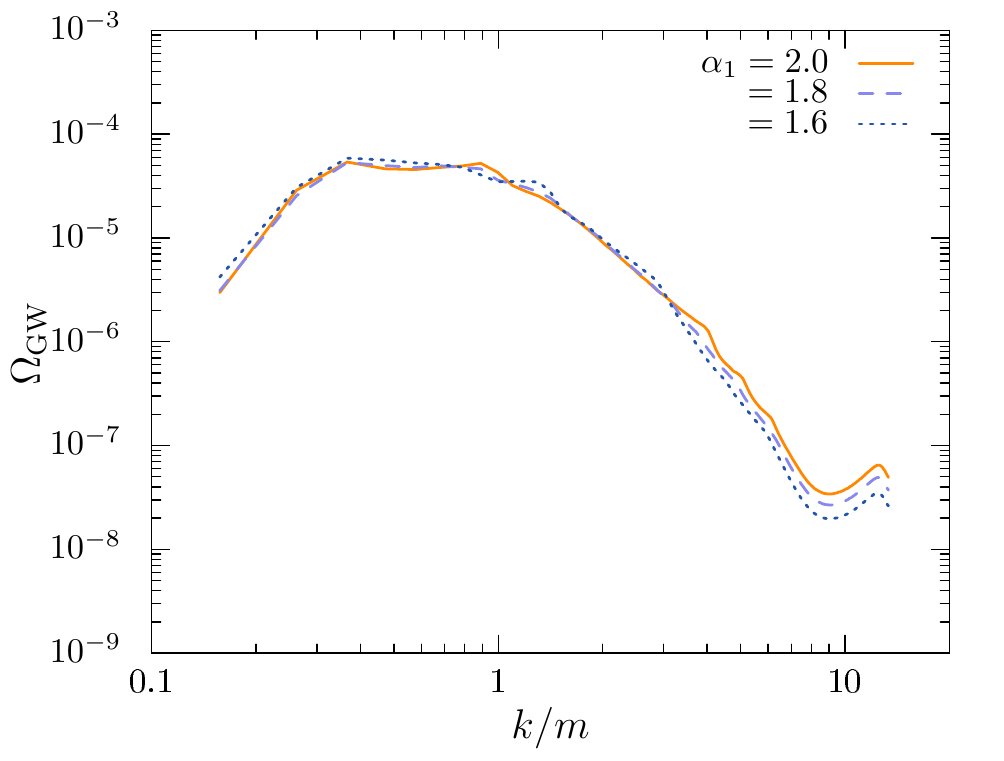}
 \includegraphics[width=.24\textwidth]{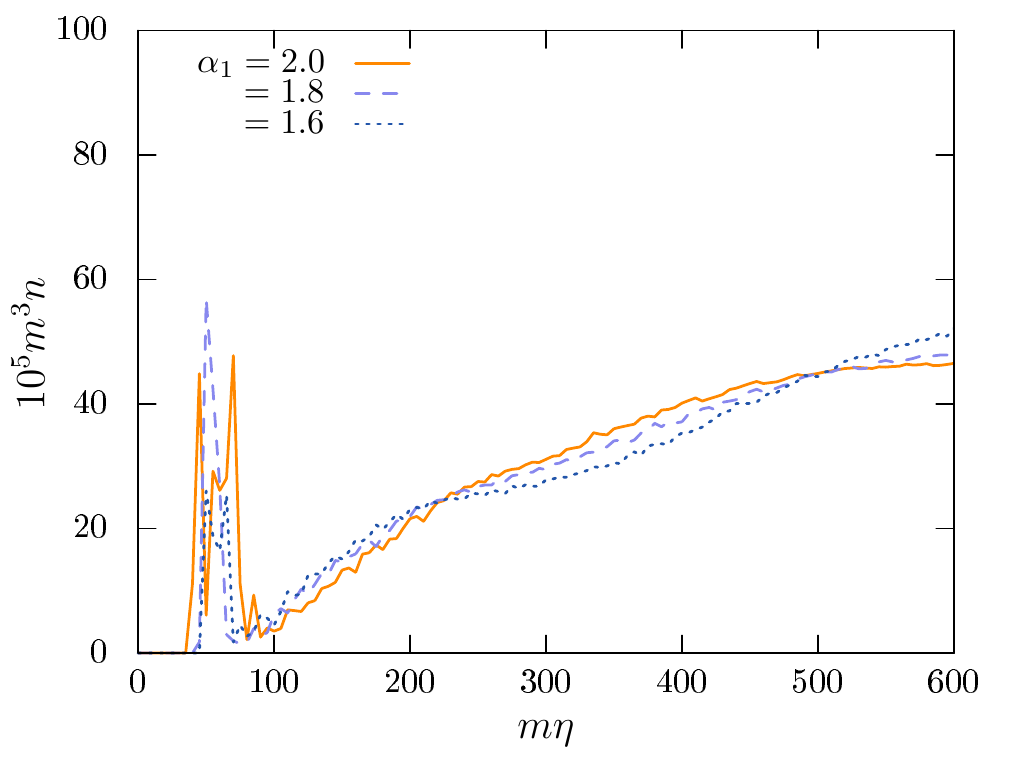}
 \includegraphics[width=.24\textwidth]{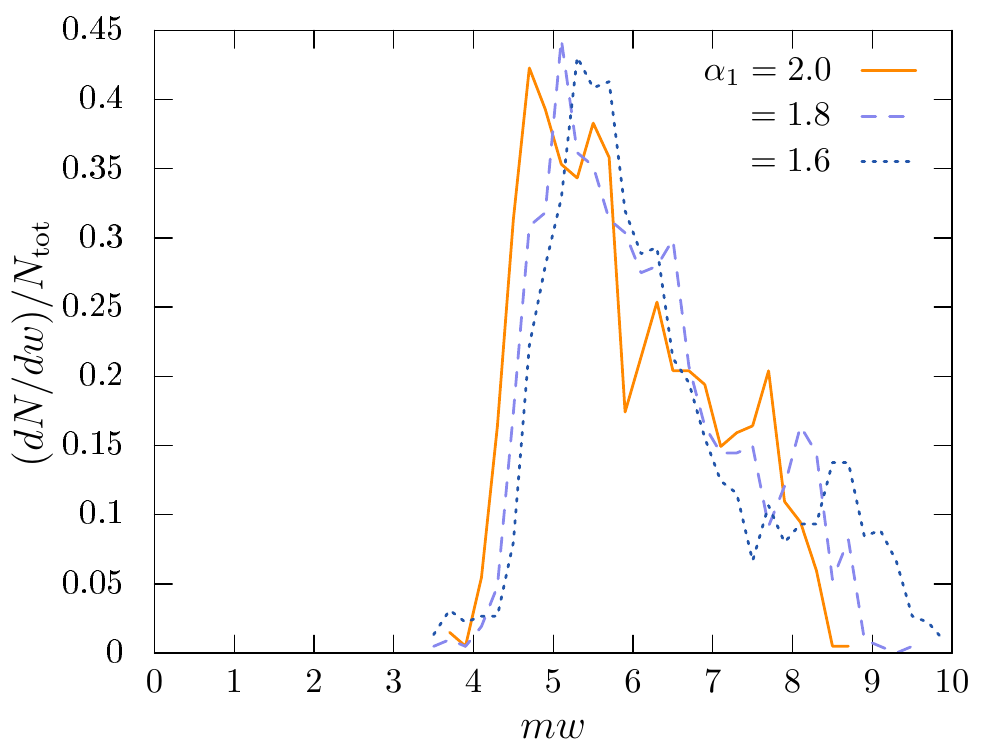}
 \includegraphics[width=.24\textwidth]{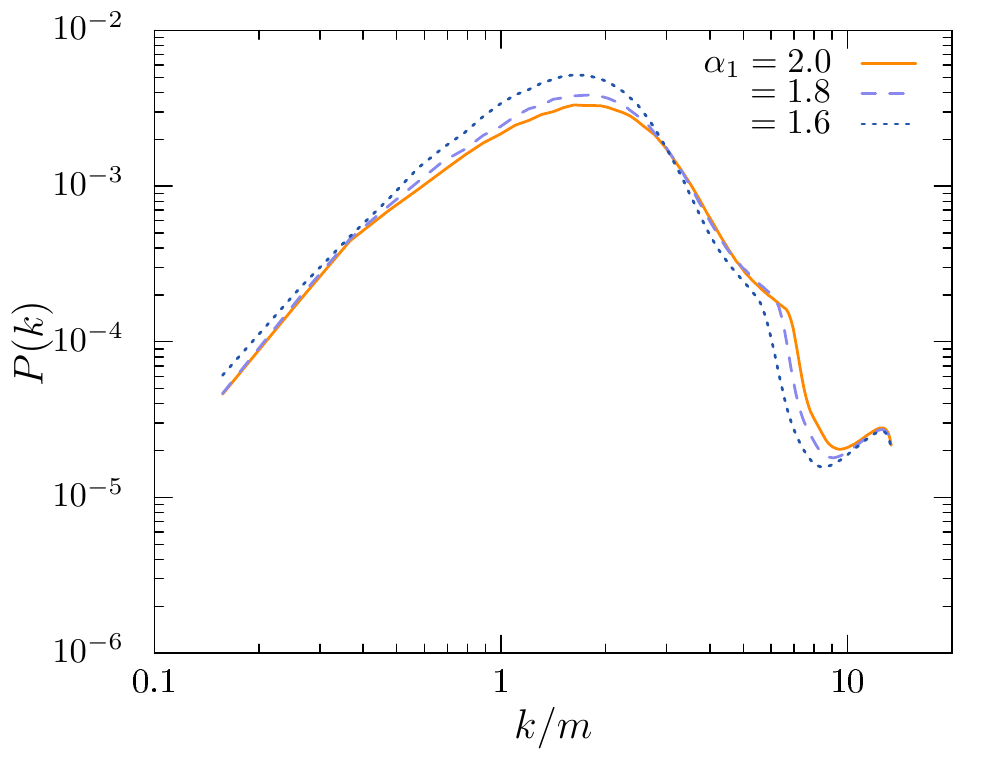}
 }
 \caption{{\it From left to right:} The gravitational wave spectrum at the end of the simulation,
the time evolution of the number density of oscillons,
the size distribution of oscillons at the final time of simulations and the power spectrum of the scalar field $P(k)$ at the end of the simulation {($\eta=600\,m^{-1}$).} All panels correspond to numerical results with $V_1(\phi)$ and parameters $\alpha_1=2, 1.8, 1.6$ (orange solid,  purple dashed and blue dotted respectively).
}
 \label{fig:V1}
 \centering{
 \includegraphics[width=.24\textwidth]{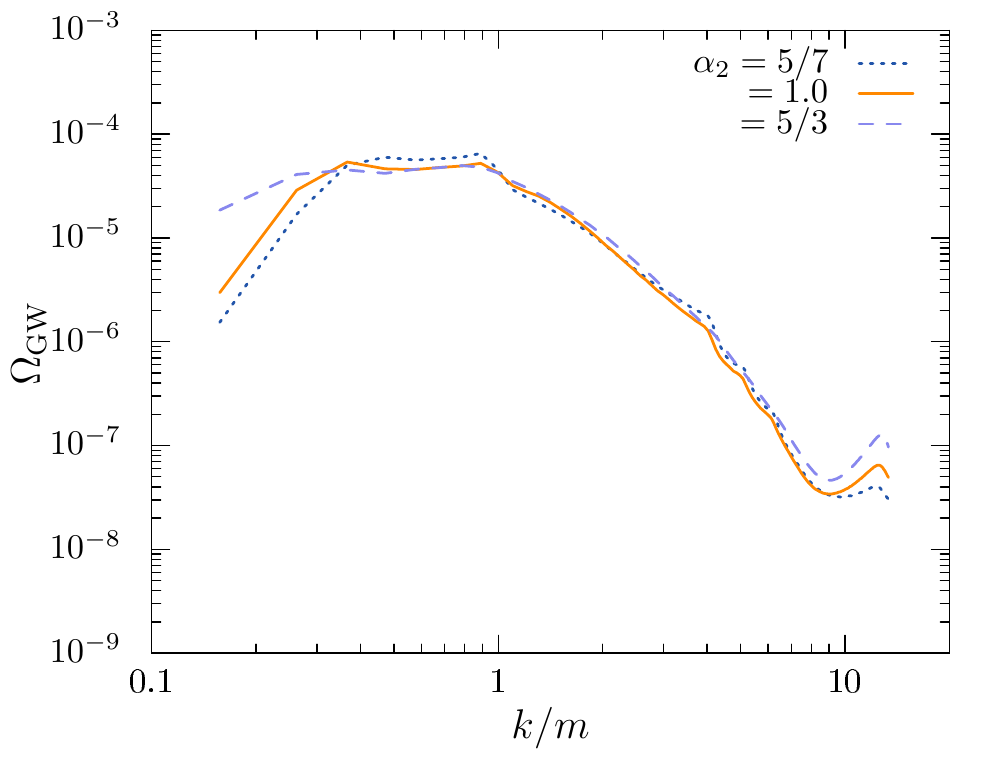}
 \includegraphics[width=.24\textwidth]{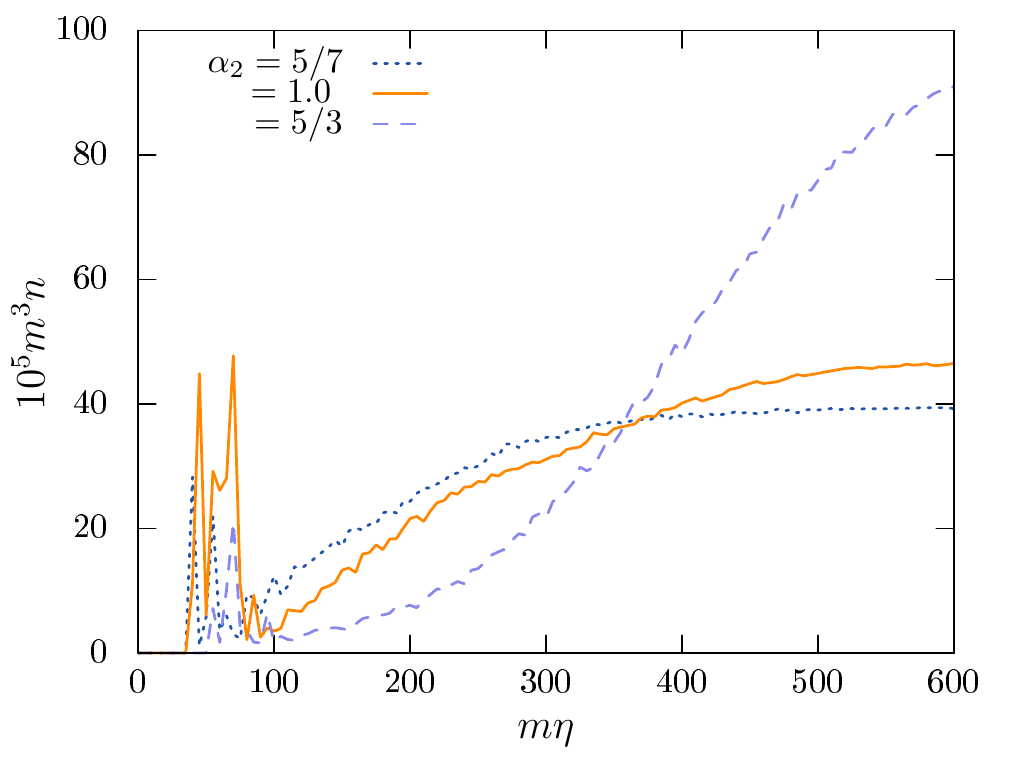}
 \includegraphics[width=.24\textwidth]{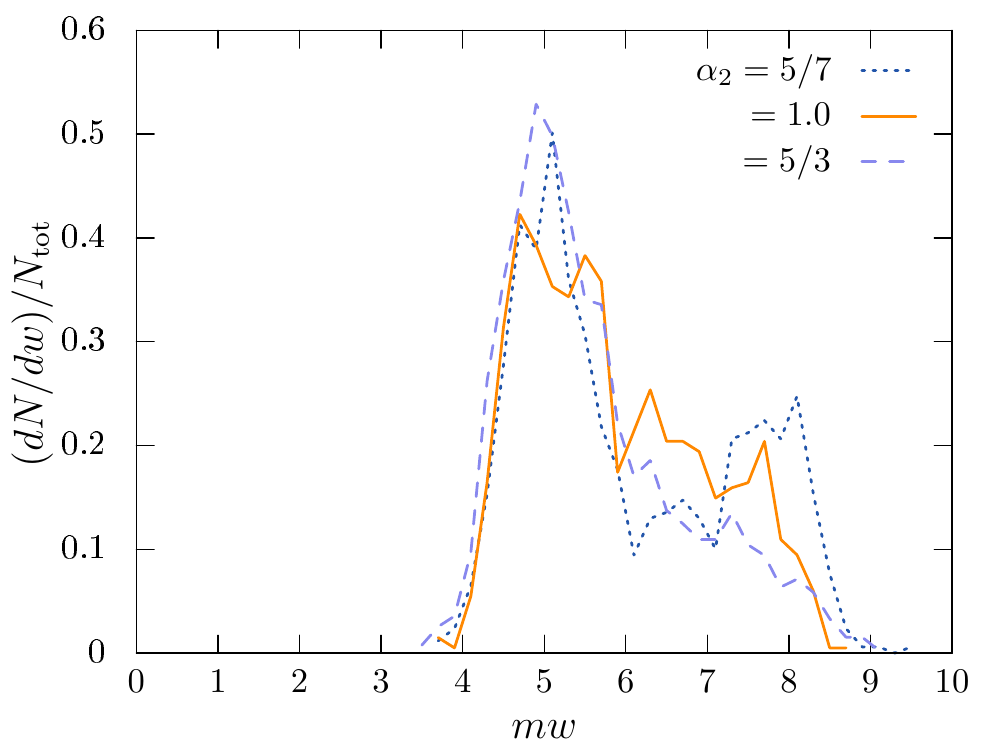}
\includegraphics[width=.24\textwidth]{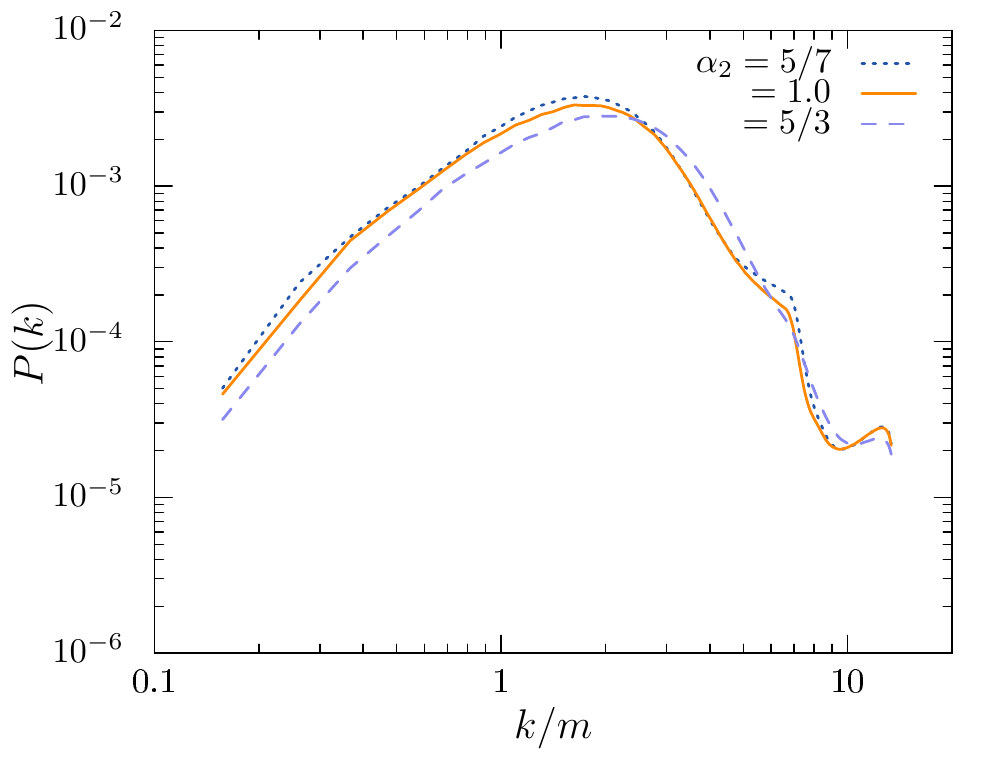}
 }
 \caption{The same figures as Fig~\ref{fig:V1}, corresponding to numerical results with $V_2(\phi)$ and parameters $\alpha_2=1, 5/3, 5/7$ (orange solid,  purple dashed and blue dotted respectively).
}
 \label{fig:V2}
 \centering{
 \includegraphics[width=.24\textwidth]{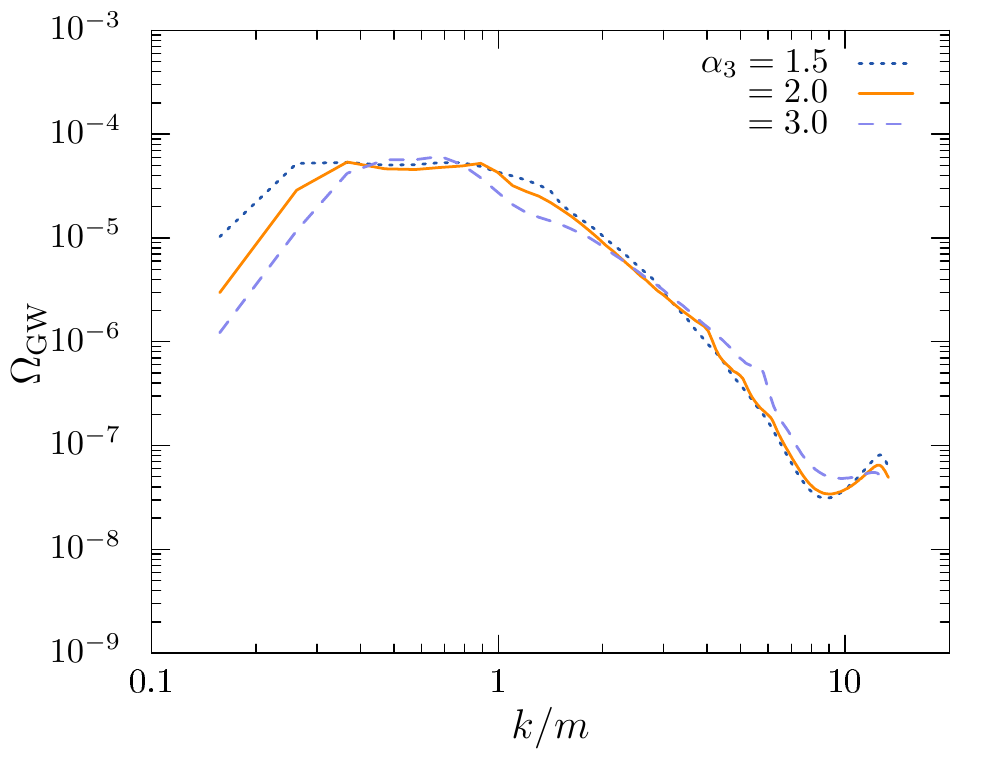}
 \includegraphics[width=.24\textwidth]{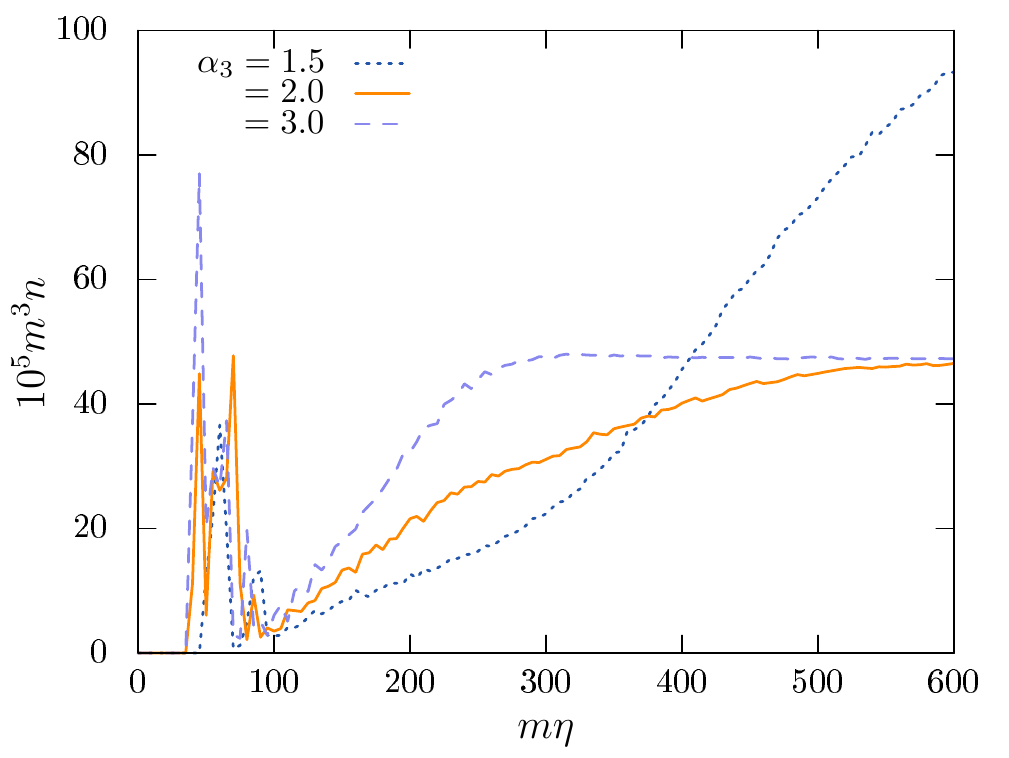}
 \includegraphics[width=.24\textwidth]{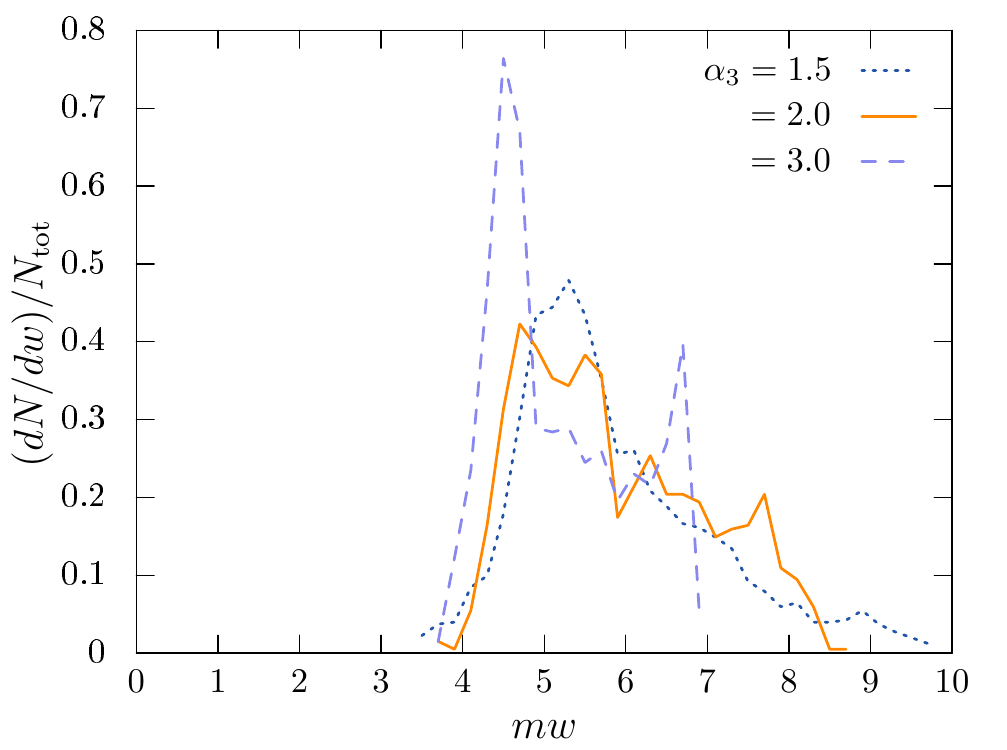}
\includegraphics[width=.24\textwidth]{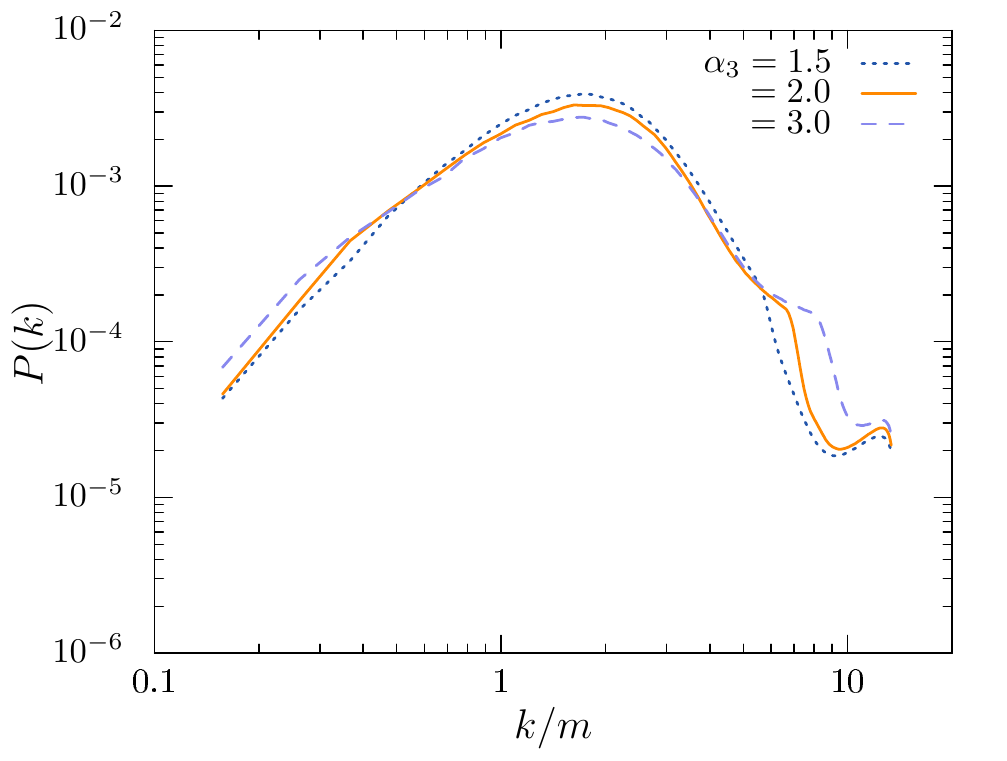}
 }
 \caption{The same figures as Fig~\ref{fig:V1}, corresponding to numerical results with $V_3(\phi)$ and parameters $\alpha_3=2, 3, 1.5$ (orange solid,  purple dashed and blue dotted respectively).
}
 \label{fig:V3}
\end{figure}

For each of the three models ($V_1, V_2, V_3$), we repeat the simulations by varying the model parameter $\alpha_n$. We show the gravitational wave spectra $\Omega_{GW}(k)$, the
time-evolution of the number density of oscillons $n$, and the size
distribution of oscillons at the final time of simulations in
the top-left panels in Figs.~\ref{fig:V1}-\ref{fig:V3}.
The size distributions are normalised by the total number of the oscillons, $N_{\rm tot} = \int\frac{dN}{dw}\,dw$, where $w$ is the physical size of oscillon. As in the case of $V_A$, we only count oscillons whose width exceeds the cutoff value  $w_c=3.5m^{-1}$.

The gravitational wave spectra are insensitive to the choice of 
$\alpha_1, \alpha_2$ and $\alpha_3$, namely, 
the asymptotic behavior of the inflaton potential
does not affect the spectra, though the amplitudes are slightly different for $k \lesssim 0.3\,m$ in the cases of $V_2$ and $V_3$. These differences come mainly from delayed growth of fluctuations and their less redshift in the cases with higher amplitudes. On the other hand, the number of oscillons {depends} on the exact value of 
 $\alpha_2$ and $\alpha_3$, while there are no differences in the results for different values of $\alpha_1$. 
We find that the oscillons can form more frequently if the
potential minimum is shallower (larger $\alpha_2$) and/or wider (smaller $\alpha_3$).
This is one of major differences from the cases
discussed in the previous subsection where we found that 
the total number of oscillons is insensitive to the
detailed shape of the potential around the origin.
Furthermore, we must note that the oscillon formation time 
is largely unchanged for various choices of $\alpha_n$, while
the potential shape around the origin, by using different approximations for the axion monodromy potential $V_A$, strongly affects it. We explain this behavior by using arguments based on linearized analysis of fluctuations in Section~\ref{subsec:VnFloq}.

From these findings, we can conclude that the 
the potential shape at $\phi \ll M$ is highly responsible for the 
growth of $\delta\phi$ fluctuations and thus  determines the oscillon formation time.
Hence small changes in the shape can result in delayed or suppressed oscillon formation.
Once the oscillons are formed, the amplitude of the scalar field becomes larger, even probing values of
$\phi \gg M$, and their macroscopic physical properties
such as the total number depend on the large-field shape of the potential.
The stability of oscillons would also depend on it (see also Ref.~\cite{Ibe:2019vyo}).
However, to see the fate of oscillons in our numerical setup,
we need extremely long computational time, since oscillons have been seen to survive for thousands of oscillation times. Furthermore, the life-time of oscillons in a realistic set-up
would also depend on how the inflaton itself decays into other particles. The existence of multi-components oscillons has been shown in certain cases (see e.g. Refs.~\cite{Graham:2006vy, Gleiser:2011xj, Sfakianakis:2012bq}), but cannot be considered to be a generic behavior.
We leave a more in-depth study of the individual oscillon properties, including their lifetime, for future work.

The size distribution is similar in  all cases, with the oscillon width $w$ ranging between $4\, m^{-1} \lesssim w \lesssim 10\, m^{-1}$. 
In the cases of the potentials $V_1$ and $V_2$, the oscillon width  $w$ shows little difference between the three values of the parameter $\alpha_1$ and $\alpha_2$ used. 
For the case of the potential $V_3$, the case of $\alpha_3=3$ shows a large``spike" at $w\simeq 4.4\, m^{-1}$ and a secondary peak at $w\simeq 6.5\, m^{-1}$, compared to the cases $\alpha_3=1.5$ and $\alpha_3=2$, which show a smoother and broader distribution of oscillon widths. This suggests that the width of the transition region between the quadratic minimum and the plateau of the potential strongly affects the oscillon shape.

The scalar field power spectrum $P(k)$ is also similar between the three cases. Slight differences of the amplitudes appear in the tail of the power spectrum $k\gtrsim 5\,m$ in the case $V_3$, in the middle of the power spectrum $k\sim m$ in the case $V_1$, and in the low-$k$ for $V_2$ and $V_3$. As for the high-$k$ tail, the case of $\alpha_3=3$ shows larger power for $k\gtrsim 5\,m$, which can be responsible for a large``spike" at $w\simeq 4.4\, m^{-1}$ and a secondary peak at $w\simeq 6.5\, m^{-1}$ in the size distribution. As for the middle-$k$, small $\alpha_1$ gives slightly larger power spectrum for $0.5\,m \gtrsim k \gtrsim 3\,m$. This feature can be explained by different growth rates (Floquet exponents), as investigated later in Section~\ref{subsec:VAappFloq}. As for the low-$k$, the case of $\alpha_2=5/3$ shows less power for $k\lesssim 2\,m$ and the cases of $\alpha_3=2,3$ show less power for $k\lesssim 0.5\,m$, which comes from the slow growth of the scalar perturbations, again as investigated later in Section~\ref{subsec:VAappFloq}. This feature can be also responsible for the slow increase of the number of the oscillons. On the other hand, the existence of the peak of the power spectrum $m\lesssim k \lesssim 2m$ appears to be rather robust.

\subsection{From  monodromy to plateau potentials}

Until now we have examined a variety of potential shapes and deformation, mostly focusing on two similarly disjoint families: the axion monodromy potential $V_A$ and its small-field deformations $V_A^{(n)}$ and plateau potentials $V_n$ with varying large-field characteristics. However, these two families can be related to each other. Fig.~\ref{fig:V1-3} shows that the potential type $V_1$ does not asymptote at a finite value for $\phi\to \infty$, but rather grows as $V_1\sim |\phi|^{2-\alpha_1}$. For values of $\alpha_1$ around $2$ we see similar behavior to the ``true" plateau potentials $V_2$ and $V_3$. However, for $\alpha_1\to 1$ this model resembles the axion monodromy potential of  Eq.~\eqref{eq:VA}, in the case of $\alpha_A=1/2$, which is where we focused our attention on.

\begin{figure}[h]
 \centering{
 \includegraphics[width=.24\textwidth]{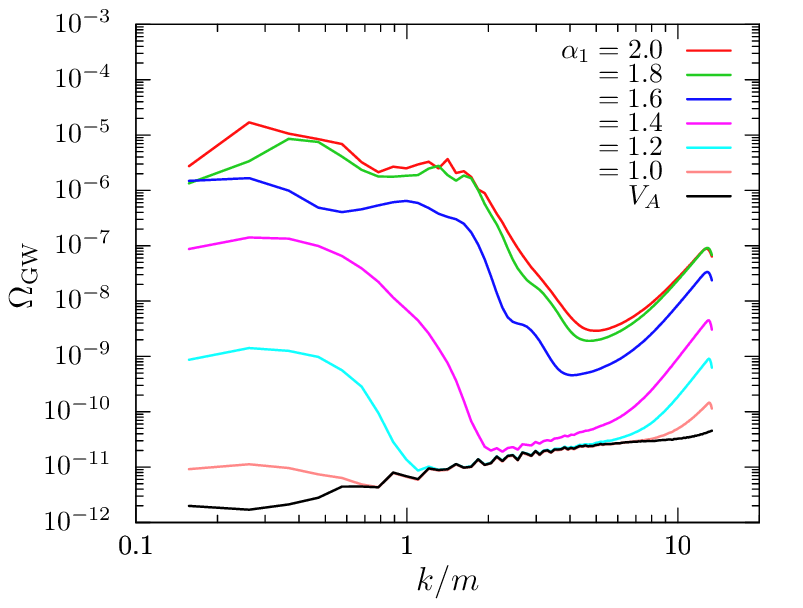}
\includegraphics[width=.24\textwidth]{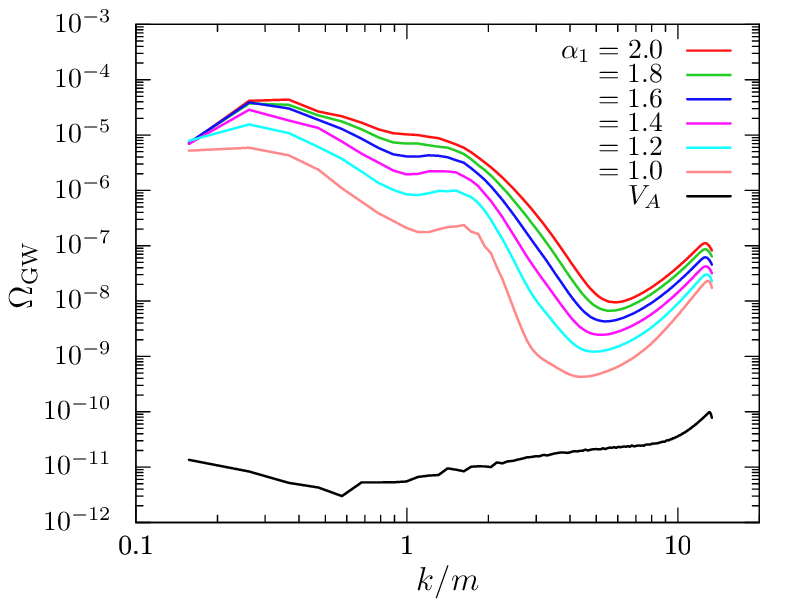}
\includegraphics[width=.24\textwidth]{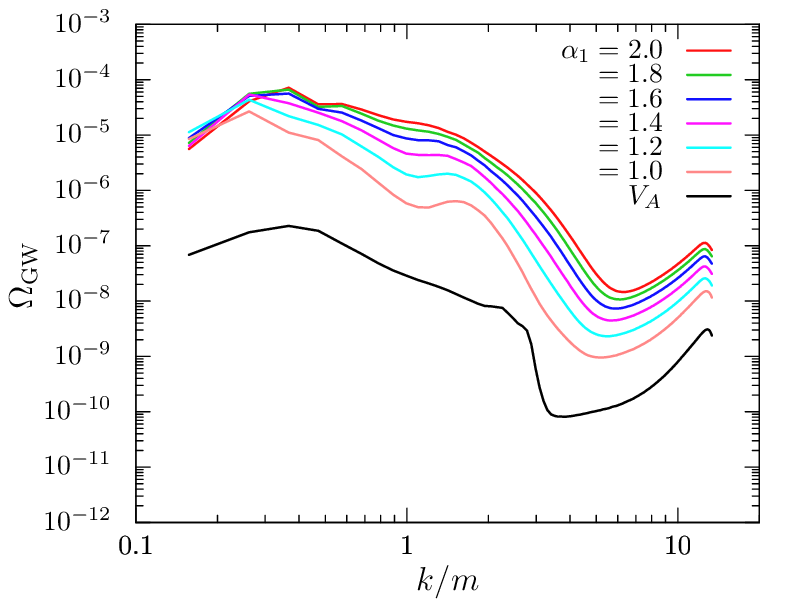}
\includegraphics[width=.24\textwidth]{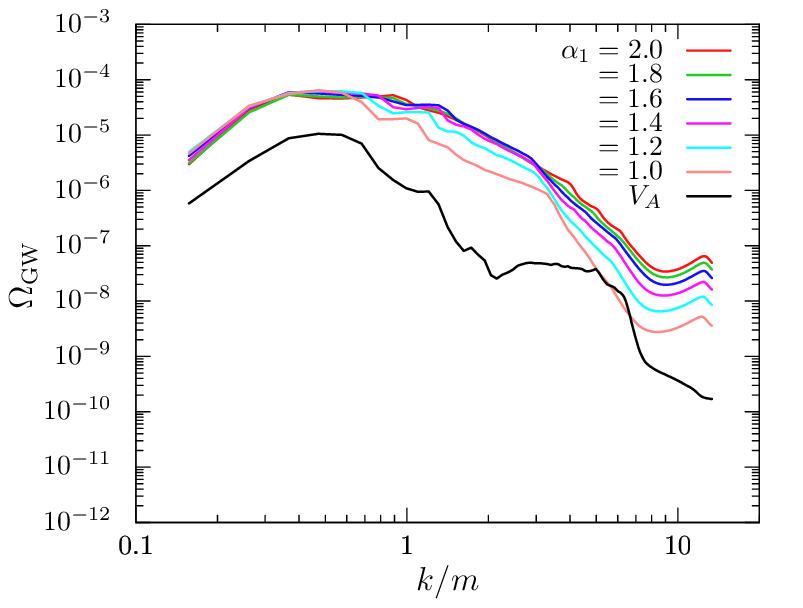}
\\
\includegraphics[width=.24\textwidth]{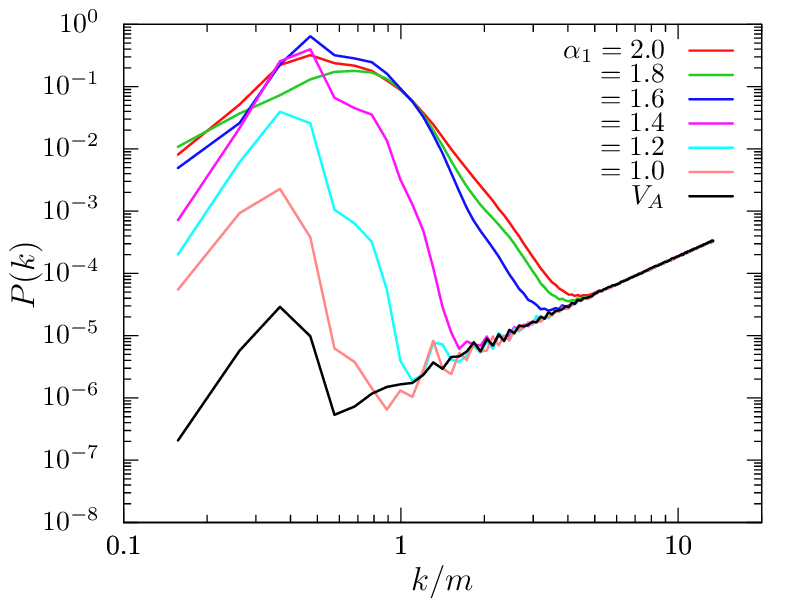}
\includegraphics[width=.24\textwidth]{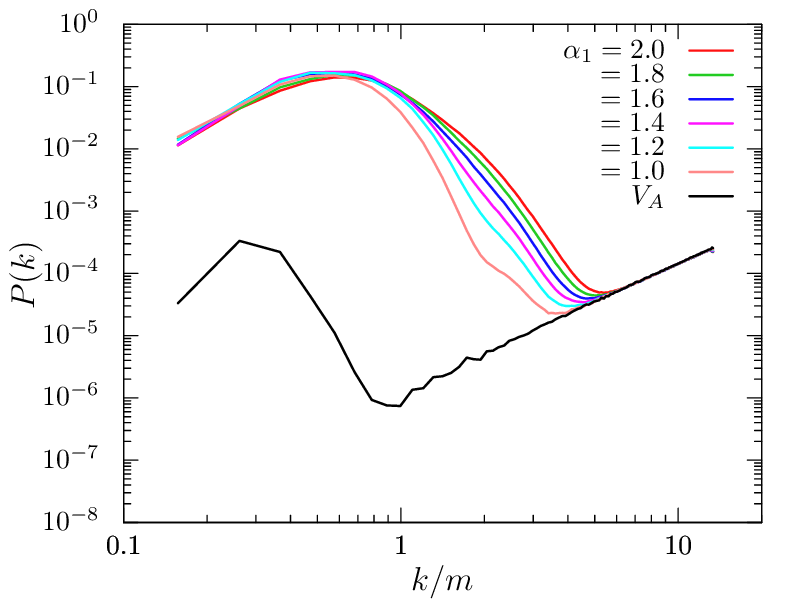}
\includegraphics[width=.24\textwidth]{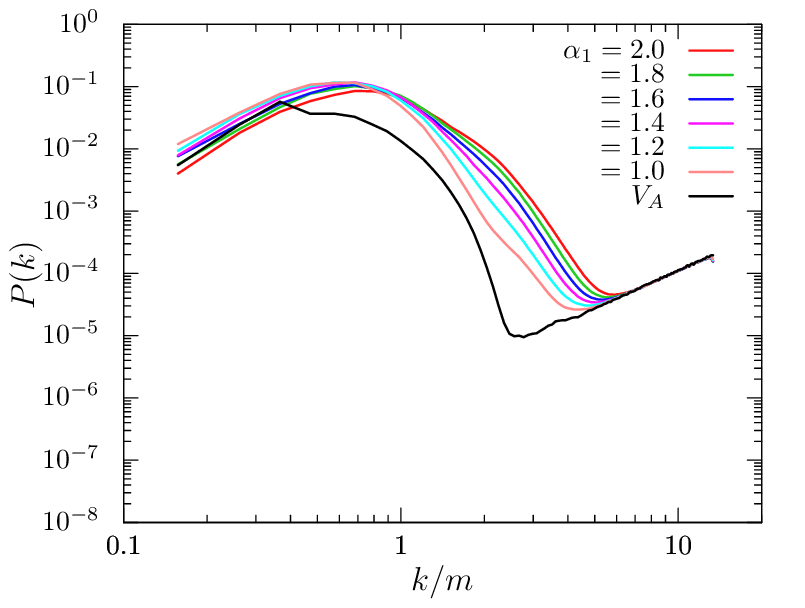}
\includegraphics[width=.24\textwidth]{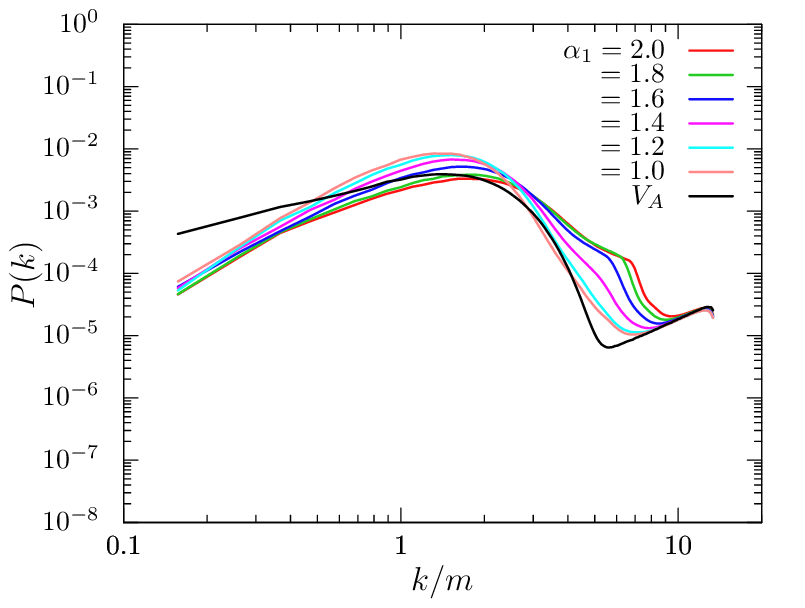}
 }
 \caption{The gravitational wave spectra (upper panels) and the corresponding scalar power spectra (lower panels) for $V_1$ with $a_1=1,1.2,1.4,1.6,1.8,2$ and $V_A$. The times correspond to $m\eta=50,100,150,600$ (left to right).}
 \label{fig:res_VA_V1}
\end{figure}

Fig.~\ref{fig:res_VA_V1} shows the time evolution of the gravitational wave spectra $\Omega_{\rm GW}$ and the scalar power spectrum of the inflaton field $\phi$. We consider several values of $\alpha_1$ ranging from $\alpha_1=2$, which is the fiducial model, to $\alpha_1=1$, which exhibits linear growth at large field values, as in $V_A$. We must note here that the linear growth of $V_1$ is not the same as that of $V_A$, namely $V_1\sim {1\over 2} m^2 M|\phi|$ as opposed to $V_A\sim m^2 M|\phi|$. This is necessary  in order for the two models to have the same mass at small field values, $m^2= \partial^2V|_{\phi=0}$. 
The first observation is in all cases of $V_1$, both the scalar and GW spectra start growing significantly earlier than in the case of $V_A$. By $\eta=100\,m^{-1}$ the scalar power spectra for $V_1$  have largely equilibrated,  regardless of the value of $\alpha_1$. The same is seen for $V_A$ at $\eta=150\,m^{-1}$. At the end of the simulation, $\eta=600\,m^{-1}$ the scalar power spectra for $V_1$ are very similar to each other. At low $k$, they are identical. However a sifference is visible at $k\sim 10\,m$, where we see more power for larger values of $\alpha_1$. On the contrary the scalar power spectrum for $V_A$ exhibits more power at low $k$ compared to $V_1$. At large $k$, the scalar power spectrum for $V_1$ with $\alpha_1=1$ is closer to that of $V_A$ than to that of $V_1$ with $\alpha_1=2$.

A more interesting observation arises for the produced GW spectrum. We see that all GW spectra for $V_1$ exhibit a two-peak structure, similar to that of $V_A$, albeit at earlier times. Specifically, the two peaks are clearly visible when $\Omega_{\rm GW}$ peaks around
$10^{-6}\lesssim \Omega_{\rm GW}|_{\rm max} \lesssim 10^{-5}$, while they are smooth for smaller values. The small values of $\Omega_{\rm GW}$, for example in the case of $V_1$ with low $\alpha_1$ at $\eta=50\,m^{-1}$ correspond to GW emission by inflaton fluctuations still being in the linear regime. When non-linearities become important and oscillons are formed, the two-peak structure appears.
The relation of the wavenumbers of the peaks and dips of the GW spectrum to the frequency content of oscillons was explained in Ref.~\cite{Zhou:2013tsa}.
However, when non-linearities occur early in our simulation, the GW spectrum evolves further towards a featureless ``equilibrated" distribution, reaching a maximum value of $\Omega_{\rm GW}|_{\rm max}\simeq 10^{-4}$. This occurs later for $V_1$ with smaller values of $\alpha_1$. For $\alpha_1=1$ the two peaks are clearly visible at $\eta=150\,m^{-1}$ but have largely disappeared for $\eta=600\,m^{-1}$.
On the contrary, the scalar power spectrum for $V_A$ enters the non-linear regime later and the features in the GW spectum appear later and remain there until the end of the simulation. The two-peak structure of the GW spectrum for $V_A$ has largely equilibrated by $\eta\simeq 400\, m^{-1} $, so we do not expect that running the simulation for much longer will result in a complete ``smearing" of the two peaks.

\begin{figure}[h]
 \centering{
 \includegraphics[width=.24\textwidth]{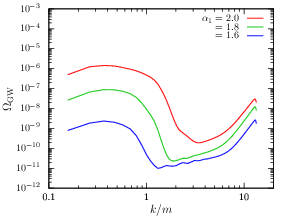}
\includegraphics[width=.24\textwidth]{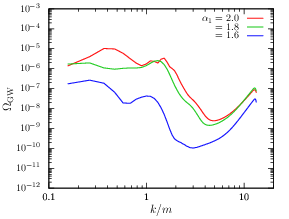}
\includegraphics[width=.24\textwidth]{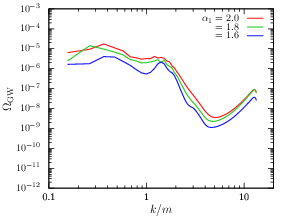}
\includegraphics[width=.24\textwidth]{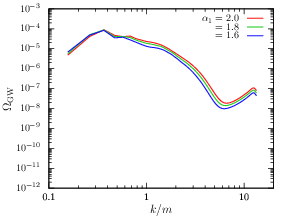}
\\
\includegraphics[width=.24\textwidth]{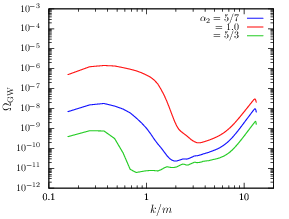}
\includegraphics[width=.24\textwidth]{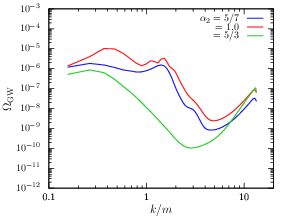}
\includegraphics[width=.24\textwidth]{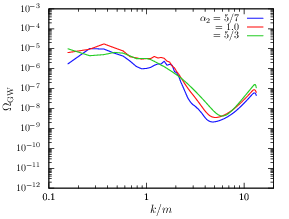}
\includegraphics[width=.24\textwidth]{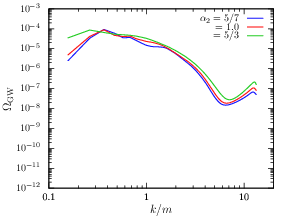}
\\
\includegraphics[width=.24\textwidth]{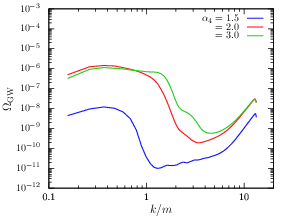}
\includegraphics[width=.24\textwidth]{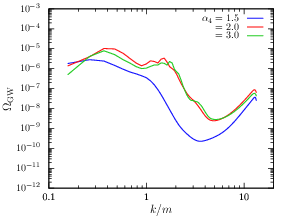}
\includegraphics[width=.24\textwidth]{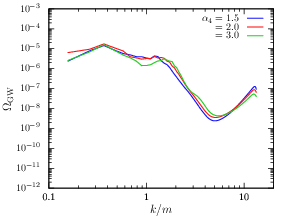}
\includegraphics[width=.24\textwidth]{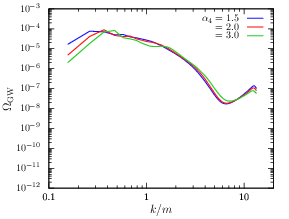}
 }
 \caption{The gravitational wave spectra for $V_1, V_2, V_3$ (upper, middle and lower panels respectively) 
 for times $m\eta = 35, 45, 55, 200$ (left to right)
}
 \label{fig:timespectra_Vn}
\end{figure}

Figure \ref{fig:timespectra_Vn} shows the time evolution of the gravitational wave spectra for the three plateau potentials $V_n$, each simulated for three values of the  corresponding parameter $\alpha_n$. We distinctly see three largely identical regions in all cases
\begin{itemize}
\item During the early period $\eta\lesssim 35\,m^{-1}$ the gravitational wave spectrum closely follows the linear scalar power spectrum of ${\cal P}(k)$, which only shows excitation for modes with comoving wavenumbers $k\lesssim m$. 
\item After the field excitations reach the point, where non-linear effects become important, we see a cascade of power in the scalar spectrum towards the UV, at $\eta\simeq 35\,m^{-1}$. The timing is different for each pair $\{ V_n, a_n\}$, but the overall behavior remains. The timing difference will be explained in Section \ref{subsec:VnFloq} with the use of Floquet theory and the analysis of linearized perturbations.
\item When oscillons are formed, their internal frequency content leads to a power deficiency (a ``dip") in the GW spectrum, leading to a two-peak shape, as discussed in Ref.~\cite{Zhou:2013tsa}. We see a two-peak structure appearing for all case of $V_n$ for $45\,m^{-1} \lesssim \eta \lesssim 55\,m^{-1} $. 
\item After this rather brief period of time, the GW spectra lose memory of the two-peak shape and exhibit a single broad peak for $k>10\,m^{-1}$. This shape is largely universal, with some minor differences between some cases, discussed in the previous section.
\end{itemize}

\section{Linear analysis}
\label{sec:linear}

The time-evolution of the power spectrum of the scalar field $\phi$ points towards the existence of instability bands, causing certain wavenumbers to undergo parametric resonance and exponential enhancement. In order to better understand the numerical results of Section~\ref{sec:results}, we perform a Floquet analysis, by neglecting the expansion of the universe and approximating the motion of the background inflaton field as being purely periodic, without any redshifting due to Hubble drag. Since we are interested in sub-horizon scales, the static universe approximation will capture the essential dynamics.

In the static universe approximation, the  equation of motion for the scalar perturbations $\delta\phi$
in Fourier space is given as
%
\begin{align}
 \ddot{\delta\phi}(t,\kk) + \left(k^2 + V''(\bar \phi)\right)\delta\phi = 0,
 \label{eq:deltaphiFloq}
\end{align}
%
where {the dots represent  derivatives with respect to cosmic time
$t$ and} the background field $\bar\phi$ satisfies 
%
\begin{align}
 \ddot{\bar \phi}(t) + V'(\bar \phi) = 0.
  \label{eq:phiFloq}
\end{align}
Eq.~\eqref{eq:deltaphiFloq} can be written as a matrix first-order equation
\begin{gather}
 {d\over dt} \left ( \begin{matrix} \delta\phi_k \\ \dot{\delta\phi_k} \end{matrix} \right )
 =
 \left ( \begin{matrix}
 0 & 1
   \\
 -(k^2+V''(\bar \phi))  & 0
   \end{matrix}
   \right )
   \left ( \begin{matrix} \delta\phi_k \\ \dot{\delta\phi_k} \end{matrix} \right )
   \label{eq:phiFloqmatrix}
\end{gather}
This equation is of the form 
\beq
\dot x(t) = {\cal P}(t)\cdot x(t)
\eeq
where ${\cal P}(t)$ is a periodic matrix, whose period is controlled by the background field motion $\bar\phi(t)$.
According to Floquet's theorem, the solution of the above equation is of the form $x(t)=e^{\mu t}Q(t)$, where $Q(t)$ is also periodic with period T. The quantity $\mu$ is the Floquet exponent. When it has a positive real part, it causes exponential enhancement of the relevant mode. In what follows, we will only focus on the real part of $\mu$ and call this the Floquet exponent.
%
We compute the instability chart by using the algorithm presented in Ref.~\cite{Amin:2014eta}. We solve Eq.~\eqref{eq:phiFloq} for the background field and compute the period of background oscillations. We then solve the fluctuation equation, Eq.~\eqref{eq:phiFloqmatrix},
using the initial conditions $\{ \delta\phi/M,\dot\delta\phi/(Mm) \}=\{1,0\}$ and $\{ \delta\phi/M,\dot\delta\phi/(Mm) \}=\{0,1\}$. Finally, we can extract the largest Floquet exponent, which signals the existence of instability bands when it has a positive real part.
The instability is controlled by two parameters, the  
amplitude of the background field $\bar \phi(t)$ and the physical wavenumber $k$.

\subsection{Small-field dependence}

\begin{figure}[h]
 \centering{
 \includegraphics[height=6cm]{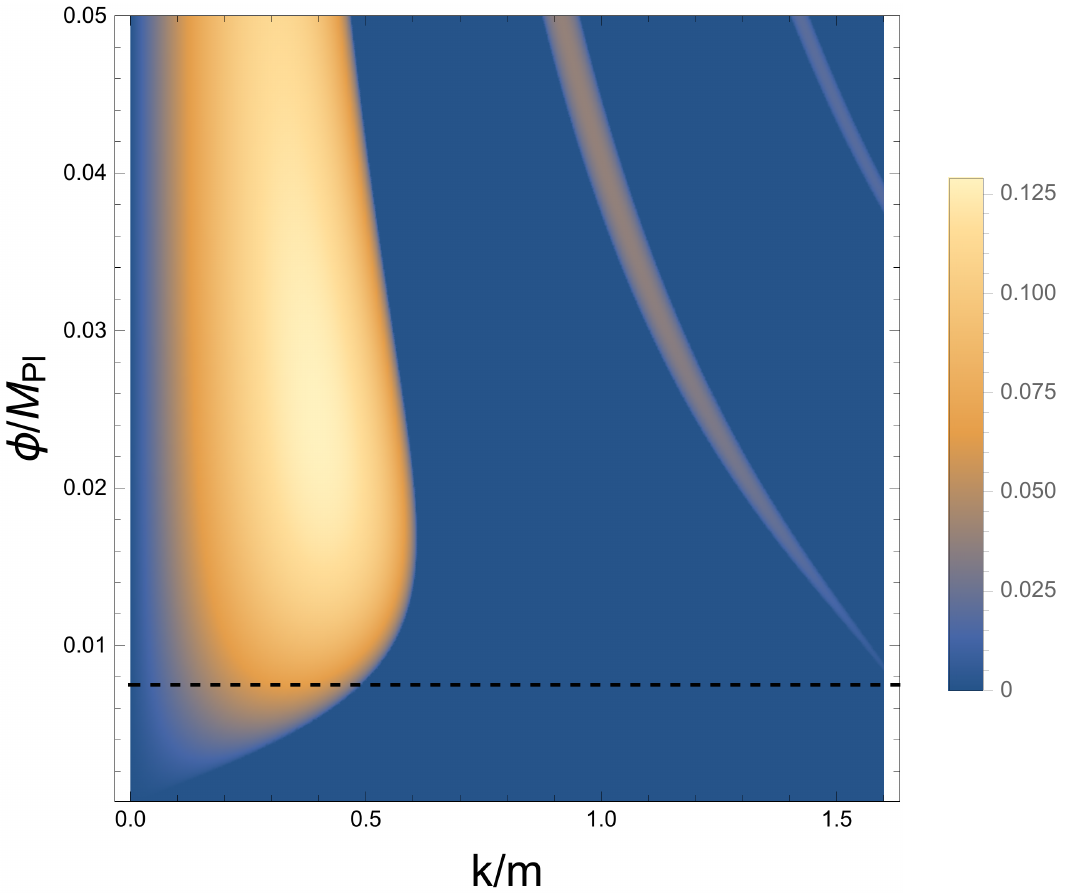}
 \includegraphics[height=6cm]{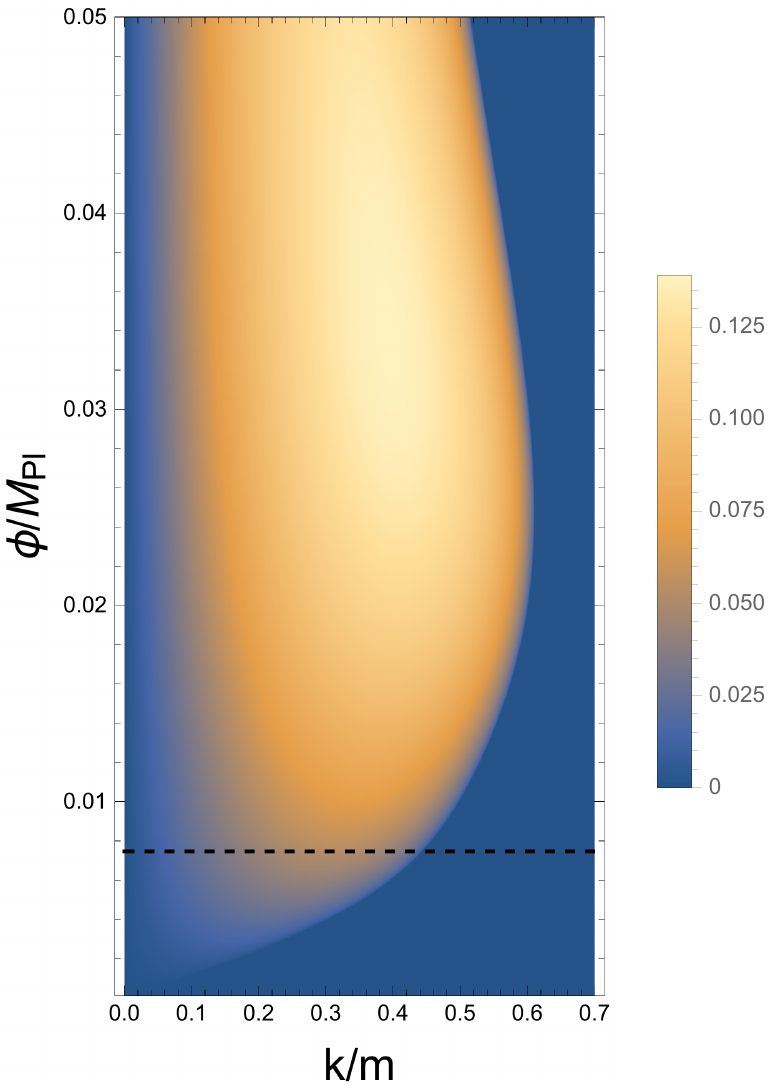}
  \includegraphics[height=6cm]{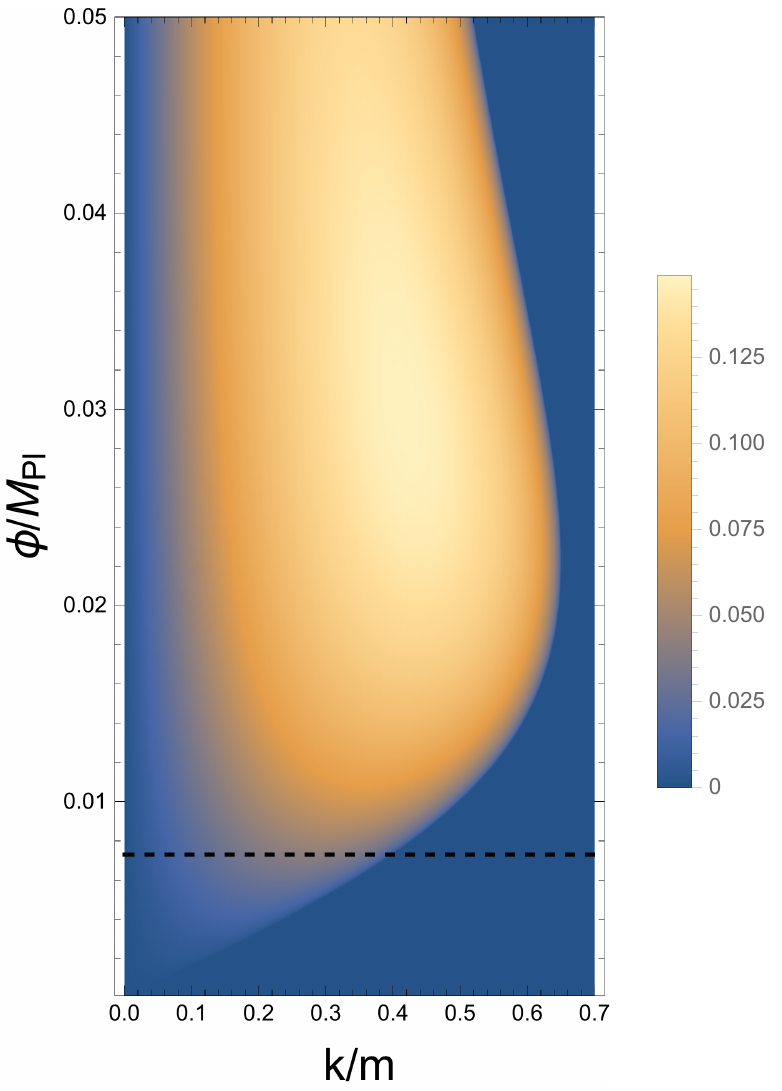}
\\
 \includegraphics[width=.23\textwidth]{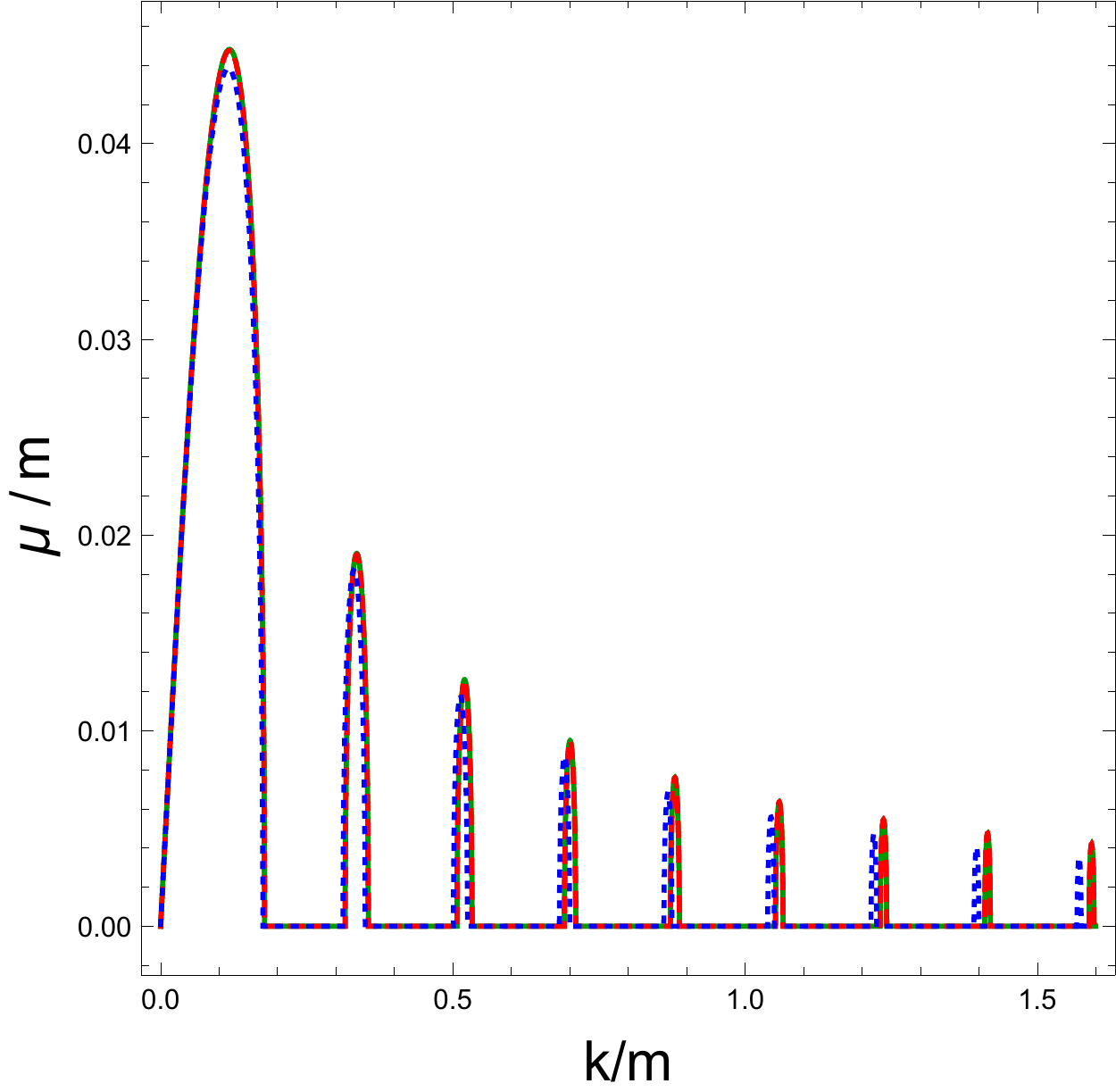}
 \includegraphics[width=.23\textwidth]{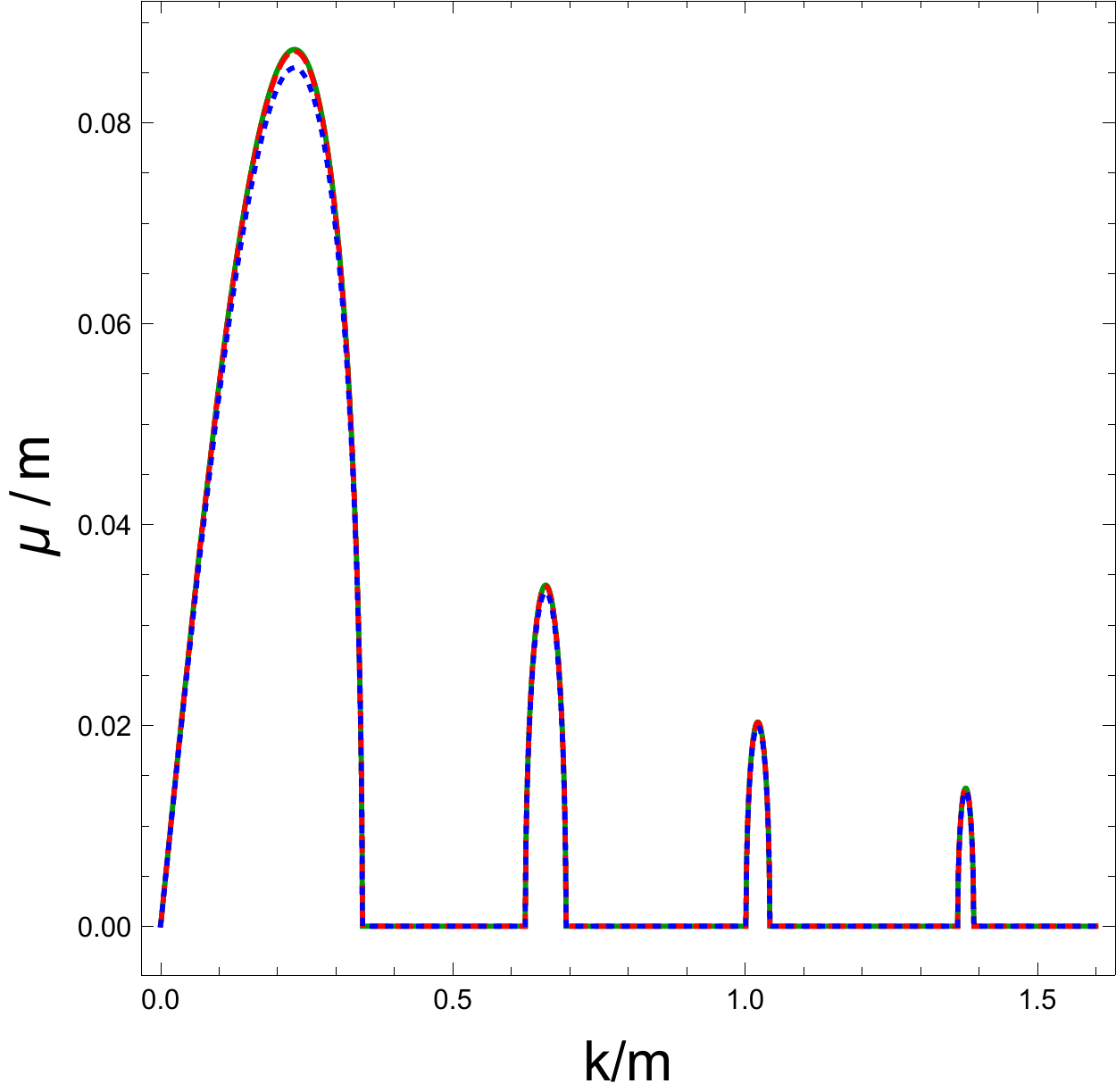}
 \includegraphics[width=.23\textwidth]{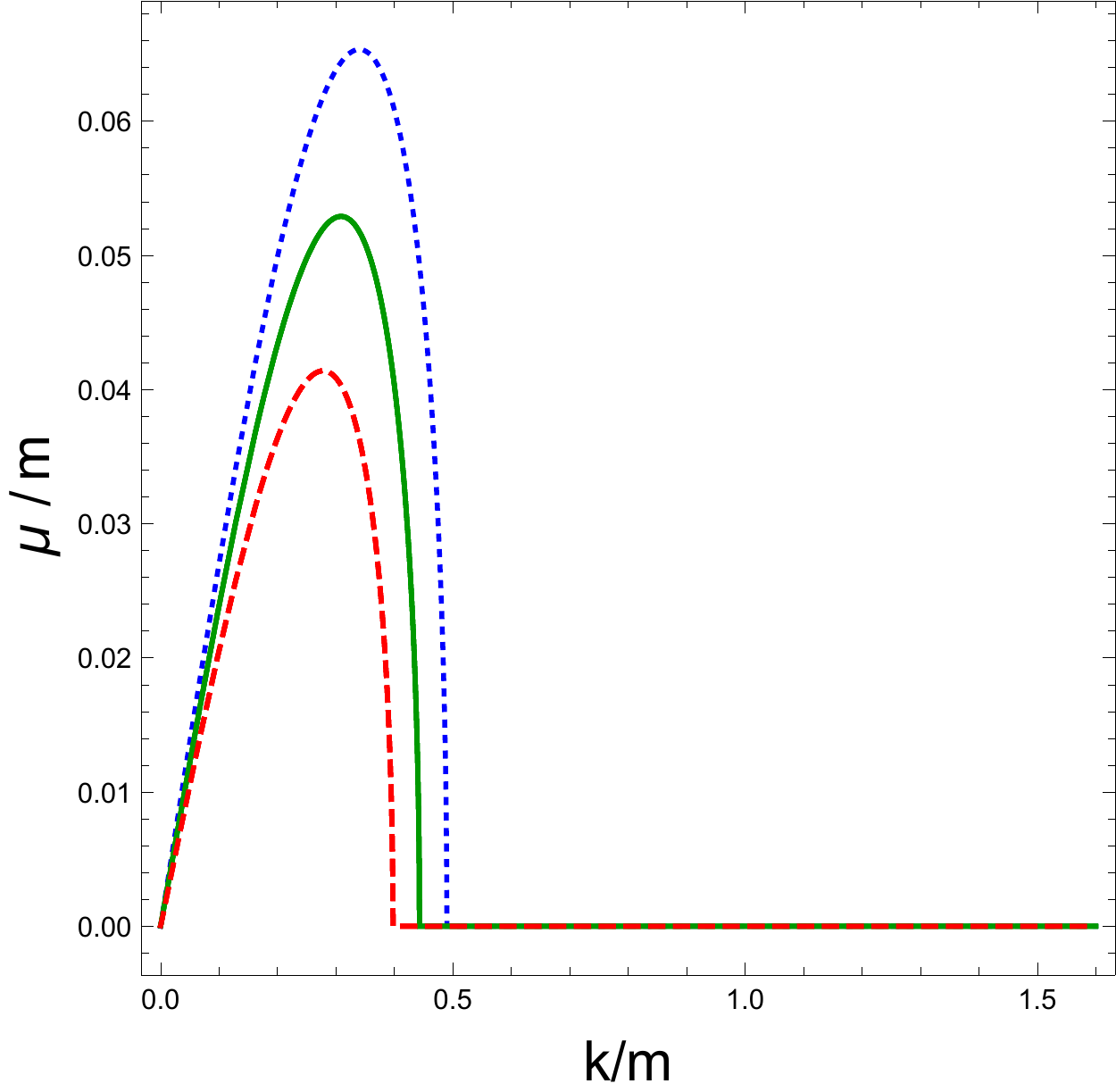}
 }
 \caption{{\it Top row, left to right:} The Floquet chart for $V_A$ and the Floquet chart focusing on the main instability band for $V_A^{(6)}$ and $V_A^{(4)}$.
 The value of the mass-scale is chosen as $M=10^{-2} M_{\rm pl}$. The field $\phi$ is measured in units of $M_{\rm pl}$ and the (physical) wavenumber $k$ is measured in units of $m$. The black dashed line shows the value of $\phi$ at the start of the simulation, given by $V=m^2M^2/4$, being $0.73\lesssim \phi/M\le0.75$ for the three potentials. 
 Inflation in all cases ends at $\phi\simeq 0.4 M_{\rm pl}$. 
 \\
 {\it Bottom row:} The Floquet exponent for the background field amplitude taken at the end of inflation (left), taken to be $\bar\phi=10M$ (center) and taken from the condition $V=M^2m^2/4$ (right). The blue-dotted, green and red-dashed curves correspond to $V_A$, $V_A^{(6)}$ and  $V_A^{(4)}$ respectively.
 }
 \label{fig:FloqVA}
\end{figure}

We start by exploring the parametric resonance behavior of the axion
 monodromy potential $V_A$ with $\alpha=1/2$, along with the
 approximations $V_A^{(4)}$ and $V_A^{(6)}$. In all three cases we choose
 $M=10^{-2}M_{\rm pl} {\,(\epsilon_G = 10^{-2})}$ and inflation ends at $\phi\simeq 0.4M_{\rm pl}$. Fig.~\ref{fig:FloqVA} shows the Floquet instability charts for each potential, as a function of wavenumber $k$ and background field amplitude $\phi$. As expected, the instability bands look identical for the exact and approximate potential for large field values, since all three potentials asymptote to $V\sim m^2 M |\phi|$ for $\phi \gg M$. However, for $\phi \lesssim M$, the main instability bands depends on the exact potential shape. For the background field amplitude taken from the relation $V =m^2M^2/4$,
 the main instability band is larger for the exact potential than for the approximate ones. This leads to a stronger instability for $V_A$ as opposed to $V_A^{(6)}$ and $V_A^{(4)}$. This explains our finding that oscillon creation occurs earlier for potentials that are closer in shape to $V_A$. However, the oscillons --when formed-- probe the potential beyond the minimum, hence all three potentials provide identical number density of oscillons, within the accuracy limits of our simulations.
 
\begin{figure}[h]
 \centering{
   \includegraphics[width=.3\textwidth]{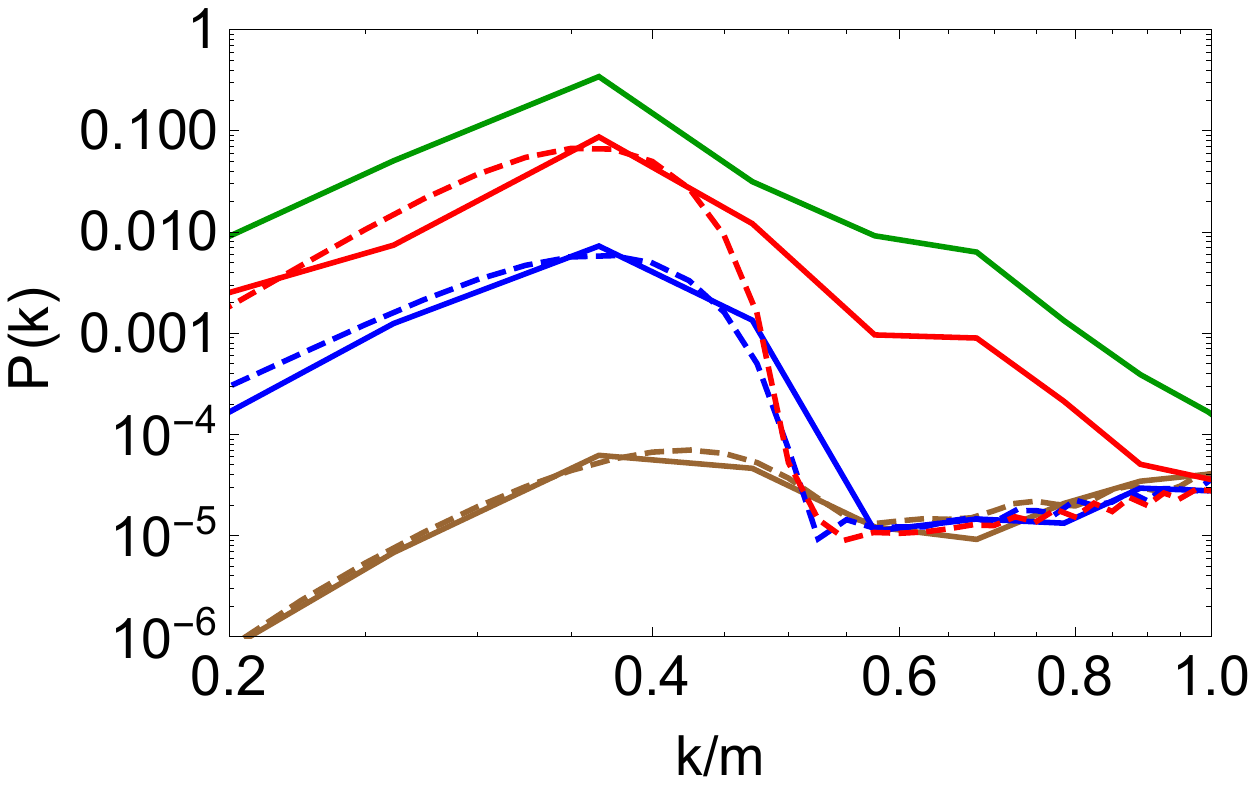}
    \includegraphics[width=.3\textwidth]{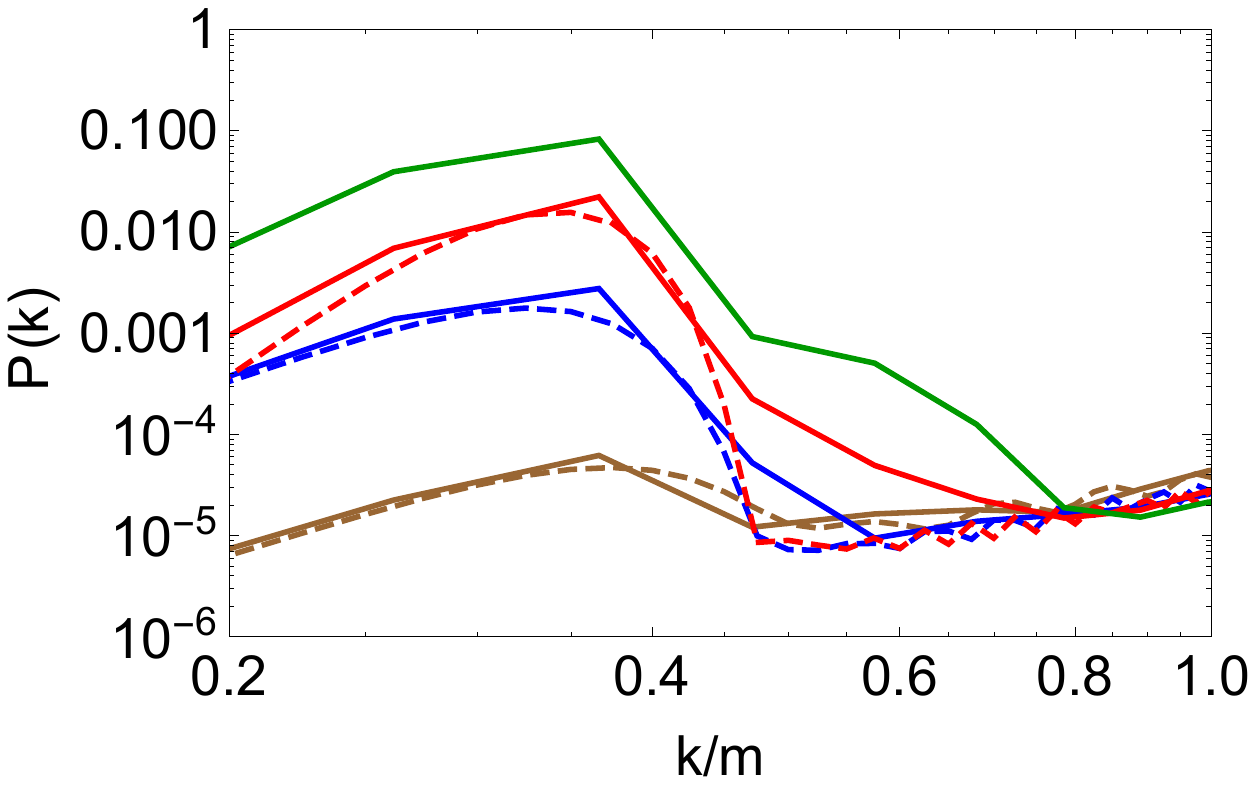}
    \includegraphics[width=.3\textwidth]{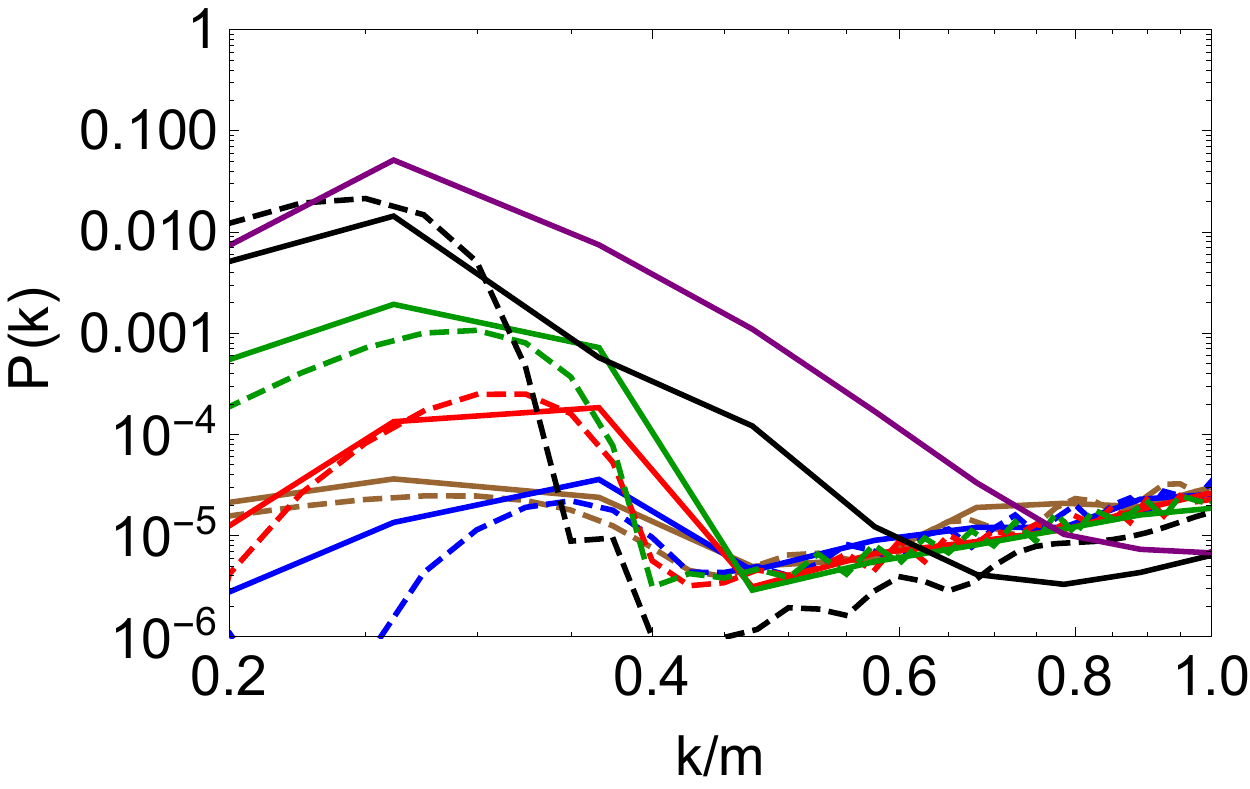}
   }
 \caption{
The power spectrum of the inflaton field computed using the linear fluctuation equations (solid) and using the full lattice simulation (dashed) for $V_A$ (left), $V_A^{(6)}$ (center) and $V_A^{(4)}$ (right). The color-coding corresponds to different times $\eta=  35, 70, 110. 120, 200, 270 $ in units of $m^{-1}$
(brown, blue, red, green, black and purple respectively).
The black and purple curves only appear on the 
 the right panel ($V_A^{(4)}$), where oscillon formation is delayed compared to $V_A$ and $V_A^{(6)}$.
Note that here we do not plot times much after oscillon formation, which are shown in Fig.~\ref{fig:VA_gw} for $V_A$. {$k$ is the comoving wavenumber with $a(\eta_0)=1$, as in the figures showing the results of the full lattice simulation.} 
 }
 \label{fig:Pk_VA}
\end{figure}

 Going one step further, we solve the linear fluctuation equation for $\delta\phi$ on a self-consistently expanding background $\{ \phi, H\}$, by neglecting non-linearities and back-reaction effects. Other than that, we are using the initial conditions and parameter values that were used in the full lattice simulation. 
Comparing the linear fluctuation spectra to the ones obtained from lattice simulations, we can see the effects of back-reaction and oscillon formation through a deviation of the full numerical spectra from the linearized ones. Fig.~\ref{fig:Pk_VA} shows the two sets of power spectra for the three cases $V_A$, $V_A^{(6)}$ and $V_A^{(4)}$. Before going into details, we can immediately see excellent agreement between the initial and final power spectra. For the full monodromy potential $V_A$, we see that the full numerical spectrum deviates from the linear approximation for $\eta\gtrsim110 \, m^{-1} $. Referring back to Fig.~\ref{fig:res_VAapp}, we see that this is approximately the time at which oscillons emerge. 
Soon after that point, the scalar power spectrum loses all similarity with the linearized approximation, which shows that it is  governed by non-linear structure formation and field self-interactions and not by Floquet theory. Furthermore, while the spectrum seems to evolve when plotted in the axes of Fig.~\ref{fig:Pk_VA}, it does not when plotted as a function of the physical wavenumber $k/a$. 
This indicates that the peak of the distribution is governed by the typical scale of inflaton fragmentation, which is related to the typical oscillon size.
Similarly the peak of the distribution does not change at late times, when rescaled by the scale-factor cubed.
We must note, that the range of time-slices plotted in Fig.~\ref{fig:FloqVA} is smaller than the one plotted in Fig.~\ref{fig:VA_gw}, because here we are interested in the beginning of oscillon formation, not their late-time behavior.
The case of $V_A^{(6)}$, shows a slightly smaller but similar growth of fluctuations as $V_A$ and a similar time of breakdown of the linear approximation. 
The case of $V_A^{(4)}$ shows significantly different dynamics, albeit reaching a similar final state. The initial parametric resonance is weaker and the range of excited wavenumbers is smaller. However,
we see a deviation of the scalar power spectrum from the linear result for $\eta\gtrsim 200 \, m^{-1}$, when the peak power spectrum of $\phi$ fluctuations reaches  values of ${\cal P}(k)={\cal O}(0.01)$, at which time the field has a large enough amplitude to start probing non-linearities\footnote{Using Parseval's theorem, we can relate the typical field displacement in real space by using the power spectrum $\langle \delta\phi^2 \rangle \propto \int d^3k P(k)$.
As $P(k)$ grows, the typical field displacement will reach $\langle \delta \phi^2 \rangle={\cal O}(M^2)$, at which point the field will probe the non-linear regime of the potential.}. While oscillon formation in this case is not as robust as in the cases of $V_A$ and $V_A^{(6)}$, the inflaton field fragments in a similar way, albeit at a later time. This leads to the final scalar and GW spectra being very similar, even though the initial Floquet charts are  different in three cases, exhibiting a maximum Floquet exponent that is more than $50\%$ larger for $V_A$ than for $V_A^{(4)}$.
In order to make sure that the difference in Floquet charts is not merely an artifact due to the different initial value of the field, we computed the maximal Floquet exponent for $0.4\le \phi/M\le 0.75$ and found that the maximum value of $\mu$ for $V_A$  is consistently $50\%$ or more larger than the maximum value of $\mu$ for $V_A^{(4)}$.

\label{subsec:VAappFloq}

\subsection{Large-field dependence}
\label{subsec:VnFloq}

We now examine parametric resonance in the three  model potentials $V_n(\phi)$ described in Section~\ref{subsec:survey}, in order to disentangle the contribution of the different potential features, such as the height of the asymptotic plateau. The field amplitude at the end of inflation $\phi_{\rm end}$, defined as the time when  $\epsilon=1$, can be analytically computed using the slow-roll equations of motion given in Appendix~\ref{sec:infl}, where  $\phi_{\rm end}$  is also computed numerically and shown in Fig.~\ref{fig:phiVend}. However, our lattice simulations are initialized at a later time, when the potential equals $V=m^2M^2/4$. For $V_1$ this corresponds to $\phi=M$ regardless of the value of the parameter $\alpha_1$.  For $V_2$ the background field value at the start of our simulations is $\phi = M/\sqrt{2-\alpha_2}$ and ranges from $0.9M$ to $1.7M$ for the values of $\alpha_2$ shown in Fig.~\ref{fig:V1-3}. For the potential $V_3$ the corresponding field value is $\phi =M/(2^{\alpha_3/2}-1)^{1/\alpha_3}$
and ranges from $0.8M$ to $1.2M$ for the parameter values shown in Fig.~\ref{fig:V1-3}.

\bigskip

\begin{figure}[h!]
 \centering{
   \includegraphics[width=.28\textwidth]{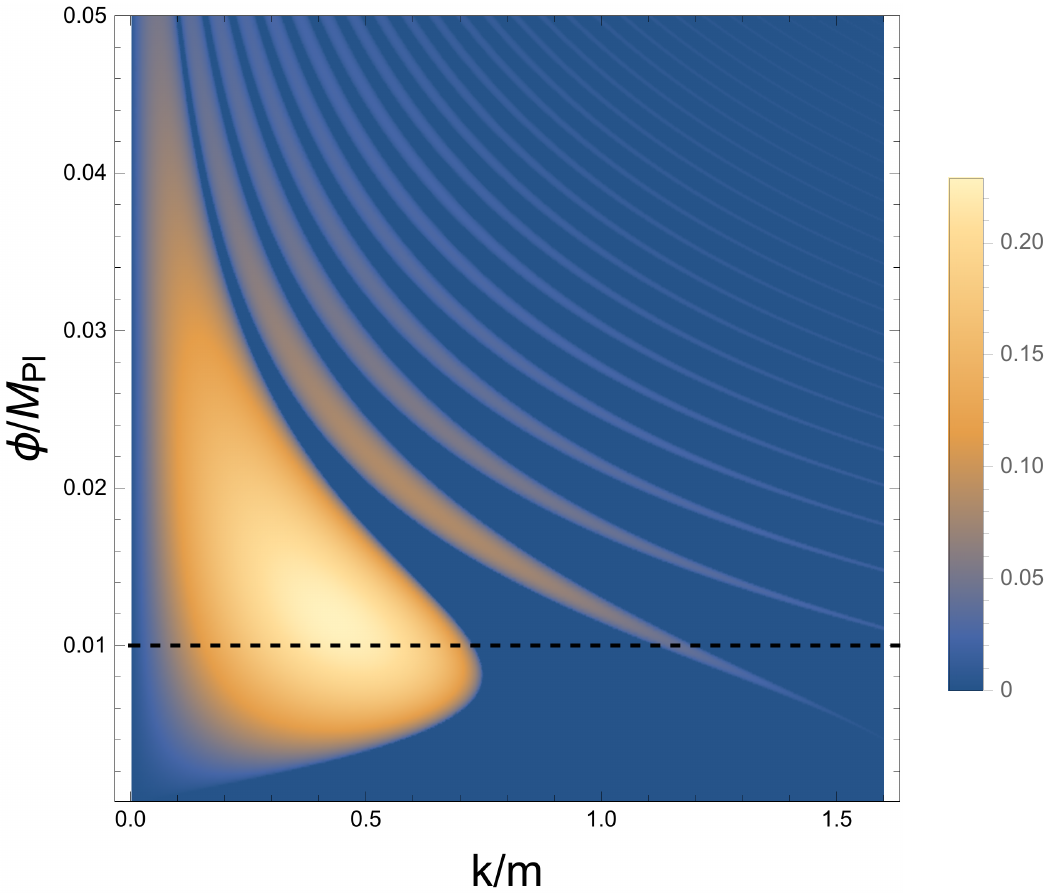}
  \includegraphics[width=.23\textwidth]{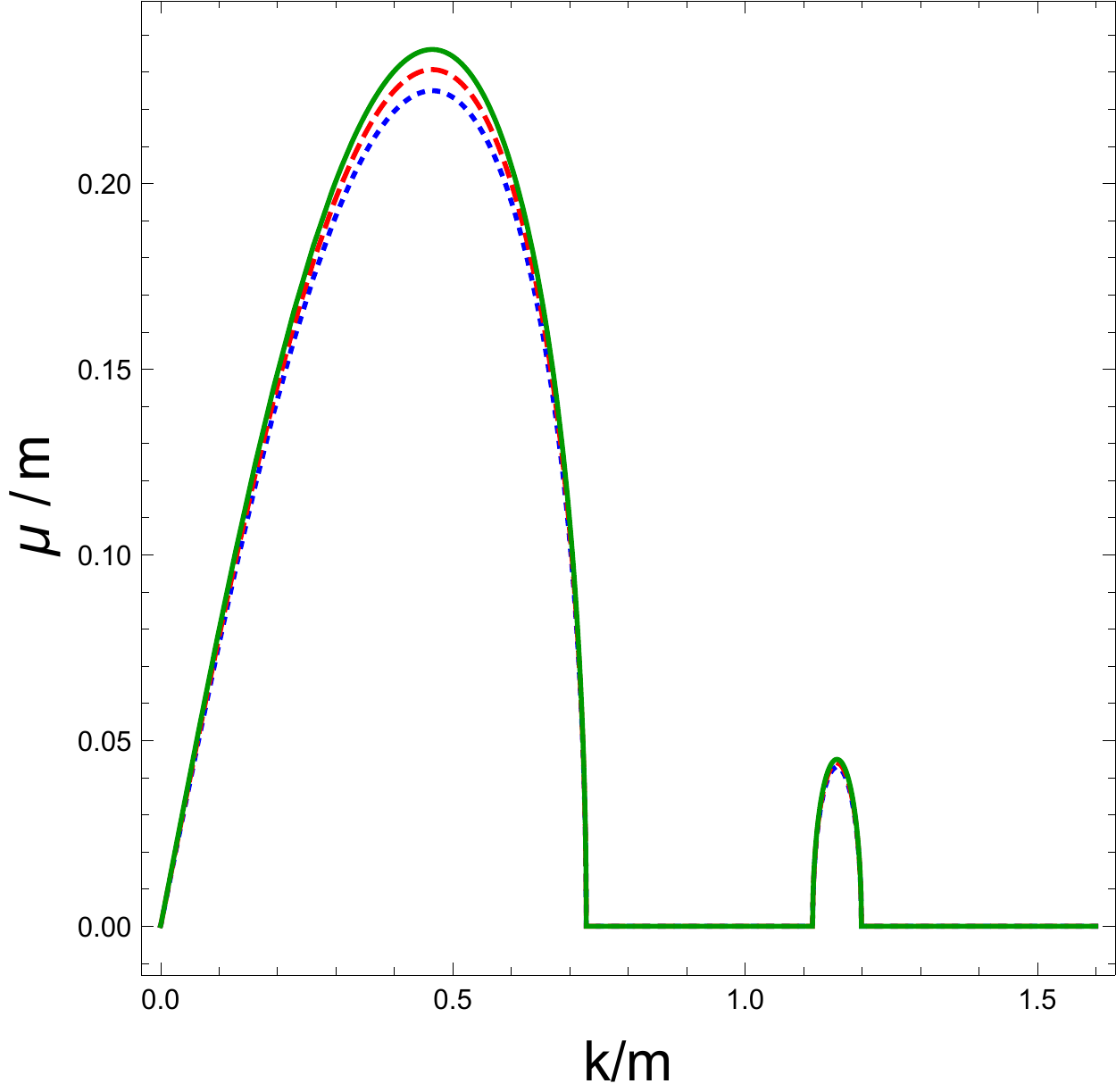}
    \includegraphics[width=.23\textwidth]{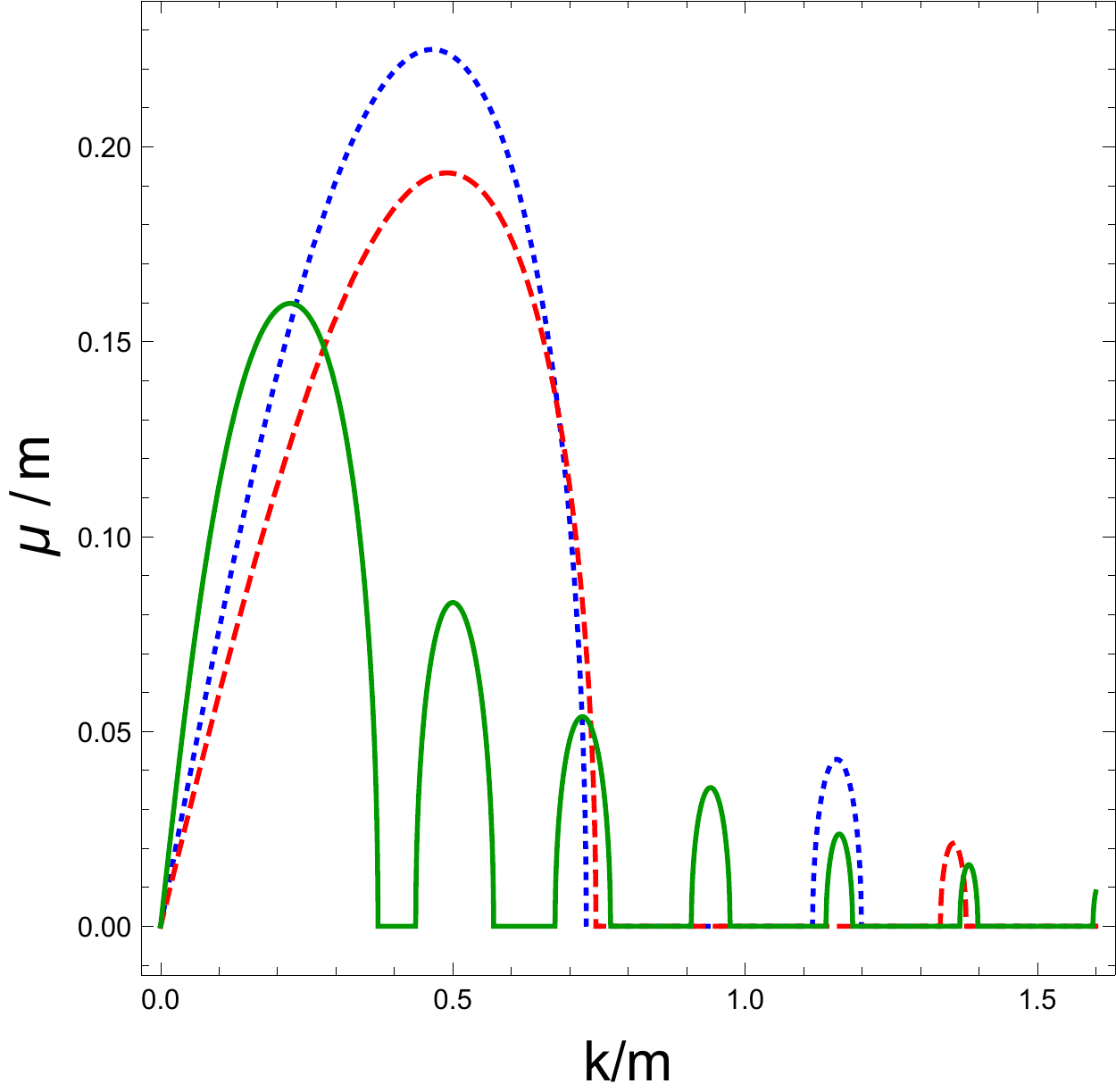}
      \includegraphics[width=.23\textwidth]{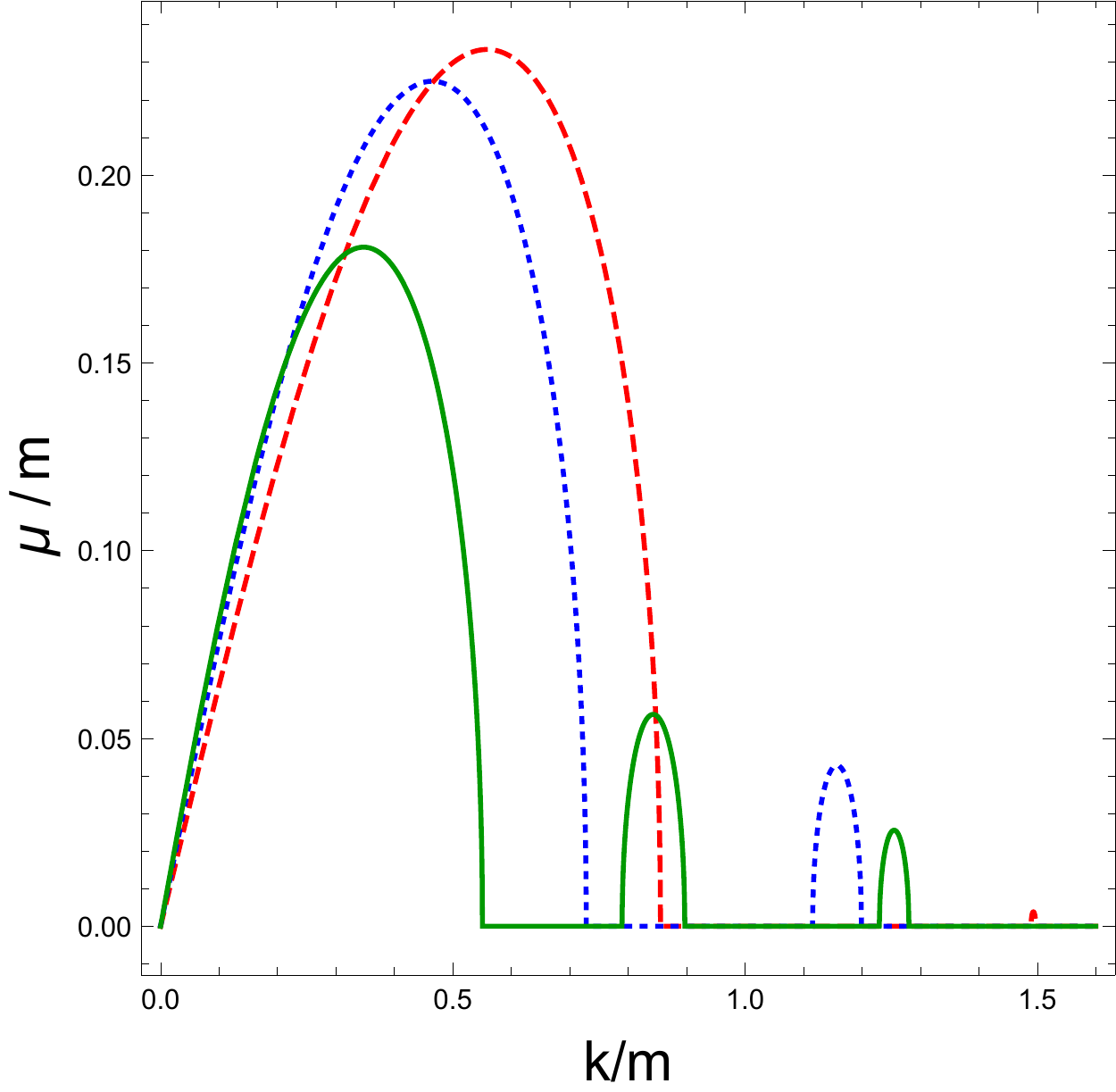}
   }
 \caption{
{\it Left:} The Floquet instability chart for the case where all there potentials $V_n$ overlap, $V_1(\phi;\alpha_1=2)=V_2(\phi;\alpha_2=1)=V_3(\phi;\alpha_3=2)
= {1\over 2} m^2 {M^2} \phi^2 / ( M^2  + \phi^2) $ and ${M=10^{-2}}\, M_{\rm pl}$. The horizontal black-dashed lines corresponds to $\phi=M$.
\\
{\it From left to right:} The instability chart for each potential, the three values of $\alpha_n$ used for our simulations, $M=10^{-2}\, M_{\rm pl}$ and the field amplitude $\phi$ given by $V=m^2M^2/4$. Color-coding goes as follows:
$\alpha_1=2, 1.6, 1.8$ blue-dotted, red-dashed and green-solid respectively; 
$\alpha_2=1, 5/7, 5/3$ blue-dotted, red-dashed and green-solid respectively; 
$\alpha_3=2, 3, 1.5$ blue-dotted, red-dashed and green-solid respectively.
The blue-dotted curves in all panels correspond to the case where all there potentials $V_n$ overlap.
We  see that the Floquet exponents for $V_n$ are significantly higher than those of $V_A$, shown in Fig.~\ref{fig:FloqVA}, while they are similar between the various cases with the exception of $V_2(\alpha_2=5/3)$ and $V_3(\alpha_3=1.5)$. This is further discussed in the main text.
 }
 \label{fig:FloqVnoneD}
\end{figure}

By using the algorithm described below Eq.~\eqref{eq:phiFloqmatrix}, we
compute the Floquet charts for the three potentials and the three
parameter values used for each potential. We must note again that there
is a ``crossover'' point, where the three potentials $V_n$ have the same
form $V_1(\phi;\alpha_1=2)=V_2(\phi;\alpha_2=1)=V_3(\phi;\alpha_3=2)=
{1\over 2} m^2 {M^2}  \phi^2 / ( M^2  + \phi^2) $. 
This can be used as the ``prototype" potential, against which to compare any modifications, according to the parameters $\alpha_1,\alpha_2,\alpha_3$.
The density plot of the instability bands for this value is shown in the far left panel of Fig.~\ref{fig:FloqVnoneD}. The qualitative form of the 2-D Floquet
charts for the other cases is similar. Instead we show the Floquet
exponent as a function of wavenumber for the starting value of the
background field $\phi$ at the start of our simulations. For most cases,
this is close to $\phi=M$, except in the case of $V_2$ with $\alpha_2=
5/3$, where the starting value is $\phi =\sqrt{3}M \simeq
1.7M$. Fig.~\ref{fig:FloqVnoneD} shows that the Floquet exponent for
almost all cases that we simulated is similar, leading to similar
initial enhancement of the fluctuations $\delta\phi$ and a similar time
of emergence for the produced oscillons. Furthermore, the Floquet
exponents are larger than {those} in the case of the axion monodromy potential $V_A$, leading to an earlier
emergence of non-linear effects, inflaton fragmentation and oscillon
formation. The case $V_2(\alpha_2=5/3)$ is different, because the $\phi$ value is significantly larger. As the universe expands and $\phi$ red-shifts, the parametric resonance structure for this case becomes similar to the others\footnote{A similar effect appears for $V_3$ with $\alpha_3=1.5$, where the initial inflaton amplitude at the start of the simulation is $\phi\simeq 1.2M$. }.

\begin{figure}[h!]
 \centering{
   \includegraphics[width=.5\textwidth]{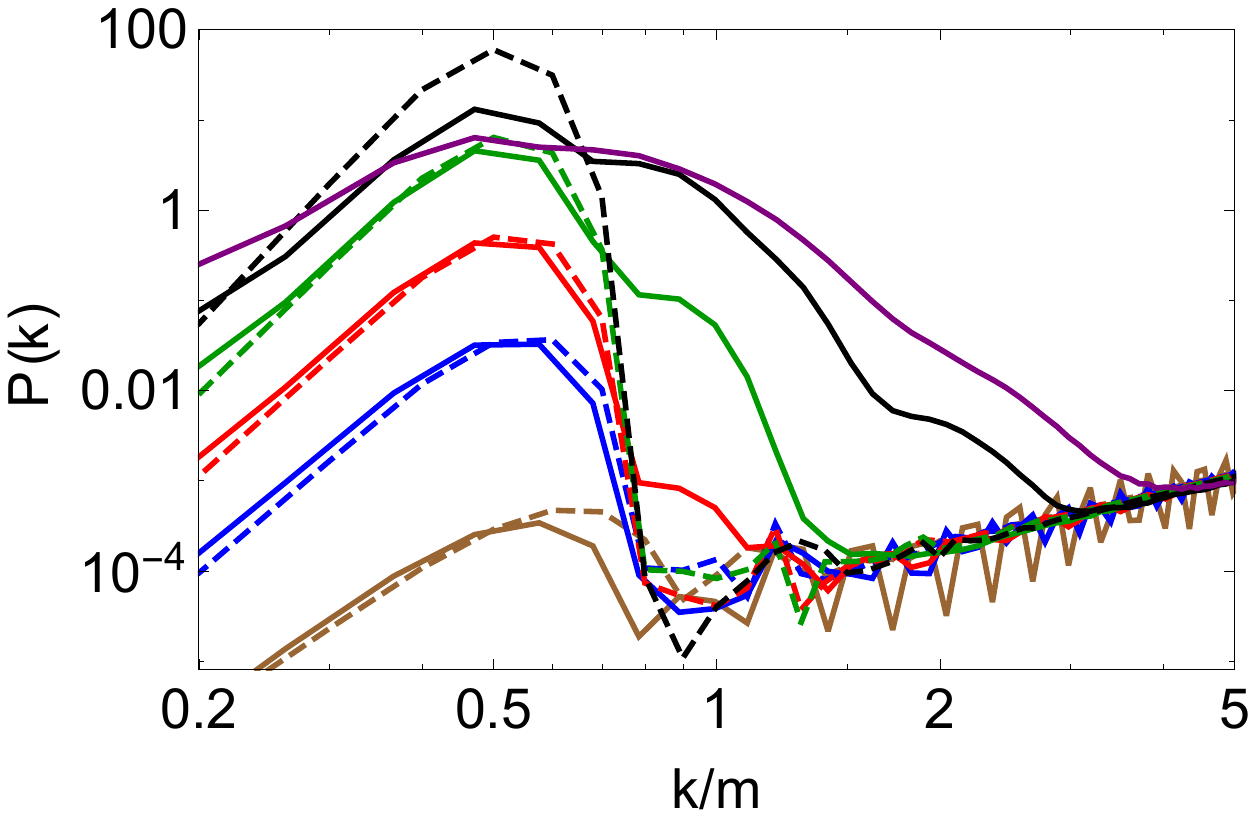}
   }
 \caption{
The power spectrum of the inflaton field computed using the linear fluctuation equations (dashed) and using the full lattice simulation (solid) for times $\eta  =  10, 20, 25, 30, 35, 45$ in units of $m$ (brown, blue, red,  green, black and purple respectively) for the case where the  potentials $V_n$ overlap. We plot earlier time-slices compared to the ones in Fig.~\ref{fig:Pk_VA}, due to the larger Floquet exponent for $V_n$ compared to $V_A$. 
We do not show the linear results when they differ greatly from the lattice simulations and their comparison does not add anything to our understanding.
 }
 \label{fig:Pk_Vn_overlap}
\end{figure}

Fig.~\ref{fig:Pk_Vn_overlap} shows the evolution of inflaton fluctuations using the linear approximation for the case where all three potentials $V_n$ overlap. We see that initially the two calculations agree very well. This starts to change at $\eta\simeq 25\,m^{-1}$, where an increase in the power spectrum at $k\simeq m$ becomes visible. This signals the onset of oscillon formation, which {has} a characteristic scale of ${\cal O}(m)$. This ``bump'' of ${\cal P}(k)$ grows with time and eventually the true spectrum, calculated using lattice results, exhibits a broad peak centered at $k\simeq m$. An important factor for oscillon formation is the existence of large initial inhomogeneities, allowing the field to locally probe the non-linearities in the potential. As far as the emission of GW's is concerned, the formation of true oscillon is not important (see Ref.~\cite{Lozanov:2019ylm}). The important factor for GW emission is the emergence of large inhomogeneities during the preheating process.

\begin{figure}[h]
 \centering{
      \includegraphics[width=.3\textwidth]{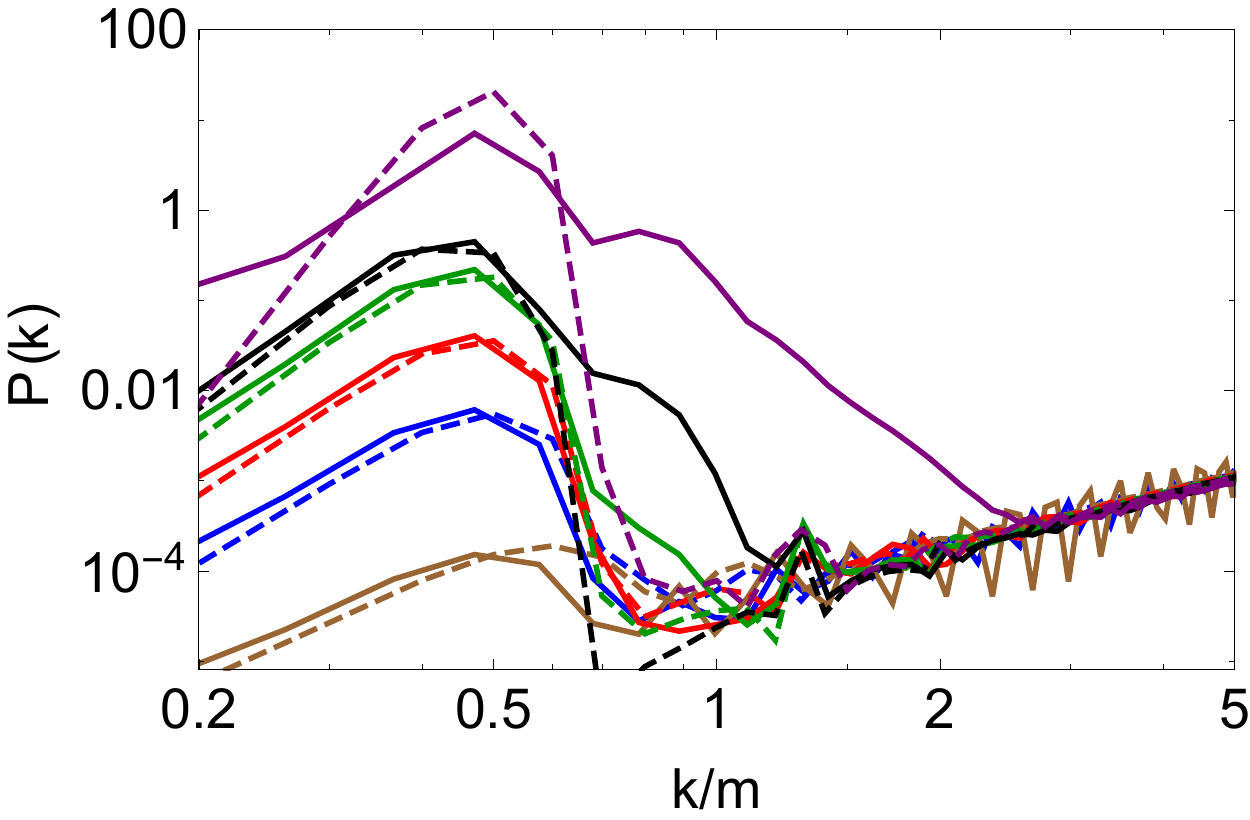}
                  \includegraphics[width=.3\textwidth]{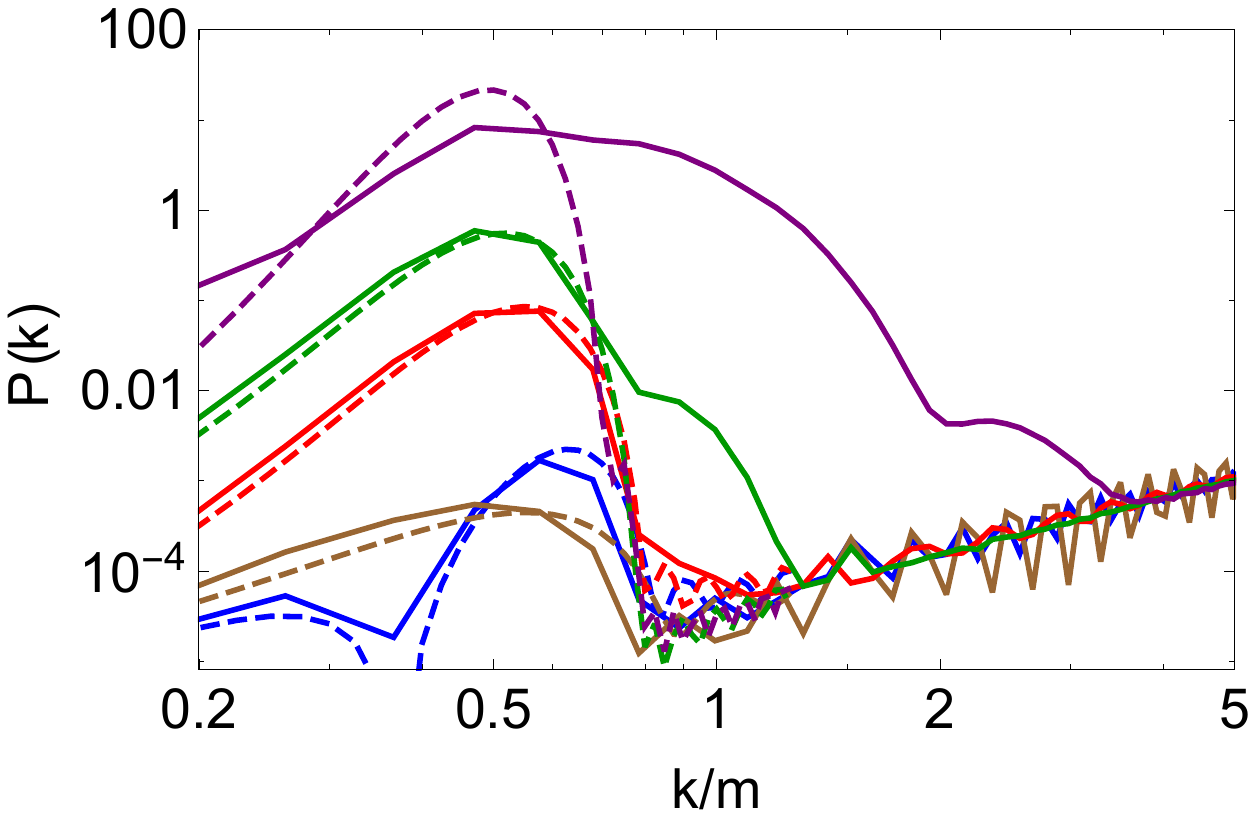}
            \includegraphics[width=.3\textwidth]{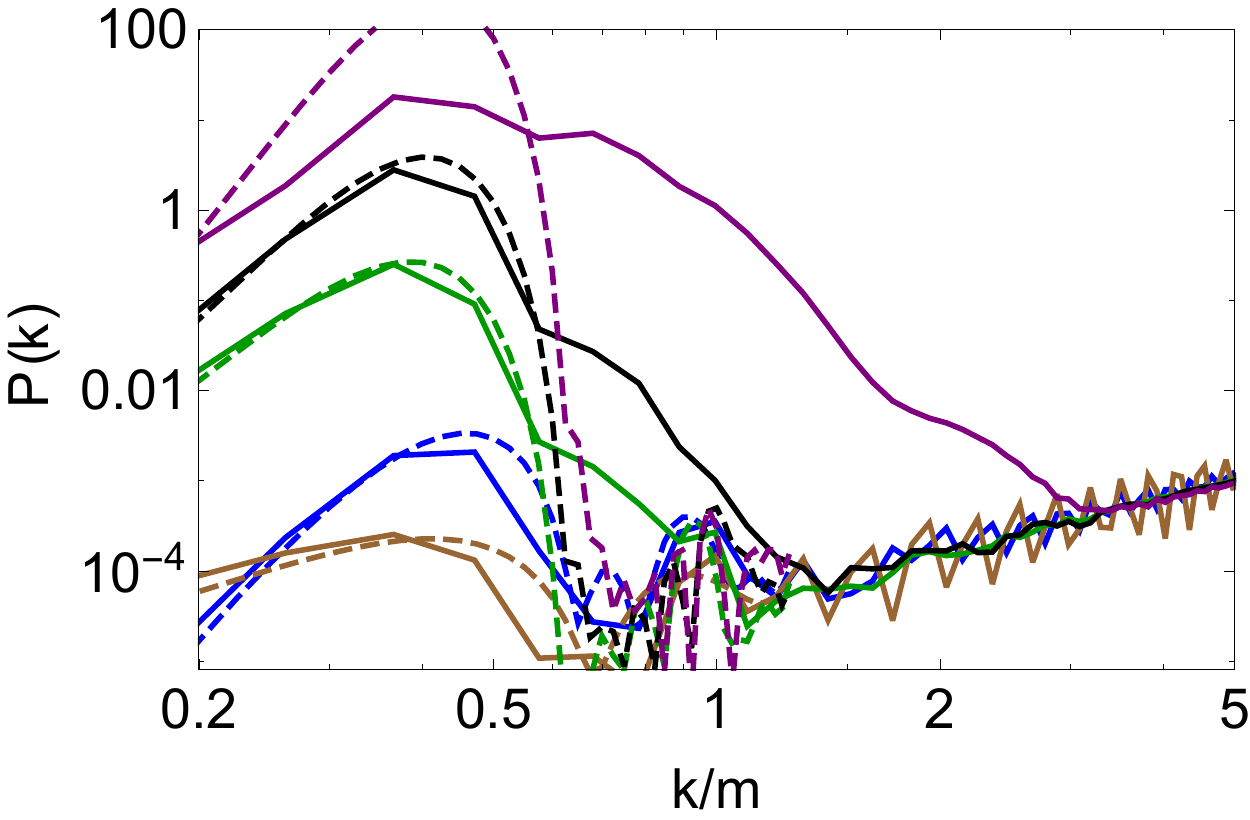}
            \\
                  \includegraphics[width=.3\textwidth]{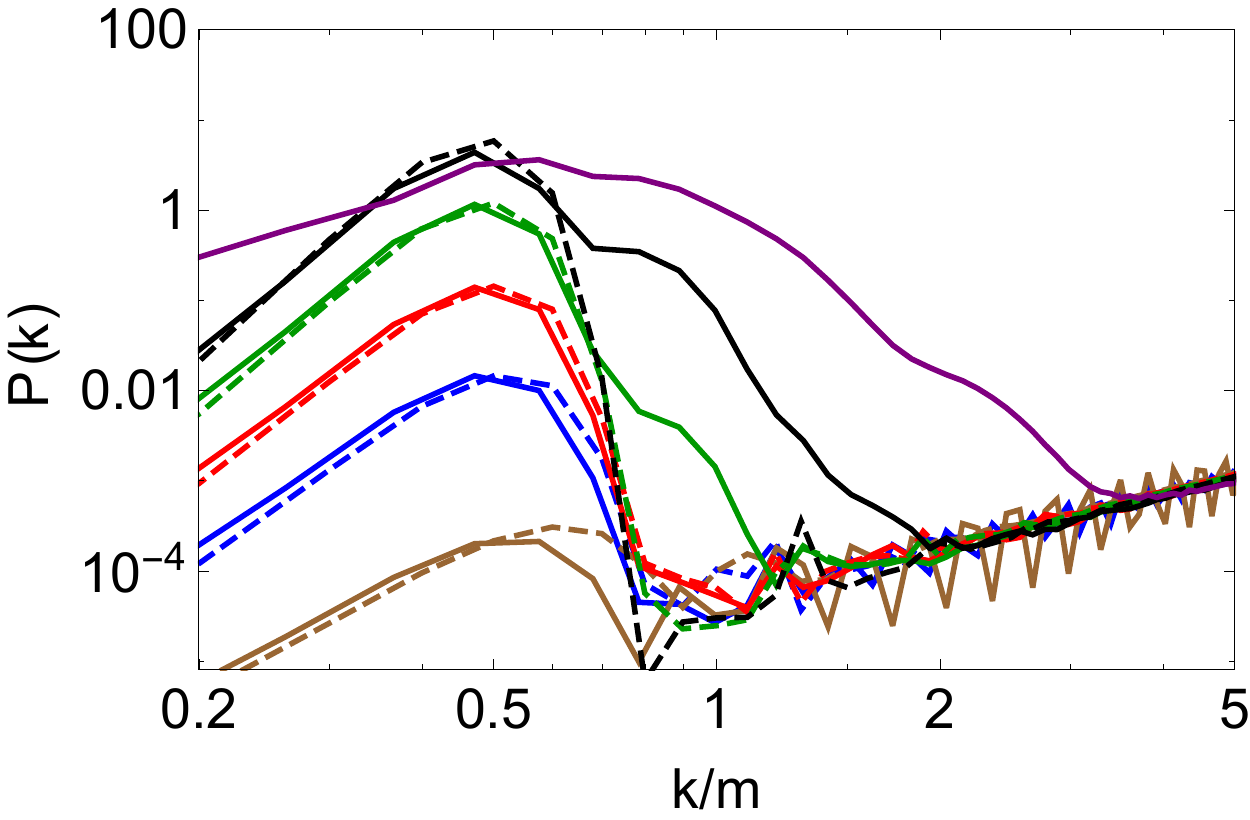}
                  \includegraphics[width=.3\textwidth]{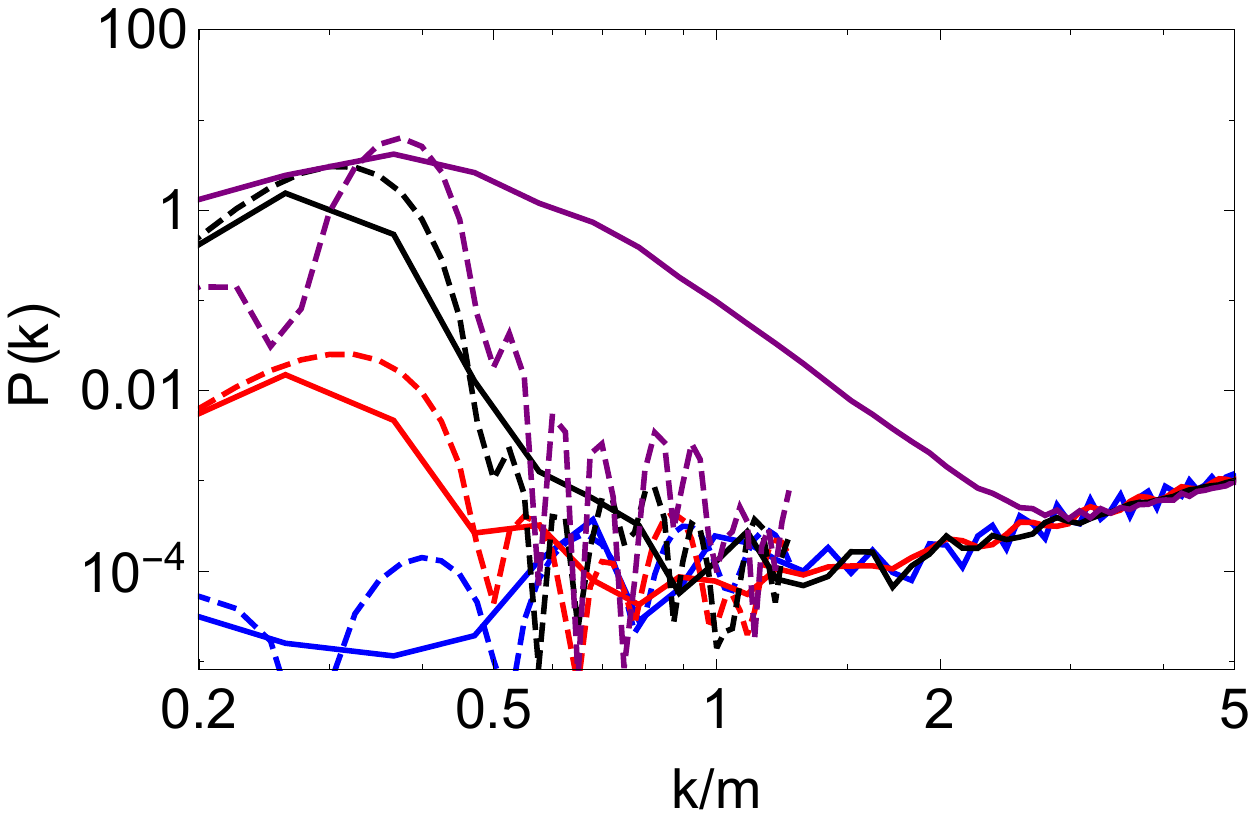}
            \includegraphics[width=.3\textwidth]{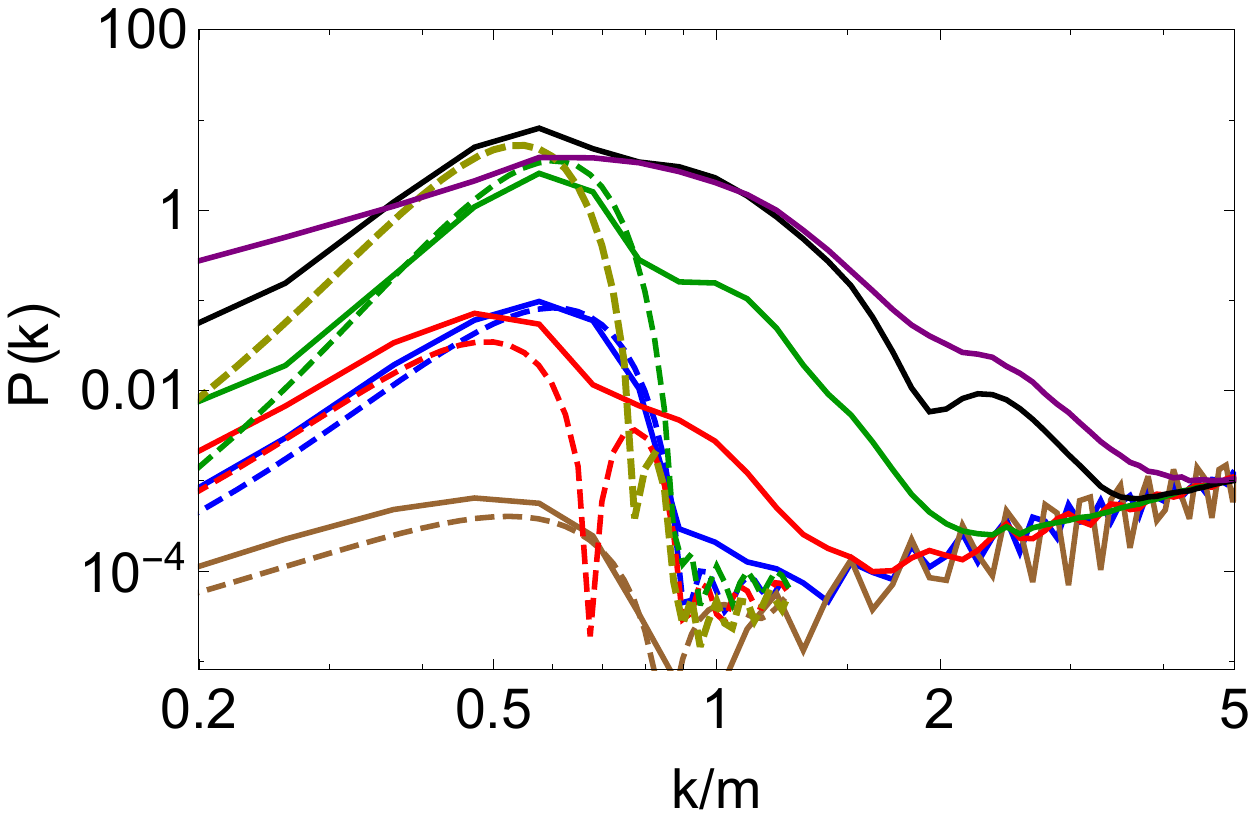}
   }
 \caption{
{\it Left:} The power spectrum of  inflaton fluctuations for $V_1$ and $n=1.6,1.8$ (upper, lower).
{\it Middle:} The power spectrum of  inflaton fluctuations for $V_2$ and $n=5/7, 5/3$ (upper, lower).
{\it  Right:} The power spectrum of  inflaton fluctuations for $V_3$ and $n=1.5, 3$ (upper, lower).
The color-coding for each time follow Fig.~\ref{fig:Pk_Vn_overlap}. We omit some time-slices for clarity.
 }
 \label{fig:Pk_Vn}
\end{figure}

Fig.~\ref{fig:Pk_Vn} shows the corresponding linear and lattice spectra for the three potentials $V_n$ and the different values of $\alpha_1,\alpha_2,\alpha_3$ that we used. For $V_1$ we see a similar growth of fluctuations among the different choices of $\alpha_1$. The slight difference in the growth rate is exactly in line with the slight difference in $\mu_k^{({\rm max})}/m$ which varies between $0.22$ and $0.24$, as shown in Fig.~\ref{fig:FloqVnoneD}. Thus we expect the field to enter the non-linear regime in all three cases at a similar time. This is however not the case for the potentials $V_2$ and $V_3$, where we see  different linear power spectra for some parameter choices. 

For $V_2$ we see that for $\alpha_2=5/3$ the emergence of non-linearity occurs at $\eta \simeq 45 \, m^{-1}$. This can be attributed to the overall smaller initial instability bands. However, this is not a result of the potential itself, but rather of the initial (larger) field amplitude, which means that initially the inflaton fluctuations probe the narrower part of the Floquet chart. As the universe expands and the field amplitude red-shifts, the instability bands grow and thus the system evolves similarly for  $\alpha_2=5/3$, like it does for  $\alpha_2=1$ and  $\alpha_2=5/7$. Again we see that the onset of back-reaction and non-linearities appears when the peak amplitude of the scalar power spectrum approximately equals unity.

A similar behavior is seen for $V_3$, where the case of $\alpha_3=1.5$ shows that the true power spectrum deviates from the linear approximation later than in the case of $\alpha_3=3$. This can again be attributed to the smaller initial Floquet bands for $\alpha_3=1.5$, shown in Fig.~\ref{fig:FloqVA}.

\subsection{Monodromy and plateau potentials}

Before we conclude the analysis of linearized fluctuations, let us discuss the parametric resonance behavior of the potentials $V_1(\alpha_1=1)$ and $V_A$, since both potentials show linear growth at large distances $V\propto |\phi|$. The main instability band of $V_1$ grows slightly for lower values of $\alpha_1$, peaking at $\mu_{\rm max}\simeq 0.25\,m$ for $\alpha_1=1$ and $\mu_{\rm max}\simeq 0.225\,m$ for $\alpha_1=2$, while the overall shape remains the same. For comparison, the maximum instability exponent for $V_A$ is $\mu_{\rm max}\simeq 0.066\,m$, three times smaller than that for $V_1$.

Comparing these results to the evolution of the GW spectrum for each case, show in Fig.~\ref{fig:res_VA_V1}, we immediately make the following correlation. In systems with large Floquet exponents, the GW spectrum evolves even after oscillon formation. This erases any memory of the oscillon structure that was present in the GW spectrum. On the contrary, systems where the Floquet exponents are lower, but still important enough to lead to inflaton fragmentation, result in a GW spectrum, which does not evolve significantly past the time of oscillon formation. This can be understood as follows, In the case of $V_n$, we see that a significant fraction of the detected overdensities change their characteristics as time progresses. There are two possible explanations for this. Either the overdensities are not true oscillons, and thus they decay or fragment, or they are oscillons which move by interacting with neighboring oscillons or overdensities. In both cases, this leads to GW production long after the oscillon formation time. Furthermore, the emitted GW's are not related to the structure of the oscillons, but to random motion and thus they have a characteristic scale of $k/a={\cal O}(m)$ and no other features. On the contrary, for $V_A$, we find overdensities that are compatible with an oscillon height-width distribution (see Appendix \ref{sec:ident}) and no other significant overdensities. Given this result, it is expected that the bulk of the energy density in the system that can source GW's is ``locked" in stationary oscillons and thus GW emission will cease after oscillon formation.

\section{Conclusion and prospects}
\label{sec:conc}

Oscillon production is ubiquitous after inflation in models with a plateau- or monodromy-type potential, as is the gravitational wave production that is associated with the fragmentation of the inflaton condensate after inflation, due to efficient self-resonance.
We performed a systematic study of oscillon potentials, computing the corresponding emergence of oscillons and the shape and amplitude of the emitted GW spectrum. More concretely, we explored the dependence of a potential on the properties of the gravitational waves during the oscillon formation processes such as their amplitudes and shapes, the power spectra of scalar perturbations, and the properties of the resultant produced oscillons such as the number densities and the size distributions. For this purpose, we have performed not only numerical simulations in the expanding universe but also a linear fluctuation analysis (Floquet analysis) in the static universe approximation to interpret the results of our simulations in terms of the corresponding Floquet charts. We focused on two main families of potentials (axion-monodromy and plateau potentials) and arranged a variety of potential types, where the small field dependence is changed with the same (large field) asymptotic behavior and the large field dependence is changed with the same small field behavior. 

We also confirmed that the growth rate of the scalar perturbations and the associated oscillon formation time are sensitive to the small-field shape of a potential. In fact, as the potential gets far away from the axion-monodromy potential in the small field region (with keeping the same asymptotic behavior), the formation times were delayed more and more
while the total number of oscillons at late times is largely insensitive to the small field behavior of potentials. On the other hand, the macroscopic physical properties of oscillons such as the total number depend on the large-field shape of a potential.

Though we have introduced a simple criterion to discriminate true oscillons from transient objects, it discrimination is quite subtle. In order to introduce more clear criterion, the lifetime of an oscillon might be the key feature. We leave a more thorough analytical and numerical investigation on longevity of created oscillons as well as on other features of oscillons through a new criterion for future work.

Finally, we have found that the shape of the spectrum and the amplitude of emitted gravitational waves are almost universal, irrespectively of the detail of potential shape. {This can be used as a smoking-gun for deducing the existence of a violent preheating phase and possible oscillon formation after inflation.} 
However, there are significant subtleties related to this issue. 
In both potential families, plateau potentials with varying large-field dependence and axion monodromy potentials with different small-field shape, the GW spectrum exhibits two peaks around the time of oscillon formation. In the case of monodromy potentials, the two-peak shape persists until the end of our simulations. In the case of plateau potentials, the two-peak structure is ``smeared" at late times and replaced by a broad peak in momentum-space. From that we conclude that potentials exhibiting efficient self-resonance will tend to give a featureless GW spectrum, while potentials that exhibit a weaker self-resonance, and thus delayed oscillon production, will tend to give a GW spectrum that encodes the characteristics of the produced oscillons, like the internal frequencies, leading to peaks and dips in power at specific wavenumbers.
If this behavior is verified for more potentials, it can act as  a smoking-gun not only for deducing the existence of a violent preheating phase  after inflation, but for inferring the strength of inflaton self-resonance and the oscillon frequency content.

Unfortunately, the typical frequencies of the emitted gravitational waves are around GHz and hence cannot be detected by the planned experiments like LISA. However, this frequency range has received increasing interest recently and hence new gravitational wave detectors for such frequency range have been proposed and developed \cite{Domcke:2020yzq,Ito:2020wxi,Ito:2019wcb,Li:2009zzy,Li:2008qr}.

That being said, oscillon formation can be the by-product not only of preheating, but of any fragmentation process of an oscillating massive scalar field with a shallow potential. This can be a modulus field or an axion in the later universe. Such a process would shift the frequency of the GW signature, possibly bringing it into the interferometer range (see e.g. Ref.~\cite{Kitajima:2018zco}).


\begin{acknowledgements}
We would like to than M. Amin and E. Copeland for useful discussions. The work of EIS was supported by the Dutch Organisation for Scientific Research (NWO) and partly supported by a fellowship from ``la Caixa'' Foundation (ID 100010434) and from the European Union's
Horizon 2020 research and innovation programme under the Marie Sk\l
odowska-Curie grant agreement No 847648. The fellowship code is
LCF/BQ/PI20/11760021.  M.\,Y. is supported in part by JSPS Grant-in-Aid
for Scientific Research Number 18K18764 and JSPS Bilateral Open Partnership Joint Research Projects. This work was supported by  Mitsubishi Foundation.

\end{acknowledgements}

\appendix

\section{Scalar and tensor spectrum}
\label{sec:spectrum}

\subsection{Scalar fluctuations}
\label{subsec:scalar}

The total power of the scalar fluctuations,
$\delta\phi(\xx,t)=\phi(\xx,t)-\overline{\phi}(t)$, in a unit volume is given as 
%
\begin{align}
 \langle\delta\phi(\xx,t)^2\rangle = 
\frac{1}{V}\int\!d^3x\,\left[\delta\phi(\xx,t)\right]^2
 = \frac{1}{V}\int\!\frac{d^3k}{(2\pi)^3}\,\left|\widetilde{\delta\phi}(\kk,t)\right|^2
 = \int\!\frac{k^2dk}{2\pi^2}\,P(k,t)
 = \int\!\frac{dk}{k}\,\mathcal{P}(k,t),
\end{align}
%
where
%
\begin{align}
 \mathcal{P}(k,t):=\frac{k^3}{2\pi^2}P(k,t)=\frac{k^3}{2\pi^2V}\int\!\frac{d\Omega_k}{4\pi}\,\left|\widetilde{\delta\phi}(\kk,t)\right|^2,
\label{eq:def_P}
\end{align}
%
is the power spectrum of the fluctuations. We evaluate the spectrum with
the discrete Fourier transformation.

\subsection{Gravitational waves}
\label{subsec:GW}

The energy density of the gravitational waves is given by
%
\begin{align}
\rho_{\rm GW}(\eta)
&= \frac{\Mpl^2}{4a^2} \langle h'_{ij}h'_{ij}\rangle_{V}
= \frac{\Mpl^2}{4a^4} \langle \gamma_{ij}\gamma_{ij}\rangle_{V}
\label{eq:defene_std}
\end{align}
%
where $\gamma_{ij}=\chi'_{ij}-\HH\chi_{ij}$, and
$\langle\cdots\rangle_{V}$ represents the average over the
spatial volume. 
The volume average can be recast as 
%
\begin{equation}
\langle \gamma_{ij}\gamma_{ij} \rangle_{V} 
= \frac{1}{L^3}\int d^3x\,\gamma_{ij}\gamma_{ij}
= \frac{1}{(2\pi)^3L^3}\int d^3k\,\widetilde{\gamma}_{ij}\widetilde{\gamma}^*_{ij},
\label{eq:volave}
\end{equation}
%
where we used the Perceval's
theorem in the last equation. The quantity $\widetilde{\gamma}_{ij}(\kk, \eta)$ is the Fourier transform of
$\gamma(\xx,\eta)$, given by
%
\begin{align}
\widetilde{\gamma}_{ij}(\kk,\eta)
     &= \int\!d^3\xx\,\gamma_{ij}(\xx,\eta) e^{-i\kk\cdot\xx},
\label{eq:fourier}
\\
\gamma_{ij}(\xx,\eta)
      &= \int\!\frac{d^3\kk}{(2\pi)^3}\,\widetilde{\gamma}_{ij}(\kk,\eta) e^{i\kk\cdot\xx}.
\label{eq:rfourier}
\end{align}
%
When working on a lattice, a discretized space, Eq.~(\ref{eq:fourier}) becomes
%
\begin{align}
\widetilde{\gamma}_{ij}(k_{pqr},\eta) &=
 \frac{L^3}{N^3}\widetilde{\Gamma}(k_{pqr},\eta), 
\label{eq:def_X}
\\
\widetilde{\Gamma}(k_{pqr},\eta) &= \sum_{\ell,m,n} \gamma_{ij}(x_{\ell mn},\eta) e^{2\pi i (p\ell+qm+rn)/N},
\label{eq:dft}
\end{align}
%
where the abbreviations $k_{pqr}$ and $x_{\ell mn}$ represent the wavenumber vector $(k^{x}_p,k^{y}_q,k^{z}_r)$
and position vector $(x_\ell, y_m, z_n)$, respectively, and $x_\ell = (L/N)\ell$,
$k^{x}_p=2\pi p/L$, and so on. Usually, since the Fast-Fourier-Transform
library calculates $\widetilde{\Gamma}$ rather than $\widetilde{\gamma}$, we have to manually multiply by the factor $L^3/N^3$ shown in Eq.~\eqref{eq:def_X}.

Combining Eqs.~(\ref{eq:defene_std}) and (\ref{eq:volave}), we have
%
\begin{equation}
\begin{aligned}
\rho_{\rm GW}(\eta) &= \frac{\Mpl^2}{32\pi^3L^3a^4}
  \int d^3k\,\widetilde{\gamma}_{ij}\widetilde{\gamma}^*_{ij},\\
  &= \frac{\Mpl^2}{32\pi^3L^3a^4}
  \int\!d\log k \int k^3d\Omega\,\widetilde{\gamma}_{ij}\widetilde{\gamma}^*_{ij},\\
\end{aligned}
\end{equation}
%
from which we can read off the energy spectrum of the produced gravitational waves,
%
\begin{equation}
S_k(\eta) = \frac{d\rho_{GW}}{d\log k} = \frac{\Mpl^2k^3}{32\pi^3L^3a^4}
   \int d\Omega\,\widetilde{\gamma}_{ij}\widetilde{\gamma}^*_{ij} \, .
\end{equation}
%
The dimensionless energy spectrum during the simulation is given by
%
\begin{equation}
  \Omega_{\rm GW}(k,\eta) \equiv \frac{S_k(\eta)}{\rho_c(\eta)}
  = \frac{k^3}{96\pi^3\HH^2L^3a^2}
    \int d\Omega\,\widetilde{\gamma}_{ij}\widetilde{\gamma}^*_{ij} \, ,
    \label{eq:defOmegaGW}
\end{equation}
%
where $\rho_c = 3\HH^2 \Mpl^2 / a^2$ is the critical density.
The last integral is replaced by the angular average of discrete
data of $\widetilde{\gamma}_{ij}$,
%
\begin{equation}
\int d\Omega\,\widetilde{\gamma}_{ij}\widetilde{\gamma}^*_{ij}
 \approx \frac{4\pi}{N_k}\sum_{{\rm all\;for}\;k=|\kk|} 
         |\widetilde{\gamma}_{ij}(\kk)|^2 \, ,
\end{equation}
%
where $N_k$ is the number of elements satisfying $k=|\kk|$. Eq.~\eqref{eq:defOmegaGW} becomes
%
\begin{equation}
\begin{aligned}
  \Omega_{\rm GW}(k,\eta) 
  &\approx \frac{k^3}{24\pi^2\HH^2L^3a^2N_k}
        \sum_{{\rm all\;for}\;k=|\kk|}|\widetilde{\gamma}_{ij}(\kk)|^2\, ,\\
  &= \frac{k^3L^3}{24\pi^2\HH^2 a^2N_kN^6}
        \sum_{{\rm all\;for}\;k=|\kk|}|\widetilde{\Gamma}_{ij}(\kk)|^2
        \, ,
\end{aligned}
\end{equation}
%
where we used Eq.~(\ref{eq:def_X}). We evaluate this equation at a
discrete wavenumber, $k_p=2\pi p/L$, so it can be more reduced to
%
\begin{equation}
\Omega_{\rm GW}(\hat{k}_p,\eta)
  = \frac{\pi p^3}{3\HH^2a^2N_kN^6}
        \sum_{{\rm all\;for}\;k=|\kk|}|\widetilde{\Gamma}_{ij}(\kk)|^2
        \, ,
 \label{eq:Omega_GW_final}
\end{equation}
%

Assuming that the generation of gravitational waves is terminated at the end
of simulation, the energy spectrum at the present time is obtained by
multiplying the computed GW spectrum with a damping factor that encodes the subsequent cosmic expansion,
%
\begin{equation}
 \Omega_{GW,0} = \Omega_{r,0}\left(\frac{g_{*,0}}{g_{*,f}}\right)^{1/3}\Omega_{GW,f},
 \label{eq:GWfactor}
\end{equation}
%
where $\Omega_{r,0}=0.916\times 10^{-4}$, the subscript 'f' represents the quantity evaluated at the end of
simulation, $\eta=\eta_f$, and we used
%
\begin{align}
  a_0^4g_{*,0}^{1/3}\rho_{{\rm rad},0} &= a_f^4g_{*,f}^{1/3}\rho_{{\rm rad},f}\, , \\
  a_0^4\rho_{{\rm GW},0} &= a_f^4\rho_{{\rm GW},f}\, , \\
  \rho_{\rm rad} &= \frac{\pi^2}{30}gT^4 \, .
\end{align}
%

In the present time, the frequency corresponding to the box size is given by
%
\begin{align}
 f_0 = \frac{k_{\rm phys,f}}{2\pi}\left(\frac{a_i}{a_f}\right)
   = \frac{1}{L}\frac{a_f}{a_0}\frac{a_i}{a_f}
   = \frac{1}{L}\frac{T_0}{T_R}\frac{a_i}{a_f}
   \, ,
\end{align}
%
where $k_{\rm phys,f}$ is the physical wavenumber at the end of
simulations, $a_i,a_f,a_0$ are the scale factor at the initial time, the
end of simulations and the present time, respectively. In addition, we
assume that the inflaton decays into radiation at $a=a_f$, leading to a reheating temperature $T_R$.

Reinstating the constants $\hbar$ and $c$, the GW frequency is written as 
%
\begin{align}
f_0 &= 
4.37 \times 10^9~{\rm Hz} \left(\frac{m}{10^{-5}\Mpl}\right)\left(\frac{\overline{L}}{100}\right)^{-1}
\notag \\ &\times\left(\frac{T_R}{10^{13}{\rm GeV}}\right)^{-1}\left(\frac{a_f/a_i}{10}\right)^{-1},
\end{align}
%
which falls outside of the  observable frequency range of ground-based interferometers for {$m\sim 10^{-5}M_{\rm Pl}$}. Future attempts of high frequency GW detection are becoming increasingly interesting, as they would open a unique observational window into preheating. {In fact, there are proposed experiments for the detection of GWs with such high frequencies \cite{Domcke:2020yzq,Ito:2020wxi,Ito:2019wcb,Li:2009zzy,Li:2008qr}, even though their sensitivity needs to be improved significantly in order to provide feasible detection opportunities.}

\section{Oscillon identification}
\label{sec:ident}

We identify oscillons in the computational domain using the following algorithm:
\begin{enumerate}
\item Once we obtain the energy density, $\rho(\xx_i,\eta)$, from the simulations, the average energy density in the computational box is computed as 
\begin{equation}
   \rho_{\rm ave}(\eta) = \frac{1}{N^3}\sum_i\!\rho(\xx_i,\eta) \, .
\end{equation}
where $N$ is the number of grid.
 \item We identify the regions where $\rho(\xx,\eta) \geq \beta\rho_{\rm ave}$ with $\beta = 20$.
 \item For each region, we find out the point, $\xx_{c}$, where the energy density is maximum, and define $\rho_{\rm core}=\rho(\xx_{c},\eta)$.
 \item For each region, we again find out the region centered around the peak location, $\xx_{c}$, where $\rho(\xx,\eta) \geq \rho_{\rm core}/e$, and then we regard the region as an oscillon  with the comoving volume $V_{\rm osc}$.
 \item Finally, we {define} the effective  physical size  {of the oscillon} as $w=a(6V_{\rm osc}/\pi)^{1/3}$, which is the diameter of a sphere whose physical volume is $a^3V_{\rm osc}$. 
\end{enumerate}

Using this algorithm and given enough data, we can construct a frequency distribution (histogram) for the
number of oscillons and  for their energy.
The distribution functions of the oscillons' size and energy can be
estimated from the computed frequency distributions. Given a distribution
function $f(x)$, the number $f_i$ in a finite bin  $x_i\leq x<x_{i+1}$, is
calculated as
%
\begin{align}
 f_i = \int_{x_i}^{x_{i+1}}\!f(x)\,dx \approx \Delta x \cdot
 f\left(x_i+\frac{\Delta x}{2}\right) \, .
\end{align}
%
If the size of bin is sufficiently small, we estimate
%
\begin{align}
 f\left(x_i + \frac{\Delta x}{2}\right) \simeq \frac{f_i}{\Delta x} \, .
\end{align}
%

The above method will identify any sufficiently large local overdensity. These might or might not be oscillons: localized long-lived scalar field structures that oscillate in time.

\begin{figure}[h]
 \centering{
   \includegraphics[width=.45\textwidth]{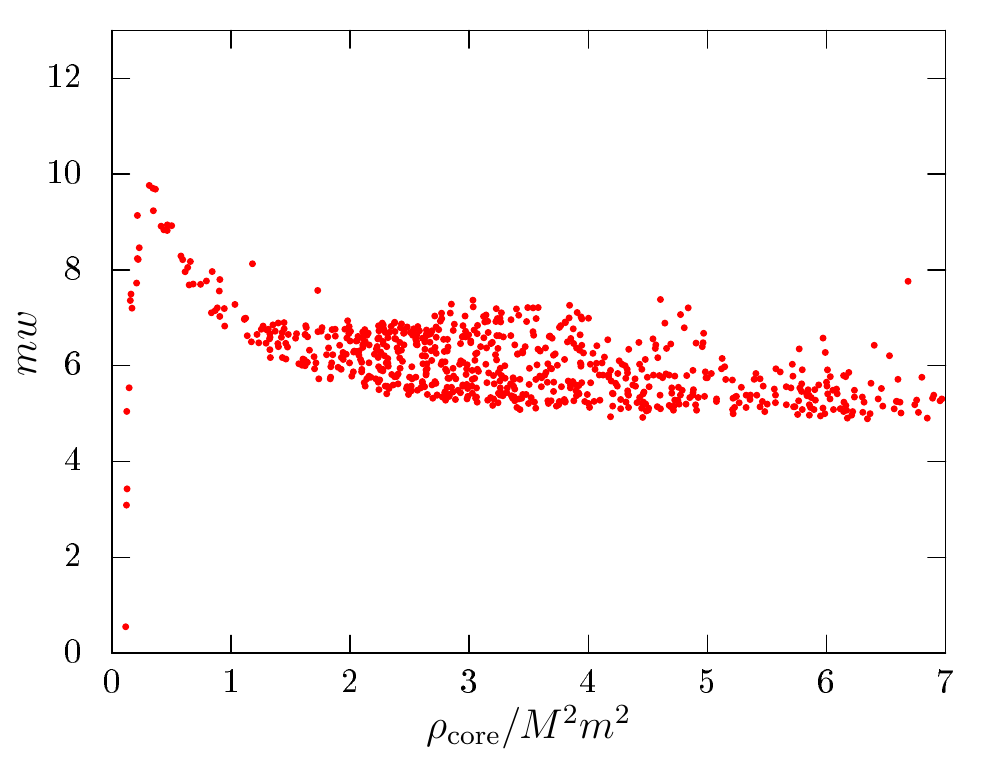}
   \includegraphics[width=.45\textwidth]{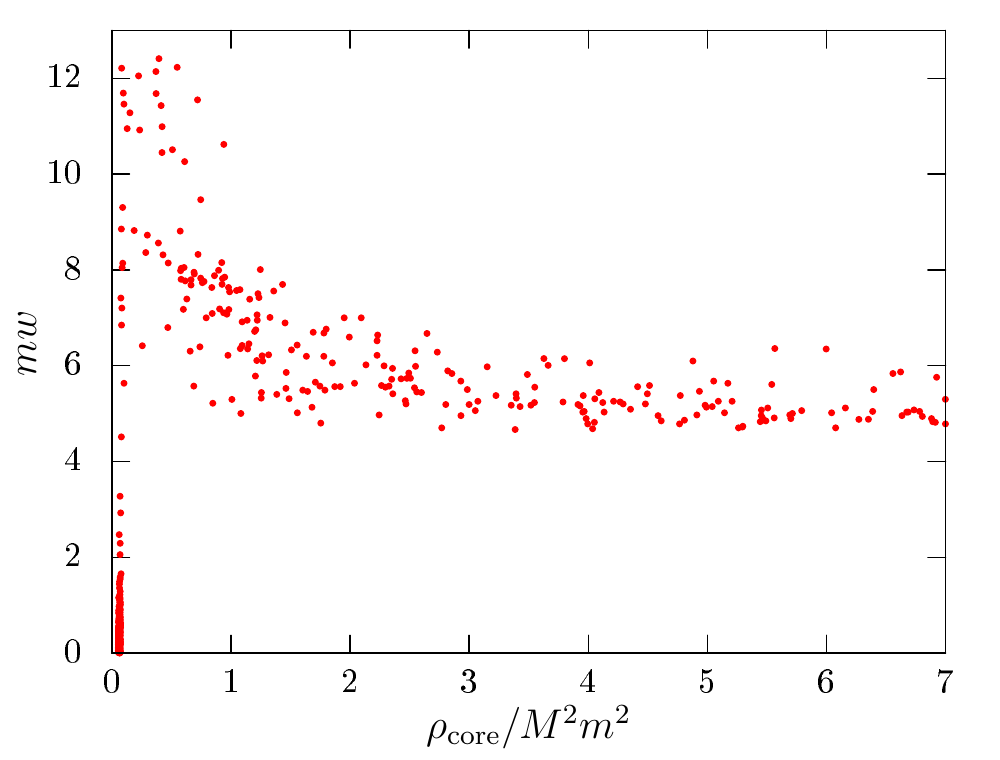}
   }
 \caption{
 The height and width of oscillons in the case of the monodromy potential $V_A$ (left) and the approximated potential $V_A^{(4)}$. While the vast majority of oscillons falls around a well defined height-width relation, we see the existence of several overdensities with small width and height, especially for $V_A^{(4)}$.  These are numerical artifacts and have been removed before we compute the number density of oscillons.
 }
 \label{fig:widthcore}
\end{figure}

Fig.~\ref{fig:widthcore} shows the overdensities identified by the procedure described above for the case of the monodromy potential.
We see that the vast majority of overdensities follow a clear relation between the oscillon width and height, where wider oscillons tend to have a smaller central density. This is reminiscent of models, where the height-width relation was derived semi-analytically. We do however see a significant amount of overdensities that have a small width and a small height, falling clearly below the ``main sequence" of oscillons. These are transient overdensities and should not be counted as oscillons.
By eliminating these overdensities, whose height-width characteristics put them below the main sequence of oscillons, we significantly reduce the numerical noise in our calculation of oscillon number density. Fig.~\ref{fig:res_VAapp} shows the resulting oscillon number, after such small overdensities have been eliminated from our count. 

For our calculation the exact stability properties of the emerging oscillons are not important, since we only require the oscillons to form and live long enough for GW's to be emitted \cite{Lozanov:2019ylm}. 
In some cases of potentials $V_n$, the distribution of oscillons that we discovered showed two somewhat disjointed regions, making the identification of real oscillons and transient overdensities even more difficult. Possible features, such as fragmentation of unstable oscillons into smaller, more stable ones, can lead to a late-time increase in oscillon number density. Furthermore, overdensities starting close to the theoretical oscillon height width curve can relax to a stable oscillon configuration over time, as  was  shown e.g. in Ref.~\cite{vanDissel:2020zje}.   

Given the rich dynamics of non-linear field theories, We leave a detailed investigation of the properties of the produced oscillons and other localized over-densities in the various potentials that we examined for future work (for  recent work on the lifetime of oscillons see Refs.~\cite{Ibe:2019vyo, Zhang:2020bec}).

\begin{figure}[h]
 \centering{
   \includegraphics[width=.45\textwidth]{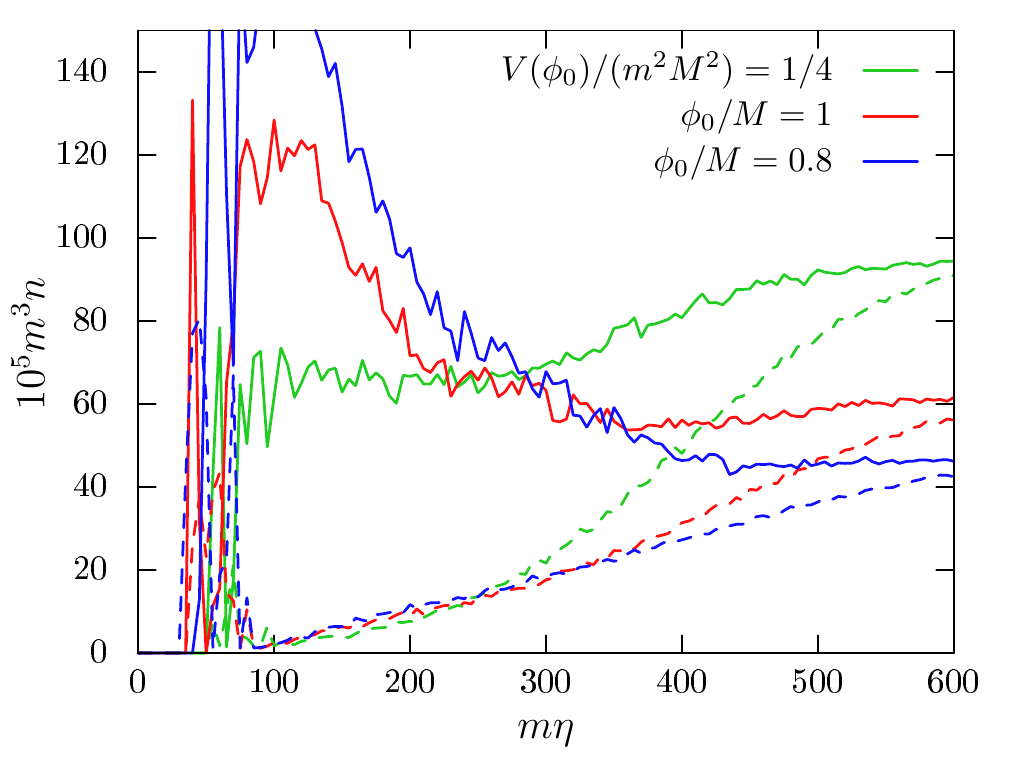}
      \includegraphics[width=.45\textwidth]{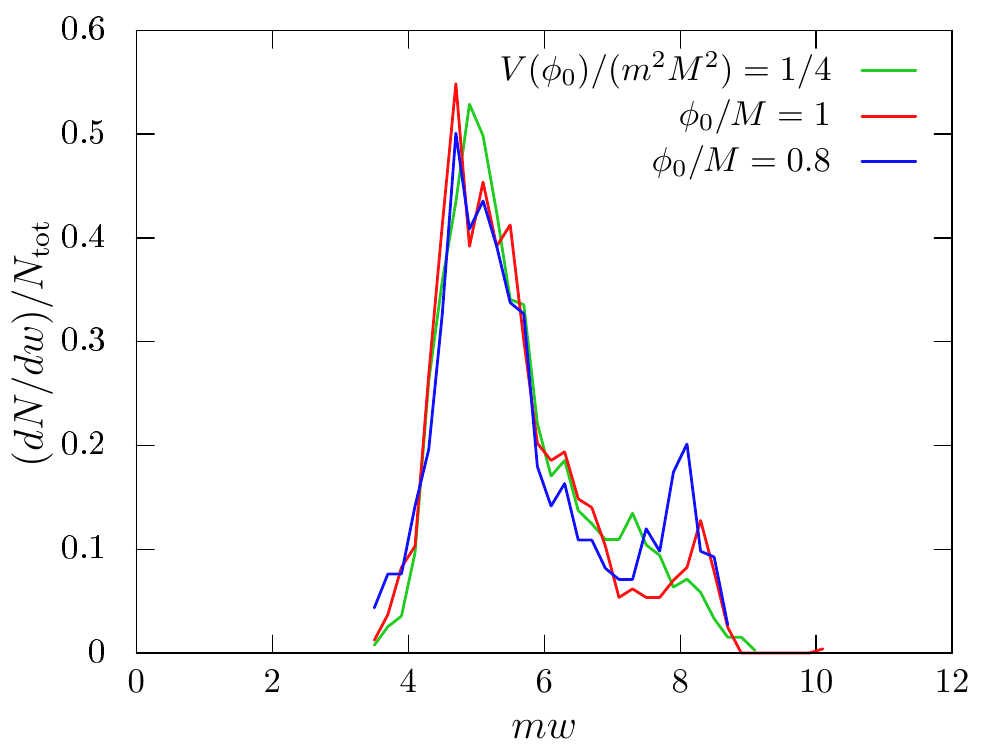}
   }
 \caption{
{\it Left:} The evolution of the number density of oscillons for the case of the potential $V_2$ with $\alpha_2=5/3$. The solid (dashed) curves correspond to counting all over-densities as oscillons (applying the selection criterion $w>w_c=3.5\,m^{-1}$). The three different colors correspond to different initial field amplitude $\phi_0$.
{\it Right:} The normalized distribution of oscillon widths for the three cases of $\phi_0$.
 }
 \label{fig:V2comp}
\end{figure}

Before we conclude, we revisit Figures \ref{fig:V2} and \ref{fig:V3}, which show the time evolution of the comoving oscillon number density $n$ for $V_2$ and $V_3$ respectively. In most cases the oscillon number density asymptotes to a constant value by the end of the simulation at $\eta=600\,m^{-1}$. However, two cases show a late-time growth of the oscillon number density: the case of $V_2$ with $\alpha_2=5/3$ and $V_3$ with $\alpha_3=1.5$. The common feature of these two cases is that the field value at the start of the simulation, defined through $V=m^2M^2/4$, is higher than the corresponding initial field value for all other cases ($\phi_0\sim M$). This means that the system must redshift more before entering the main instability band. Two interesting observations arise. One is that when the simulation starts at a lower field value $\phi=M$ or $\phi=0.8M$ the time evolution of the oscillon number density exhibits a smaller initial growth of $n$ (which is due to less transient overdensities). The second concerns the final state of the oscillons; even though the total number density is different,  the distribution of oscillon widths is nearly identical for all cases. Furthermore, the solid and dashed curves, corresponding to counting all over-densities or using a width cutoff criterion, converge at late times.
Overall, while the time evolution of oscillons depends both on the cutoff criterion and on the initial field value, their distribution is identical, since it depends solely on the structure of the scalar potential.


\section{Inflationary dynamics}
\label{sec:infl}

For completeness, we present the basic results for the inflationary evolution for the potentials $V_n$ with $n=1,2,3$. 

We start with $V_1$, which is not a plateau-type potential, since it grows as $V_1 \sim \phi^{2-\alpha_1}$ for large values of $\phi$, where $\alpha_1$ is a dimensionless parameter. In order for the potential to be monotonically decreasing towards the origin, we must choose $0\le \alpha_1  \le 2$. 
The first slow-roll parameter is
\beq
\epsilon 
\simeq {M_{\rm pl}^2\over 2} \left ({ V_{1,\phi}\over V_1} \right)^2
\simeq 
{M_{\rm pl}^2  (2-\alpha_1)^2\over 2\phi^2} 
\, ,
\eeq
where the last expression holds for $\phi \gg M$ and $\alpha_1 < 2$. 
The number of $e$-folds of inflation is easily computed as
\beq
N = \int H \, dt 
\simeq 
{1\over M_{\rm pl}^2 }\int {V_1\over V_{1,\phi} }d\phi
\simeq {(\phi /M_{\rm pl})^2 \over (4-2\alpha_1)}\,,
\label{eq:V6efolds}
\eeq
where the integral can be performed analytically for $\phi \gg M$ and $\alpha_1 < 2$. 
Overall, under some assumptions, the slow-roll parameter becomes
$
\epsilon \simeq {(2-\alpha_1)/( 4N)}$,which we can use to read off the mass-scale $m$, such that the scalar power spectrum has the correct amplitude,
\beq
A_s = {H^2\over 8\pi^2 M_{\rm pl}^2 \epsilon} \simeq 
 {m^2 \over M^2} {M^3\over M_{\rm pl}^3} { N^{3/2}\over 6\sqrt{2}\pi^2}\,.
 \label{eq:AsV1}
\eeq
 Using $m=10^{-2}M$ and $M=10^{-2}M_{\rm pl}$, Eq.~\eqref{eq:AsV1} leads to $A_s={\cal O}\left (10^{-9}-10^{-10}\right )$, which is close to the observed value of $A_s\simeq 2\times 10^{-9}$.

The potential $V_2$ changes its behavior from quadratic to flat close to the scale $M$.
   The first slow roll parameter is
 \beq
 \epsilon \simeq
 \frac{2  M_{\text{pl}}^2}{\phi ^2 \left(1+ \alpha_2  \phi ^2/M^2\right)^2}\,.
  \eeq
 The total number of $e$-folds is given by
\begin{eqnarray}
\nonumber
N = \int H \, dt \simeq 
\frac{1}{8} {\phi ^2\over M_{\rm pl}^2} \left(\frac{\alpha_2  \phi ^2}{M^2}+2\right).
\label{eq:V7efolds}
\end{eqnarray}

Using the relation between the field amplitude $\phi$ and the number of $e$-folds before the end of inflation $N$, the power spectrum can be written as
\beq
A_s \simeq {m^2\over M^2}{ M^3\over M_{\rm pl}^3}\, {1\over \sqrt{\alpha_2}}{ N^{3/2}\over 3\sqrt{2}\pi^2}\,. 
\label{eq:AsV2}
\eeq
 By using $m=10^{-2}M$ and $M=10^{-2}M_{\rm pl}$, the amplitude of the power spectrum becomes again $A_s\simeq 10^{-9}$ which is of the correct order as the observed value of $A_s\simeq 2\times 10^{-9}$. 

We conclude with  the potential $V_3$. The first slow-roll parameter is
\beq
\epsilon \simeq \frac{2 {M_{\rm pl}}^2} {\phi^2 \left(  \left(\frac{\phi }{M}\right)^{{\alpha_3}}+1 \right)^2}
\,,
\eeq
and the number of $e$-folds of inflation is easily computed as
\beq
N = \int H \, dt 
\simeq
\frac{\phi ^2 \left(2+{\alpha_3}+2 \left(\frac{\phi }{M}\right)^{{\alpha_3}}\right)}{4 ({\alpha_3}+2) {M_{\rm pl}}^2} 
\, .
\label{eq:V3efolds}
\eeq
The amplitude of the scalar power spectrum is
\beq
A_s  \simeq 
{1\over 48\pi^2} \left [
2^{\alpha_3} (2+\alpha_3)^{2+2\alpha_3}
\right ]^{1\over 2+\alpha_3}
N^{\frac{2 (\alpha_3+1)}{\alpha_3+2}}
\, {m^2\over M^2}
\left ({M\over M_{\rm pl}}\right )^{2(4+\alpha_3)\over 2+\alpha_3}.
\eeq
 Using similar parameters as before, $m=10^{-2}M$ and $M=10^{-2}M_{\rm pl}$, we again get $A_s ={\cal O}(10^{9})$.

\begin{figure}[h]
\centering
\includegraphics[width=0.45\textwidth]{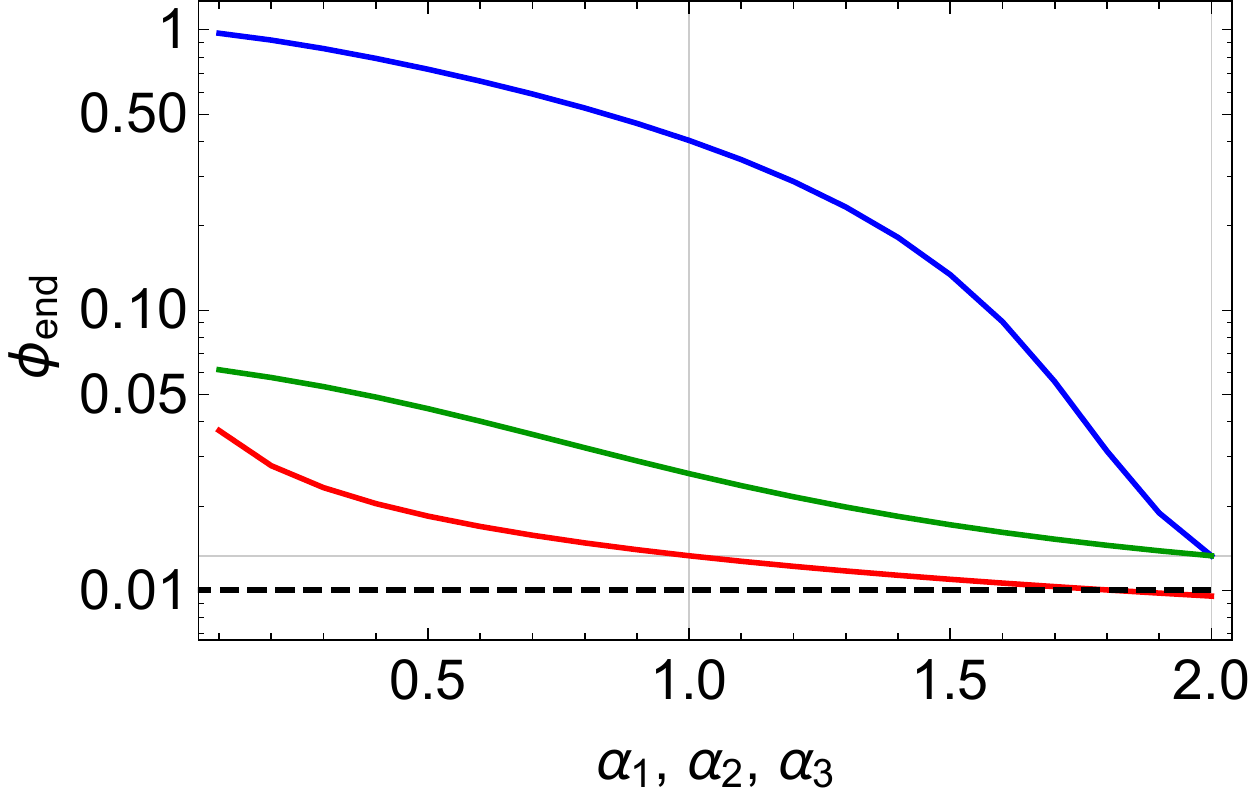}
\includegraphics[width=0.45\textwidth]{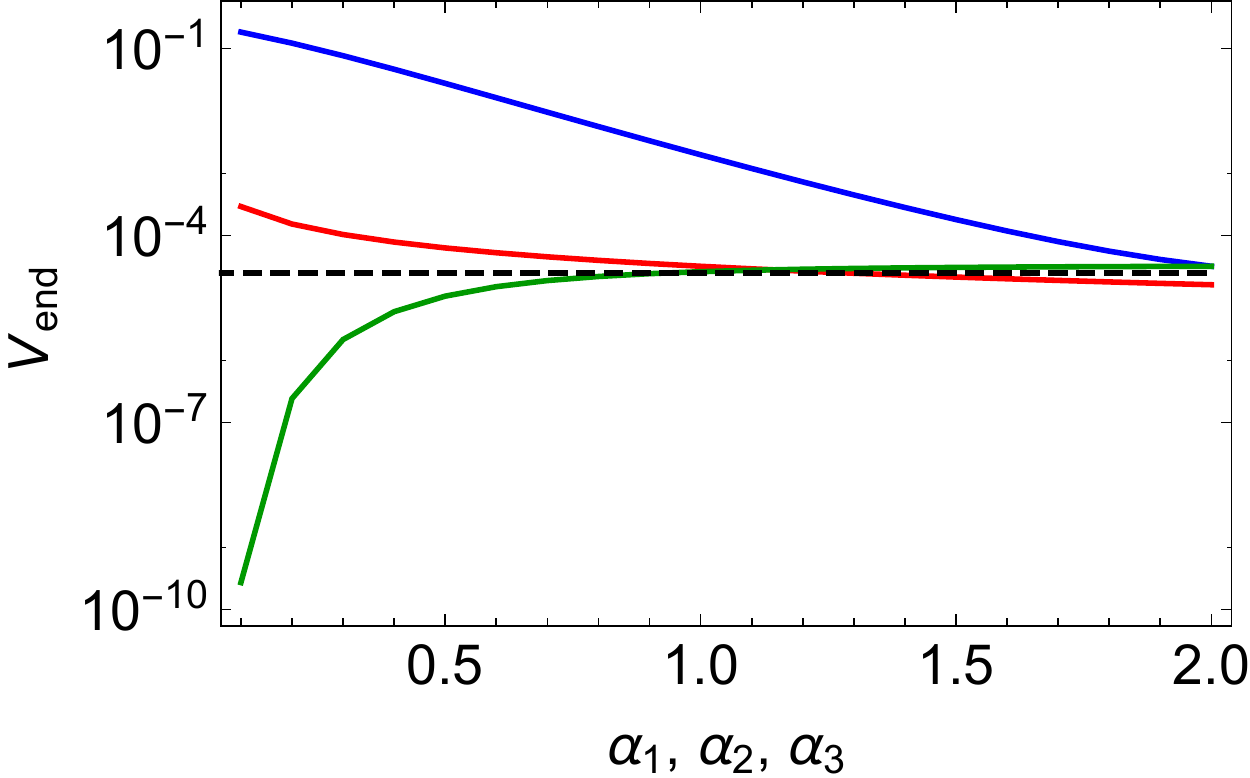}
\caption{
{\it Left:} The field value at the end of inflation $\phi_{\rm end}$, in units of $M_{\rm pl}$ as a function of $\alpha_n$. The blue, red and green curves correspond to $V_1,V_2,V_3$ respectively. The horizontal black-dashed line corresponds to the value of $M =10^{-2} M_{\rm pl}$.
{\it Right:} The value of the potential at the end of inflation in units
 of $m^2M_{\rm pl}^2$ for the same color-coding. The horizontal black-dashed line corresponds to $V=m^2 M^2/4$.
}
\label{fig:phiVend}
\end{figure}

We can now summarize the three potentials $V_1,V_2,V_3$. Fig.~\ref{fig:phiVend} shows the field and the potential at the end of inflation, as computed by numerically solving the background equation of motion.

\end{document}